\documentclass[11pt]{article}
\usepackage{multicol}      
\usepackage{lipsum}  
\usepackage{booktabs,longtable,pdflscape,array}
\usepackage{mathtools}
\usepackage{float}   
\usepackage{booktabs}
\usepackage{array}
\usepackage{tabularx}
\newcolumntype{Y}{>{\centering\arraybackslash}X}
\usepackage{booktabs}
\usepackage{array}
\usepackage{booktabs}
\usepackage[left=1.5cm,right=1.5cm]{geometry}
\usepackage[numbers,square]{natbib}
\usepackage[T1]{fontenc}
\usepackage{tikz-cd}
\usepackage[british]{babel}
\usepackage{booktabs}
\usepackage{layout}
\usepackage{amsmath}
\usepackage{amssymb,amstext, amsthm, simplewick, amsfonts}
\usepackage{framed}
\usepackage{amsfonts}
\usepackage{color} 
\usepackage{braket}
\usepackage{multicol} 
\usepackage{epsfig}
\usepackage{cancel}
\usepackage{caption}
\usepackage{epsf}
\usepackage{feynmf} 
\usepackage{frontespizio}
\usepackage{rotating}
\usepackage{subfigure}
\usepackage{pstricks}
\usepackage{type1ec}
\usepackage{lettrine}
\usepackage{bbold}
\usepackage{calligra}
\usepackage{tikz}
\usepackage{float}
\usepackage{subfigure}
\usepackage{slashed}
\usepackage{mathrsfs}
\usepackage{curve2e}
\usepackage{setspace}
\usepackage{indentfirst}
\usepackage{emptypage}
\usepackage{relsize}
\usepackage{mathrsfs}
\usepackage{stackengine}
\usepackage{calc}
\usepackage{hyperref}
\hypersetup{
	colorlinks,
	citecolor=green,
	filecolor=black,
	linkcolor=blue,
	urlcolor=black
}
\newcommand{\na}{\nabla}
\newcommand{\del}{\delta}

\newcommand{\3}[1]{C_{
		\ifthenelse{\equal{\ThreePt}{\empty}}{#1}{
			\ifthenelse{\equal{#1}{\empty}}{\ThreePt}{\ThreePt,#1}}}}
\newcommand{\redef}[1]{{C'}_{
		\ifthenelse{\equal{\ThreePt}{\empty}}{#1}{
			\ifthenelse{\equal{#1}{\empty}}{\ThreePt}{\ThreePt,#1}}}}
\newcommand{\ren}[1]{C_{
		\ifthenelse{\equal{\ThreePt}{\empty}}{#1}{
			\ifthenelse{\equal{#1}{\empty}}{\ThreePt}{\ThreePt,#1}}}}
\newcommand{\sd}[1]{D_{
		\ifthenelse{\equal{\ThreePt}{\empty}}{#1}{
			\ifthenelse{\equal{#1}{\empty}}{\ThreePt}{\ThreePt,#1}}}}

\newcommand{\G}{\Gamma}

\renewcommand{\a}{\alpha}
\newcommand{\m}{\mu}

\newcommand{\vf}{\varphi}

\newcommand{\pa}{\partial}						

\def\sq{\,\raise.5pt\hbox{$\nbox{.09}{.09}$}\,}
\def\sqb{\,\raise.5pt\hbox{$\overline{\nbox{.09}{.09}}$}\,}

\numberwithin{equation}{section} 
\makeatletter
\@addtoreset{equation}{section}
\newcommand{\bea}{\begin{eqnarray}}
\newcommand{\eea}{\end{eqnarray}}
\makeatother
\newcommand{\beqa}{\begin{eqnarray}}
	\newcommand{\eeqa}{\end{eqnarray}}

\newcommand{\nn}{\nonumber}
\let\a=\alpha   \let\b=\beta      \let\d=\delta
         
    \let\k=\kappa  \let\l=\lambda  \let\m=\mu
\let\n=\nu           \let\p=\pi

\newcommand{\bann}{\begin{eqnarray*}}
\newcommand{\eann}{\end{eqnarray*}}

\def\nbox#1#2{\vcenter{\hrule \hbox{\vrule height#2in
			\kern#1in \vrule} \hrule}}

\newcommand{\cA}{\mathcal{A}}

\newcommand{\cL}{\mathcal{L}}

\newcommand{\cS}{\mathcal{S}}

\newcommand{\bmi}{\begin{minipage}}
\newcommand{\emi}{\end{minipage}}

\newcommand{\psl}{p \! \! \!  /}
\newcommand{\qsl}{q \! \! \!  /}
\newcommand{\lsl}{l \! \! \!  /}

\let\G=\Gamma  \let\D=\Delta

\newcommand{\bs}[1]{\boldsymbol{#1}}

\newcommand{\be}{\begin{equation}}
	\newcommand{\ee}{\end{equation}}
\newcommand{\sdfrac}[2]{\mbox{\small$\displaystyle\frac{#1}{#2}$}}

\renewcommand{\Re}{\textrm{Re}}

\newcommand{\beq}{\begin{equation}}
	\newcommand{\eeq}{\end{equation}}

\newcommand{\ThreePt}{\empty}
\usepackage{accents}
\makeatletter
\newcommand{\xLine}[2][]{\ext@arrow 0359\Rightarrowfill@{#1}{#2}}
\makeatother
\xdefinecolor{darkgreen}{RGB}{102, 204, 70}
\xdefinecolor{darkblue}{RGB}{0, 0, 153}
\usepackage{amssymb}
\usepackage{pifont}
\newcommand{\bes}{\begin{subequations}}
	\newcommand{\ees}{\end{subequations}}

\usepackage{tikz-feynman}
\usetikzlibrary{arrows}
\usetikzlibrary{snakes}
\usetikzlibrary{decorations,decorations.pathmorphing,decorations.markings}

\tikzset{graviton/.style={decorate, decoration={snake}, double}}
\tikzset{gluon/.style={decorate, decoration={coil, segment length=8, aspect=1.2, amplitude=3 }}}

\vspace{0.3cm}

\begin{document}
	\begin{center}
		\vspace{1.5cm}
	\begin{center}
	\vspace{1.5cm}
	\vspace{0.5cm} 
	{\Large \bf  Anomaly-mediated Scalar Gravitational  Interactions   \\ }
\vspace{0.5cm} 
{\Large \bf  and the Coupling of Conformal Sectors \\}

\vspace{0.2cm}
	\vspace{1.3cm}

{\large \bf $^{(1,2)}$Claudio Corian\`o, $^{(1)}$Stefano Lionetti, $^{(1)}$Dario Melle and $^{(1)}$Leonardo Torcellini \\}

\vspace{0.3cm} 
	
	\vspace{0.3cm}
		\vspace{1cm}

{\it  $^{(1)}$Dipartimento di Matematica e Fisica, Universit\`{a} del Salento \\
and INFN Sezione di Lecce,Via Arnesano 73100 Lecce, Italy\\
National Center for HPC, Big Data and Quantum Computing\\}

{\it $^{(2)}$ CNR-nanotec, Lecce \\}

\vspace{0.5cm}

	\end{center}

\begin{abstract}

We investigate the anomaly-induced activation of a gauge-invariant scalar degree of freedom in General Relativity, the conformalon mode, directly at the level of \(2\to2\) scattering amplitudes. The analysis couples anomalous three-point functions of conformal sectors, involving gravitons \((TTT)\) and Abelian gauge currents \((TJJ)\), through single-graviton exchange derived from the quadratic expansion of the Einstein--Hilbert action. Unlike related treatments based on the nonlocal anomaly action, these interactions are suppressed by the Planck scale. We show that the conformalon, invariant under linearized diffeomorphisms, admits an interpretation as an effective scalar correction to scattering amplitudes, both in virtual exchange channels and in effective real-emission processes. Around flat space, this behaviour arises from anomaly-induced nonlocal massless insertions on the external graviton and photon legs of the three-point functions, sewn through the scalar component of the graviton propagator in de Donder gauge. The resulting anomaly-mediated \(4\)-point interaction reduces to contact terms, with the Planck mass setting the suppression scale. The construction consistently matches the spin decomposition of flat-space conformal Ward identities in momentum space, which determine the vertices, with the corresponding spin decomposition of the graviton propagator. In the eikonal limit, these interactions generate contact corrections to the leading logarithmic phase in impact-parameter space. We further show that anomaly-mediated \(2\to2\) graviton amplitudes associated with the virtual exchange of such modes exhibit a characteristic double-copy structure.
\end{abstract}

	\end{center}
	

\newpage
\tableofcontents

\section{Introduction}

The standard cosmological model, $\Lambda$CDM, combines a cosmological constant and cold dark matter with Einstein's General Relativity (GR), and provides an excellent description of the Cosmic Microwave Background, large-scale structure and many other observations. At the same time, persistent tensions in parameters such as the Hubble constant $H_0$, the clustering amplitude $S_8$, and the baryonic acoustic oscillations (BAO)-inferred sound horizon suggest that the standard picture may be incomplete. If these discrepancies are not entirely due to unresolved systematics, they may point either to extensions of the cosmological sector or to modifications of gravity itself. This motivates a closer examination of theoretically consistent alternatives to GR and of new observational probes capable of testing them \cite{CosmoVerseNetwork:2025alb}.

A broad class of extensions of GR introduces additional propagating degrees of freedom beyond the massless spin--2 graviton. Such fields can affect the expansion history, the growth of structure, and the properties of gravitational radiation. In particular, scalar gravitational waves may arise whenever the gravitational sector contains a dynamical scalar mode in addition to the two tensor polarizations of GR \cite{Eardley1973,Lee1974,Will1977DipoleRadiationBinaryPulsar,Gong2018,Alves2010,Hyun2019}.
In Einstein gravity, diffeomorphism invariance together with the Einstein--Hilbert action imply that only transverse and traceless tensor perturbations propagate, corresponding to the familiar plus and cross polarizations.

 Gravitational-wave astronomy has made these issues directly testable. Observations by LIGO and Virgo \cite{Aasi:2015cqb,Acernese:2014hva,Abbott:2016blz,Abbott:2016nmj,Abbott:2016pxj,Abbott:2017vtc,Abbott:2017oio} have already yielded stringent tests of GR in the strong-field regime, including bounds on the graviton mass and constraints on deviations from the post-Newtonian predictions of Einstein gravity. In this context, the polarization content of gravitational radiation provides a particularly clean probe of additional propagating modes \cite{Callister:2017ocg}.\\
 Observationally, scalar gravitational waves are constrained both by binary pulsar timing and by interferometric measurements of gravitational-wave polarizations \cite{DamourEspositoFarese1996,AntoniadisEtAl2013,ShaoEtAl2017PRX,ZhuEtAl2019J1713Symmetries,Callister:2017ocg}. Present data are compatible with purely tensorial radiation, but they still leave room for subleading scalar components. A detection of such modes would provide direct evidence for additional gravitational degrees of freedom, whereas increasingly strong null results further sharpen the empirical success of GR.\\
Kinematically, scalar waves differ sharply from tensor waves: a massless scalar produces an isotropic transverse \emph{breathing} deformation, while a massive scalar can also excite a longitudinal component. These signatures provide, at least in principle, a way to distinguish scalar radiation from the purely tensorial signal of GR.
In a generic metric theory of gravity, gravitational waves can indeed excite several independent components of the spacetime curvature compatible with wave propagation \cite{Eardley1973,Lee1974,Will1977DipoleRadiationBinaryPulsar,Gong2018,Alves2010,Hyun2019,Kausar2016,Isi2023,Capozziello2021,Shankaranarayanan2019}. Their effect on freely falling test particles is governed by the geodesic deviation equation
\begin{equation}
\frac{d^{2}\xi^{i}}{dt^{2}} = - R_{0i0j}\,\xi^{j},
\label{geodev}
\end{equation}
so that the polarization content of a plane wave can be classified in terms of the independent components of $R_{0i0j}$. The $E(2)$ classification of null momenta shows that a general metric theory can contain up to six independent polarizations \cite{Eardley1973}: two tensor modes, two vector modes, and two scalar modes. For propagation along the $z$ direction, the tensor modes are the usual $h_+$ and $h_\times$ of GR; the vector modes, often denoted $h_x$ and $h_y$, generate transverse shear distortions; and the scalar sector contains a transverse breathing mode $h_b$ together with a longitudinal mode $h_L$. GR is the special case in which only the two tensor polarizations are dynamical.

Extra degrees of freedom arise naturally once higher--derivative corrections are included. Such terms have been studied both as possible ingredients of a perturbatively renormalizable extension of gravity \cite{tHooftVeltman1974,DeserNieuwenhuizen1974,DeserVanNieuwenhuizenTsao1976,Stelle1977,Stelle1978} and as an effective framework for early-universe cosmology. In the latter context, higher-curvature interactions can generate an inflationary phase without introducing elementary scalar inflaton fields by hand, and they also provide a mechanism for its termination. This addresses the so-called graceful-exit problem: an inflationary model is physically viable only if the quasi-de Sitter accelerated expansion is not eternal, but evolves smoothly into the standard decelerated Friedmann--Robertson--Walker evolution, allowing reheating and the onset of the radiation-dominated era. A paradigmatic example is the Starobinsky model,

\begin{equation}
S = M_P^2 \int d^4x \sqrt{-g}\, f(R) + S^{(m)}(g,\psi),
\label{eq:starobinsky_action}
\end{equation}
with
\begin{equation}
f(R)=R+\frac{R^2}{M^2},
\label{eq:fR_starobinsky}
\end{equation}
where \(\psi\) collectively denotes the matter fields and \(S^{(m)}\) is their action. The Einstein--Hilbert term,
\begin{equation}
S_{EH}\equiv M_P^2 \int d^4x \sqrt{-g}\,R,
\label{eq:EH_term}
\end{equation}
is therefore supplemented by a quadratic curvature correction capable of driving inflation \cite{Starobinsky1980}. In the metric formulation, this theory can be recast, via a field redefinition and a conformal transformation to the Einstein frame, as general relativity plus an independent massive scalar degree of freedom (the scalaron). By contrast, in Palatini formulations the \(R^2\) term does not by itself introduce the same propagating scalar degree of freedom, and inflation typically requires additional scalar fields  \cite{Antoniadis:2018ywb}. Despite their phenomenological appeal, higher--derivative corrections raise well-known issues of unitarity and ghost propagation, and are best interpreted within an effective-field-theory framework. In ordinary GR this program has been proposed, for instance, in \cite{Donoghue1994GRasEFT}.

\subsection{Scalar dilaton-like couplings to the anomaly}
In the present work, however, our emphasis is different: we focus on scalar modes generated by the conformal anomaly, namely by the quantum breaking of classical conformal symmetry in matter sectors coupled to gravity.
Classically conformal matter does not source a scalar mode at tree level, but at the quantum level the trace anomaly generates an effective scalar coupling even for classically conformal fields \cite{Capper:1974ic,Duff:1977ay,Mottola:2016mpl}. In many theories this scalar couples to the trace of the energy--momentum tensor,
\begin{equation}
\mathcal{L}_{\text{int}} \sim \varphi \, T^\mu_{\ \mu},
\end{equation}
so that sources with nonvanishing trace can emit scalar radiation. In the anomaly-induced case, by contrast, the scalar degree of freedom is more naturally interpreted as an effective excitation associated with the intrinsically nonlocal structure \(R\,\Box^{-1}\), which singles out and activates a particular scalar component of the metric. In this respect it is conceptually distinct from an ordinary local propagating field. We recall that a genuine asymptotic mode is associated with a field which, in the absence of external sources, obeys an independent hyperbolic equation of motion and is therefore able to propagate freely to infinity. Our discussion is precisely aimed at clarifying this distinction and at analyzing the subtle behaviour of the anomaly-triggered scalar excitation generated by the conformal anomaly.\\
This specific coupling emerges in perturbative analyses of conformal matter coupled to gravity through anomalous vertices such as $TJJ$ and $TTT$, where $T$ denotes the stress energy tensor and $J$ are Abelian currents. The analysis can be extended to non-Abelian currents but it will be discussed elsewhere. \\
For the \(TJJ\) vertex, first analyzed in free-field theories such as QED and QCD \cite{Giannotti:2008cv,Armillis:2009pq,Armillis:2010qk}, the anomaly manifests as a single massless exchange (a massless pole) in the conformal limit, when all the fermion masses are set to vanish. The residue of this interaction is nonzero on the light cone and affects, in our construction, both conformal sectors. The role of linearized gravity is to transfer this interaction.\\
 The anomaly-induced action  can be recast in a symmetric form involving a single dynamical scalar field, the conformalon $\varphi$ by the inclusion of a Weyl invariant term \cite{Mottola:2016mpl}, which is allowed by the solution. From this perspective, anomaly-induced gravity may be regarded as a particular realization of Sakharov's induced-gravity idea \cite{Sakharov1967,Visser:2002ew}, in which the scalar sector emerges from the nonlocal effective action generated by the anomaly \cite{Mottola:2016mpl,Mottola:2010gp}. 
\\
While the framework discussed in \cite{Mottola:2016mpl} is based on the nonlocal anomaly-induced action with gravity treated as a background field, our analysis is carried out directly within the dynamical framework of GR. The discussion is largely formal, since in our setting the Planck-scale suppression of anomaly-mediated effective interactions is entirely natural within a perturbative treatment. Our aim is therefore not immediate phenomenological applicability, but rather a detailed investigation of how trace-anomaly contributions to three-point functions interface with tree-level graviton exchange in quantum gravity.

\subsection{The activation of extra modes in gravity}
A static source with $T\neq 0$ can generate only a near-zone scalar field; genuine radiation requires time-dependent scalar multipoles, $\dot{M}_\phi\neq 0$, $\ddot{\mathbf D}_\phi\neq 0$, $\dddot{Q}^{\,\phi}_{ij}\neq 0$, $\ldots$. In ordinary scalar--tensor systems, dipole radiation may dominate over the tensor quadrupole channel \cite{Will2014LRR,Eardley1975}, although it vanishes when the scalar charges of the two bodies are equal. By contrast, the anomaly-induced scalar ($\varphi$) is not sourced by body-dependent scalar charges but by the anomaly functional itself. In the local form of the anomaly action,
\begin{equation}
S_{\rm anom} \sim 
\int d^4x \sqrt{-g}\,
\varphi \left(E-\frac{2}{3}\Box R\right)
+\ldots ,
\end{equation}
the field $\varphi$ couples universally to curvature invariants, and its equation of motion is sourced directly by the terms appearing in the trace anomaly \cite{Riegert:1984kt,Fradkin:1983tg,Mottola:2006ew,Deser:1976yx}. The corresponding scalar excitation is therefore tied to metric perturbations and spacetime curvature rather than to independent scalar charges carried by compact objects.\\
 For these reasons, more broadly, anomalies and the breaking of conformal symmetry have also been invoked in early-universe physics, including inflationary dynamics, matter-antimatter asymmetry, and alternative cosmological scenarios \cite{Starobinsky1980,Mavromatos:2024szb,Basilakos:2019acj,Alexander:2004us,McFaddenSkenderis2010,McFaddenSkenderis2011,HinterbichlerKhoury2012,Rubakov2014,HinterbichlerKhouryLevyMatas2013,BarsSteinhardtTurok2014,BarsSteinhardtTurok2013,Shaposhnikov:2018xkv,Shaposhnikov:2022zhj}. They can be  envisioned as transient phenomena, before that other sources of symmetry breakings set in, through other Higgs-like scalars. \\
The activation of additional degrees of freedom, which in this case may become dynamical, can only be discussed at a qualitative level within a quantum-gravity setting, given the well-known limitations of ordinary quantum-field-theoretic methods even in the analysis of the simplest curved spacetimes.\\
In effective field theory, scalar interactions can modify amplitudes and produce infrared enhancements even when the ordinary tensor contribution from the metric is suppressed. This is especially relevant in processes controlled by anomalous couplings, where a scalar channel can dominate the low-energy behavior despite being absent in classical Einstein gravity.\\
 We recall that infrared corrections to GR of the type discussed in \cite{Mottola:2016mpl} are induced by conformal quantum sectors, while the gravitational dynamics itself remains classical, as already emphasized. In this context, the anomaly-induced action plays a central role, since it resums these effects into an effective description and is related to stress-energy-tensor correlators with an arbitrary number of insertions, generated after integrating out the conformal matter sector.

Quantum effects related to the anomaly are described by effective actions which are characterised by the presence of specific invariants of mass dimension four.   
Quadratic invariants such as $R^2$, the Gauss--Bonnet density $E_4$, and the Weyl invariant $C^2$ arise generically in effective actions and as counterterms in the renormalization of conformal quantum field theories coupled to gravity \cite{tHooft:1974toh,Duff:1993wm,Capper:1975ig,Buchbinder:1992rb,Birrell:1982ix}. In four dimensions, $E_4$ is topological at the classical level, but it reappears naturally in the anomaly-induced effective action \cite{Coriano:2021nvn,Coriano:2020ees}. Within this framework, the anomaly-associated scalar mode has a special status and should be distinguished from the more familiar scalar degree of freedom encountered, for instance, in $f(R)$ gravity.\\
Its simplest realization occurs in conformal free-field theories such as QED and QCD coupled to external gravity, where the relevant anomalous interaction is encoded in the $TJJ$ and $TTT$ correlators, for example, that will be the subject of our investigation \cite{Coriano:2011ti,Giannotti:2008cv}.  
Related parity-odd correlators, that have been studied extensively \cite{Bonora:2015nqa,Armillis:2010pa,Abdallah:2023cdw,Bastianelli:2019zrq,Larue:2023qxw} may also play a role in early universe cosmology, enhancing the generation of chiral currents and generating possible coupling to gravitational waves \cite{Coriano:2023cvf,delRio:2021bnl}.  The conformal anomaly sector may also provide a bridge to dark conformal sectors coupled to ordinary matter only through gravity, a point that we will also address in our analysis. \\ 
In this work, the nonlocal conformal-anomaly action enters only in the analysis of \(2\to2\) amplitudes with four external gravitons, where it determines the structure of the two gravitationally coupled \(TTT\) vertices. In this case, the resulting amplitude exhibits an interesting double-copy-like structure, expressed in terms of local \(F^2\)-type interactions.

\subsection{Content of this work}
The aim of this work is to clarify the role of scalar contributions in virtual gravitational exchange and to distinguish ordinary gauge-dependent scalar pieces from the genuinely anomaly-induced scalar channel. This issue is especially subtle in gravity, since tree-level amplitudes in covariant gauges may contain spin--0 exchanges that do not correspond to physical propagating modes, while the trace anomaly introduces a nonlocal scalar contribution of a different origin.\\
The paper is organized as follows. In Sections 2 and 3 we review induced gravity in the presence of visible or dark conformal sectors and summarize the structure of the nonlocal anomaly action \cite{Sakharov1967,Visser:2002ew,Mottola:2016mpl,Mottola:2010gp}. Section 4 is devoted to the \(TJJ\) correlator, both in its general conformal form and in its perturbative realization \cite{Bzowski:2013sza,Armillis:2009pq}, with special emphasis on its anomaly pole and on the associated spin decomposition. In Sections 5 and 6 we move to four-point functions and analyze the graviton propagator and tree-level graviton exchange in the de~Donder gauge, isolating the spin--0 contribution. Section 7 then discusses scalar gravitational scattering in the case of explicit, non-anomalous breaking, showing the absence of a genuine scalar exchange in ordinary four-point functions.\\
The central part of the paper begins in Section 8, where we study the Fierz--Pauli Lagrangian in the presence of the anomaly and show how the anomaly activates a distinguished scalar response channel while leaving the TT and vector sectors unaffected. This analysis is closely related to the role played by anomaly poles in the solution of anomalous conformal Ward identities \cite{Coriano:2023hts,Coriano:2023cvf,Coriano:2023gxa}. Section 9 applies this result to the conformal limit of the photon four-point function, where the anomaly-induced interaction can be represented in terms of an effective scalar exchange. Section 10 extends the analysis to the massive fermion case and discusses how explicit mass corrections modify the anomaly form factor and its discontinuity structure.\\
In Sections 11 and 12 we turn to graviton scattering. We first analyze anomaly corrections to graviton--graviton scattering at tree level, using the anomalous part of the \(TTT\) vertex as studied both perturbatively and through the anomaly action \cite{Coriano:2017mux}, and then discuss the corresponding double-copy-like structure and scalar emission channels. Section 13 is devoted to the anomaly correction to the eikonal phase, while Section 14 briefly comments on the extension of the analysis to dark sectors. Our conclusions are presented in Section 15.

\section{Sakharov's induced gravity for conformal and visible/dark sectors}
The role of conformal symmetry and its anomalous breaking cannot be understated. A spacetime which is conformal before inflation allows a wider evolution compared to an ordinary De Sitter space, due to the presence of a dynamic conformal factor.\\
The integration of quantum matter fields, either visible or dark, generically induces an effective gravitational action. 
If the degrees of freedom that are integrated out belong to a dark sector, gravity mediates the interaction between the visible and dark sectors through corrections to the Einstein--Hilbert (EH) action. 
When the integrated-out matter sector is non-conformal, the induced corrections typically include local terms—possibly reproducing the EH action itself—together with higher-curvature operators accompanied by dimensionful coefficients. 
This mechanism realizes Sakharov’s idea of induced gravity~\cite{Sakharov1967} (see~\cite{Visser:2002ew} for a review). \\
At the opposite extreme, when conformal matter (visible or dark) is integrated out, the dominant contribution originates from the breaking of Weyl invariance through the trace anomaly, inducing a scaleless and intrinsically nonlocal effective action $\mathcal{S}_{anom}$, defined in \eqref{Snonlsq}. 
Around flat spacetime, this action can be formally expanded in terms of nonlocal structures of the form $\Box^{-1} R\, $, where $R$ is the scalar curvature \cite{Giannotti:2008cv,Coriano:2022ftl}. As we are going to clarify, this interaction appearing on each external graviton leg 
of a correlator, identifies a transverse projector $\pi^{\mu \nu}$, function of the graviton momentum, which is sandwiched by the scalar component $P^{(0\text{-}s)}_{\mu\nu,\alpha\beta}$ of the graviton propagator in the De Donder gauge. The same projector 
\beq
P^{(0\text{-}s)}_{\mu\nu,\alpha\beta}
=
\frac13\,\pi_{\mu\nu}\pi_{\alpha\beta},
\eeq
acting on the covariant decomposition of the metric 
will extract the only gauge invariant scalar component of the metric, denoted as $\equiv h - \Box w$ in \cite{Mottola:2016mpl}, and denoted as $\Phi_{cov}\equiv \varphi$ in our notations, extracted in a covariant decomposition of its fluctuations around flat space, that plays 
a key role in the analysis of the anomaly induced action.  
$\mathcal{S}_{anom}$ defines the backreaction on the metric, bringing in additional contributions to the gravitational dynamics, supplementing the local EH term 
\beq
\mathcal{S}=\mathcal{S}_{EH} +\mathcal{S}_{anom}.
\eeq
These aspects will be discussed in detail with reference to a diagrammatic expansion presented in the sections below that should, as we hope, clarify the way in which anomaly interactions are mediated 
in GR at quadratic level.
If metric fluctuations are quantized and normalized as fields of mass dimension one,
\begin{equation}
g_{\mu\nu} = \eta_{\mu\nu} + \kappa\, h_{\mu\nu} \, ,
\end{equation}
with $\kappa^{2} = 16\pi G_{N}$, the resulting gravitational interaction mediating the anomaly is mediated by $\Phi_{cov}$ and strongly suppressed below the Planck scale.\\
The two sectors are interpolated by the ordinary two-point function of gravitons as determined from the EH action. 
This analysis, however, is clearly not exhaustive. 
 If the full gravitational action is assumed to combine the Einstein--Hilbert (EH) action with the anomaly--induced (Riegert) action, one may write
\begin{equation}
\label{total}
\mathcal{S}_g \equiv \mathcal{S}_{\mathrm{EH}} + \mathcal{S}_{\mathrm{anom}} \, .
\end{equation}
The anomaly--induced contribution $\mathcal{S}_{\mathrm{anom}}$ admits a local formulation through the introduction of an auxiliary scalar degree of freedom, the conformalon $\varphi$ \cite{Mottola:2016mpl}. In this representation the interaction generated by $\mathcal{S}_{\mathrm{anom}}$ is scaleless and resums infrared contributions that are not contained in the Einstein--Hilbert sector. The presence of both terms in \eqref{total} therefore corresponds to two distinct dynamical regimes. The first is governed by $\mathcal{S}_{\mathrm{EH}}$, whose effects are suppressed by the Planck scale but become relevant as the characteristic energy approaches that scale. The second originates from $\mathcal{S}_{\mathrm{anom}}$, which encodes the resummation of infrared effects and can become important in backgrounds with sufficiently large curvature.

In the present analysis we restrict our attention to the dynamics generated by $\mathcal{S}_{\mathrm{EH}}$. This sector can be consistently studied in the vicinity of flat space within a framework entirely based on free--field realizations of the corresponding interactions. In particular, the four--point functions constructed in this work arise from the coupling of ordinary conformal sectors---characterized by well--defined three--point correlators, such as those appearing in QED---with the interactions mediated by $\mathcal{S}_{\mathrm{EH}}$ at leading order in the gravitational coupling, as illustrated in Fig.~1. In this regime the relevant dynamics is described by the (massless) Fierz-Pauli action obtained by expanding the Einstein--Hilbert action around flat spacetime.
 The gravitational coupling of vertices affected by the conformal anomaly allows for a clean disentanglement of the standard spin--2 graviton exchange from the anomaly-induced scalar channel, which turns physical.  
 
\begin{figure}[t]
\begin{center}
\includegraphics[scale=1,angle=0]{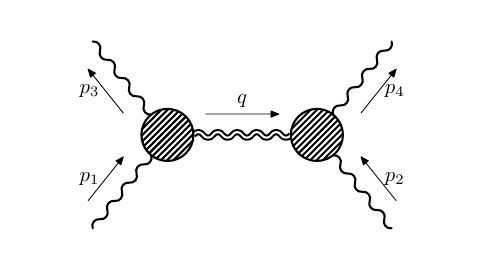}
\includegraphics[scale=1,angle=0]{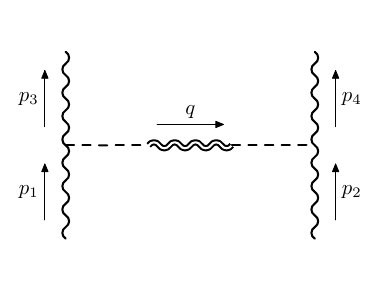}
\caption{$t$-channel scattering of two spin-1 states $p_1+p_2 \to p_3 + p_4$, mediated by the exchange of a single graviton (left). The two conformal sectors, represented by the darker blobs, may both belong to the visible sector or alternatively to one visible and one dark sector. The anomaly-mediated interaction between the two conformal sectors, responsible for the exchange of a scalar gravitational component, is shown on the right. This contribution arises from the combination of the external $R\Box^{-1}$ mixing with the spin-0 component of the graviton propagator.}
\end{center}
\label{fig1x}
\end{figure}

\section{The anomaly and the nonlocal action: review} 
In order to set the stage for our discussion in a consistent manner, we briefly review the construction of anomaly--induced actions as solutions of a variational problem, before turning to their simplest realizations at the level of three--point functions.\\
The fundamental identity characterising the anomaly is that of the one-point function, evaluated in a generic metric background, which takes the general form
\be
 g_{\m\n} \big\langle T^{\m\n}(x)\big\rangle_g \equiv 
\mathcal{A}(x)\ee
with
\be
\mathcal{A}(x)= T_{cl} +  b\, C^2 +
 b' \big(E - \tfrac{2}{3}\sq R\big) + b'' \sq R + \sum_i \beta_i \,\cL_i .
\label{trace1}
\ee

The symmetric energy-momentum-stress tensor $T^{\m\n}(x)$ coupling to the gravitational metric field 
$g_{\m\n}(x)$ is defined by the metric variation
\be
T^{\m\n}(x)\equiv \frac{2}{\sqrt{-g(x)}}\frac{\ \del{\rm S_{cl}}[\bar\Phi, g] }{\del g_{\m\n}(x)}
\ee
where ${\rm S_{cl}} [\bar\Phi, g]$ is the classical action of the fields, denoted generically by $\Phi = \{\Phi_i\}$. Here ${\rm S_{cl}}[\bar\Phi, g]$ is viewed 
as a functional also of the general curved spacetime background metric $g$. From a 
phenomenological perspective, $\Phi$ can be either visible or dark. If visible, in the sense that they interact ordinarily with fields of the Standard Model, their interactions with gravity will be taken as mass suppressed, but this assumption is not the only one. For example, if the interaction is present before the inflationary epoch, they may be simply the result of a sector coupled to gravity via a portal.

Here $T_{cl}$ is the trace expected from the non-invariance of the classical action itself
\beq
T_{cl} = \frac{2\,g_{\m\n}(x)}{\sqrt{-g(x)}}\left\langle\frac{\ \delta{\rm S_{cl}}[\bar \Phi, g] }{\delta g_{\m\n}(x)}\right\rangle_{\!g}
\label{cltrace}
\eeq
absent any anomalies, and the additional finite terms in (\ref{trace1}) are the quantum anomalous contributions. 
These are given in terms of the dimension-four curvature invariants 
\bes
\bea
&&E \equiv\, ^*\hskip-.5mm R_{\m\a\n\b}\,^*\hskip-.5mm R^{\m\a\n\b} =
R_{\m\a\n\b}R^{\m\a\n\b}-4R_{\m\n}R^{\m\n} + R^2 \\
&&C^2\equiv C_{\m\a\n\b}C^{\m\a\n\b} = R_{\m\a\n\b}R^{\m\a\n\b} -2 R_{\m\n}R^{\m\n}  + \tfrac{1}{3}R^2 \label{EFdef}
\eea\ees
which are the Euler-Gauss-Bonnet invariant and square of the Weyl conformal tensor respectively (with $R_{\m\a\n\b}$ the Riemann curvature tensor,
$^*\hskip-.5mm R_{\m\a\n\b}$ its dual, together with the Ricci tensor $R_{\mu\nu}$ and Ricci scalar $R$).
\(\mathcal{L}_i\) include, in general, several background (classical) operators 
\begin{equation}
\mathcal{L}_i \in
\left\{
F_{\mu\nu}F^{\mu\nu},
\;
R\,\phi^2,
\;
(\nabla \phi)^2,
\;
\phi^4,
\;
\dots
\right\},
\label{eq:Li-examples}
\end{equation}
as well as Yukawa terms and gauge-scalar mixing terms, 
depending on the matter content of the theory. They arise when the quantum theory is embedded in more general backgrounds, with extra classical gauge and diff-invariant terms of dimension-four other than pure curvature invariants. Their coefficients \(\beta_i\) are the corresponding beta functions of the renormalized couplings. Thus, $\beta_i \mathcal{L}_i$
encodes running couplings and explicit scale breaking, rather than new gravitational dynamics.\\
The coefficients $b$, $b'$, and $\beta_i$ depend on the number of light fields in the theory and contain no suppression by the Planck scale (see \cite{Coriano:2020ees} for a review), while $b''$ is 
prescription-dependent. As already mentioned above, unlike local curvature corrections arising in conventional effective field theory expansions, the anomaly gives rise to a non-local effective action with logarithmic scaling at large distances, making it \emph{marginally relevant} in the infrared \cite{Mottola:2016mpl}. Its effects are cumulative on macroscopic scales, even in spacetimes of small curvature. In the presence of a dark sector, gravitationally coupled to a visible one, $b$, $b'$, get separate contribution from the multiplicities of massless conformal fields present in both sectors. \\
 One expects such contributions to manifest themselves in the form of scalar gravitational waves directly generated by the anomaly, which could have been present in the early Universe at energy scales that may be regarded as pre--inflationary and could therefore leave an imprint on the stochastic gravitational--wave background. 
  \subsection{The nonlocal action}
No local covariant action exists whose variation reproduces \eqref{trace1}. The anomaly can be integrated to construct a nonlocal effective action, a procedure that relies on the existence of conformally covariant operators, most notably the Paneitz operator \cite{Paneitz2008}, whose Green’s functions exhibit the characteristic nonlocal nature of conformal symmetry breaking by the anomaly. 
The anomaly contribution can be generated by the nonlocal Riegert action, which in its compact form reads \cite{1984PhLB..134...56R} \cite{AntoniadisMottola1992}

\bea
&&\hspace{-8mm}\cS_{\rm anom}^{^{NL}}[g] =\sdfrac {b'}{8\,}\!\int \!d^4x\,\sqrt{-g_x}\int\! d^4x'\,\sqrt{-g_{x'}}\, 
\bigg[\big(E - \tfrac{2}{3}\sq R\big) +  \sdfrac{b}{\,b'}\,C^2\bigg]_{x}\!
D_4(x,x')\bigg[\big(E - \tfrac{2}{3}\sq R\big) +  \sdfrac{b}{\,b'}\,C^2\bigg]_{x'}\nn\\
&&\hspace{2cm} = \sdfrac {1}{8b'}\int d^4x \int d^4x'\, \cA(x)\, D_4(x,x')\, \cA(x')
\label{Snonlsq}
\eea
where
\begin{equation}
\Delta_{4}
= 
\Box^{2} 
+ 2 R^{\mu\nu}\nabla_{\mu}\nabla_{\nu}
-\frac{2}{3} R\, \Box
+\frac{1}{3} (\nabla^{\mu} R)\nabla_{\mu}.
\label{eq:PaneitzOp}
\end{equation}
The inverse Paneitz operator \(G_{4}(x)\equiv \Delta_{4}^{-1}(x)\) \cite{Paneitz2008} renders the action \eqref{Snonlsq} intrinsically nonlocal and is responsible for the massless scalar pole characterizing anomalous correlators. The combination $1/4(E-2/3 \Box R)$ defines the first realization of the 
$Q$-curvature operators whose siblings in higher dimensions have been investigated in the context of conformal geometry. 
The anomaly interaction mediated by this action has been explicitly verified at the level of three-point functions, in particular for the \(TTT\) correlator \cite{Coriano:2017mux} against the perturbative description. A comparison with perturbative flat-space results for four-point functions, such as \(TTJJ\), however indicates that additional Weyl-invariant terms are required in order to ensure consistency with the hierarchical conformal Ward identities (CWIs) \cite{Coriano:2022jkn}.

Since the analysis of conformal interactions on curved backgrounds is technically challenging—owing to the absence of a conventional S-matrix— here we focus instead on flat spacetime, where both perturbative and non-perturbative formulations are available and the conformal constraints are fully under control \cite{Bzowski:2013sza,Bzowski:2018fql,Bzowski:2017poo,Coriano:2017mux,Coriano:2018bbe,Coriano:2023gxa,Coriano:2023hts}. This approach allows one to trace, in a controlled way, the activation of a virtual scalar mode in gravitational interactions induced by the trace anomaly. The construction is rigorously defined, since each intermediate step in the derivation of the four–point functions can be reproduced within a standard perturbative framework. In particular, the structure of the relevant conformal three–point functions in momentum space can be reconstructed explicitly using free field theories. This ensures that the correlators entering the analysis admit a fully perturbative realization, while still capturing the anomaly–induced scalar dynamics. Therefore, our analysis combines three-point functions obtained from the solution of the CWIs with single-graviton exchange in the \(s\), \(t\), and \(u\) channels, revealing anomaly-driven corrections that excite a scalar state and modify the standard spin-2 exchange. This analysis will be presented in Section~11, where we first briefly review the derivation of the anomaly-induced part of the $TTT$ correlator and then turn to the computation of the four-graviton amplitude in the presence of anomaly-mediated interactions.

\section{The $TJJ$ correlator and anomaly mediation}

In this section we start investigating the role of trace anomaly interactions and their coupling via a gravitational exchange. The construction of the coupling of two conformal sectors starts from each 3-point vertex, assuming that this is determined from conformal invariance. We start from the $TJJ$ correlator, which allows the description of the photon-photon-graviton vertex in flat space. The vertex will be used in the later sections in the study of 4-point functions.

The structure of these correlators is fully constrained by conformal symmetry and can be obtained by solving the corresponding conformal Ward identities. The resulting solutions are expressed in terms of Appell hypergeometric functions \cite{Coriano:2013jba}, which enter the independent form factors appearing in the tensorial decomposition of the correlator in the form of 3K integrals \cite{Bzowski:2013sza}. These are parametric integrals of three Bessel functions. The general solution is determined up to a finite set of constants and is most naturally analyzed in momentum space. In this representation, the dynamics of the anomaly interaction are encoded in massless exchanges \cite{Giannotti:2008cv,Coriano:2025fom,Armillis:2009pq}, which exhibit a nonvanishing residue on the light-cone. In the present analysis we restrict our attention to the Abelian case, relevant for spin-1 gauge fields, whether belonging to the visible or to a dark sector. The conformal constraints take the form of a dilatation Ward identity

\begin{equation}
0=\left[\sum_{j=1}^3\Delta_j-(n-1)d-\sum_{j=1}^2\,p_j^\a\sdfrac{\partial}{\partial p_j^\alpha}\right]\braket{{T^{\mu_1\nu_1}(p_1)\,J^{\m_2}(p_2)\,J^{\mu_3}(\bar p_3)}},
\label{DilatationTJJ}
\end{equation}
where $\Delta_j$ are the scaling dimensions of $T$ ($\Delta_1=4$)  and $J$ $(\Delta_2=\Delta_3=3)$ as for conserved currents.\\
Other constraints are the Lorentz ones 
\begin{align}
0&=\sum_{j=1}^{2}\left[p_j^\nu\sdfrac{\partial}{\partial p_{j\mu}}-p_j^{\mu}\sdfrac{\partial}{\partial p_{j\nu}}\right]\braket{{T^{\mu_1\nu_1}(p_1)\,J^{\m_2}(p_2)\,J^{\mu_3}(\bar p_3)}}\notag\\
&\qquad+2\left(\delta^{\nu}_{\a_1}\d^{\mu(\mu_1}-\delta^{\mu}_{\alpha_1}\delta^{\nu(\mu_1}\right)\braket{{T^{\nu_1)\alpha_1}(p_1)\,J^{\m_2}(p_2)\,J^{\mu_3}(\bar p_3)}}\notag\\
&\qquad+\left(\delta^{\nu}_{\a_2}\d^{\mu\mu_2}-\delta^{\mu}_{\alpha_2}\delta^{\nu\mu_2}\right)\braket{{T^{\mu_1\nu_1}(p_1)\,J^{\alpha_2}(p_2)\,J^{\mu_3}(\bar p_3)}}\notag\\
&\qquad+\left(\delta^{\nu}_{\a_3}\d^{\mu\mu_3}-\delta^{\mu}_{\alpha_3}\delta^{\nu\mu_3}\right)\braket{{T^{\mu_1\nu_1}(p_1)\,J^{\mu_2}(p_2)\,J^{\alpha_3}(\bar p_3)}}\label{RotationTJJ},
\end{align}
where, in our convention 
\begin{equation}
\d^{\n(\m_1}\,\braket{{T^{\nu_1)\alpha_1}\,J^{\m_2}\,J^{\mu_3}}}\equiv\sdfrac{1}{2}\big(\d^{\n\m_1}\,\braket{{T^{\nu_1\alpha_1}\,J^{\m_2}\,J^{\mu_3}}}+\d^{\n\n_1}\,\braket{{T^{\mu_1\alpha_1}\,J^{\m_2}\,J^{\mu_3}}}\big),
\end{equation}
 and the special conformal Ward identities
\begin{align}
0&=\sum_{j=1}^{2}\left[2(\Delta_j-d)\sdfrac{\partial}{\partial p_j^\k}-2p_j^\a\sdfrac{\partial}{\partial p_j^\a}\sdfrac{\partial}{\partial p_j^\k}+(p_j)_\k\sdfrac{\partial}{\partial p_j^\a}\sdfrac{\partial}{\partial p_{j\a}}\right]\braket{{T^{\mu_1\nu_1}(p_1)\,J^{\m_2}(p_2)\,J^{\mu_3}(\bar p_3)}}\notag\\
&\qquad+4\left(\d^{\k(\mu_1}\sdfrac{\partial}{\partial p_1^{\a_1}}-\delta^{\k}_{\alpha_1}\delta^{\l(\mu_1}\sdfrac{\partial}{\partial p_1^\l}\right)\braket{{T^{\nu_1)\alpha_1}(p_1)\,J^{\m_2}(p_2)\,J^{\mu_3}(\bar p_3)}}\notag\\
&\qquad+2\left(\d^{\k\mu_2}\sdfrac{\partial}{\partial p_2^{\a_2}}-\delta^{\k}_{\alpha_2}\delta^{\l\mu_2}\sdfrac{\partial}{\partial p_2^\l}\right)\braket{T^{\mu_1\nu_1}(p_1)\,J^{\alpha_2}(p_2)\,J^{\mu_3}(\bar p_3)}.
\label{SCWTJJ}
\end{align} 
We have denoted with $\bar p_3$ the momentum of one of the current, taken as a function of the other two incoming momenta $p_1$ and $p_2$ as  $\bar p_3=-p_1 -p_2$. Momentum differentiation should be performed only with respect to $p_1$ and $p_2$ and implicitly on $\bar p_3$ as a function of the first two. 
The former constraints are accompanied by conservation Ward identities
\begin{align}
\label{tr}
p_{1\n_1}\braket{T^{\mu_1\nu_1}(p_1)\,J^{\m_2}(p_2)\,J^{\m_3}(p_3)}_q&=4\,\big[\d^{\m_1\m_2}p_{2\l}\braket{J^\l(p_1+p_2)\,J^{\m_3}(p_3)}_q-p_2^{\m_1}\braket{J^{\m_2}(p_1+p_2)\,J^{\m_3}(p_3)}_q\big]\notag\\
&+4\,\big[\d^{\m_1\m_3}p_{3\l}\braket{J^\l(p_1+p_3)\,J^{\m_2}(p_2)}_q-p_3^{\m_1}\braket{J^{\m_3}_q(p_1+p_3)\,J^{\m_2}(p_2)}_q\big].
\end{align}
with the 2-point function of two conserved vector currents $J_i$ $(i=2,3)$  in any conformal field theory in $d$ dimensions given by 
\beqa
\langle J_2^\alpha(p)J_3^\beta(-p) \rangle =\delta_{\Delta_2\, \Delta_3}\left(c_{123} \Gamma_J \right)\pi^{\alpha\beta}(p) (p^2)^{\Delta_2-d/2},
\qquad \Gamma_J=\frac{\pi^{d/2}}{ 4^{\Delta_2 -d/2}}\frac{\Gamma(d/2-\Delta_2)}{\Gamma(\Delta_2)},
\eeqa
with $c_{123}$ an overall constant and $\Delta_2=d-1$ are the ordinary scaling dimensions of conserved currents. In the coupling of the two conformal sectors, once we take the on-shell limit of the spin-1 external states, then one finds that the spin-1 part of the propagator decouples in the amplitude. In other terms the conservation Ward identity can be taken as a homogeneous equation. 
Concerning the $JJ$, this is affected by a divergence that can be identified, for example, just by a simple free field theory analysis with $n_f$ fermions and requires the introduction of the counterterm
\beq
\mathcal{L}_{\textrm{q\, count}}=-\frac{1}{(d-4)}\frac{g_s^2}{16 \pi^2}\frac{2}{3} n_f
F^{a\, \mu\nu}F^{a}_{\mu\nu}.
\eeq
$n_f$ denotes the number of flavours running in the loop of the $JJ$ and $d$ the spacetime dimensions. Free field theory parallels in every single step the reconstruction of this correlator, otherwise performed by the general approach developed in \cite{Bzowski:2013sza}.
Indeed, in the case of the $TJJ$, using a free-field theory realization with $n_f$ fermion flavours, the general solution of the conformal Ward identities in the Abelian case takes the form 
\begin{align}
\!\big\langle
T_{\mu_1\nu_1}(p_1)\,
J_{\mu_2}(p_2)\,
J_{\mu_3}(p_3)
\big\rangle
&=
\langle
t_{\mu_1\nu_1}(p_1)\,
j_{\mu_2}(p_2)\,
j_{\mu_3}(p_3)
\rangle
\nonumber\\
&\quad
+
2\,\mathscr{T}_{\mu_1\nu_1}{}^{\alpha}(p_1)
\Big[
\delta^{\mu_3}{}_{[\alpha}\,p_{3\beta]}\,
\langle
J_{\mu_2}(p_2)\,
J^{\beta}(-p_2)
\rangle
\nonumber\\
&\qquad\qquad\qquad
+
\delta^{\mu_2}{}_{[\alpha}\,p_{2\beta]}\,
\big\langle
J_{\mu_3}(p_3)\,
J^{\beta}(-p_3)
\big\rangle\!
\Big]
\nonumber\\
&\quad
+
\frac{1}{d-1}\,
\pi_{\mu_1\nu_1}(p_1)\,
\mathcal A^{\mu_2\mu_3},
\label{eq:TJJreconstructed2}
\end{align}
with
\begin{equation} \label{a:T}
\mathscr{T}_{\mu\nu \alpha} (\bs{p}) = \frac{1}{p^2} \left[ 2 p_{(\mu} \delta_{\nu)\alpha} - \frac{p_\alpha}{d-1} \left( \delta_{\mu\nu} + (d-2) \frac{p_\mu p_\nu}{p^2} \right) \right]
\end{equation}	
and
\[\label{Idecomp}
\delta_{\mu(\alpha}\delta_{\beta)\nu}=\Pi_{\mu\nu\alpha\beta}(\bs{p})+\mathscr{T}_{\mu\nu (\alpha}(\bs{p})\,p_{\beta)}+\frac{1}{d-1}\pi_{\mu\nu}(\bs{p})\delta_{\alpha\beta}.
\]
The contribution of the anomaly to the expansion of the correlator is given by
\beq
\mathcal{A}_q^{\alpha\beta} = -{1 \over 3} \, \,  \, \, \frac{g_s^2}{16\pi^2} \,  2n_f u^{\alpha \beta}(p_1,p_2),
\label{anoms}
\eeq 
The correlator is decomposed into a transverse traceless part $\langle
t_{\mu_1\nu_1}(p_1)\,
j_{\mu_2}(p_2)\,
j_{\mu_3}(p_3)
\rangle$, a longitudinal one (proportional to $JJ$) and a trace part.
The trace Ward identity includes the anomaly contribution 
\begin{equation}
\label{x2}
g_{\mu\nu} \langle T^{\mu\nu}(p_1)J^{\alpha }(p_2)J^{\beta }(p_3)\rangle_q=\beta(e) u^{\alpha \beta  },
\end{equation}
with 
\bea
&&u^{\alpha\beta }(p,q) = -\frac{1}{4}\int\,d^4x\,\int\,d^4y\ e^{ip\cdot x + i q\cdot y}\ 
\frac{\delta^2 \{F_{a\mu\nu}F^{a\mu\nu}(0)\}} {\delta A_{\alpha}(x) \delta A_{\beta}(y)}\vline_{A=0} \,
\label{locvar}
\eea
explicitly given by
\beq\label{uab}
u^{\a \b } (p,q)\equiv (p\cdot q) g^{\alpha\beta} - p^\beta q^\alpha.
\eeq
The pole in the trace sector is part of the $\pi^{\mu\nu}$ projector. 
This tensor structure summarizes the conformal anomaly contribution, being the Fourier transform of the anomaly term at $O(e^2)$. Notice that the spin decomposition in \eqref{eq:TJJreconstructed2} is 
characterised by a spin-2 part, coupling to the $P^{(2)}$ component of the graviton propagator, and a spin-0 
part associated to the anomaly, with an interpolating $P^{(0)}$ projector. 
The transverse traceless (spin-2) gravitational sector, is related to the $tjj$ part of the correlator, the spin-1 $(jj)$ part decouples due to energy momentum conservation once we couple two $TJJ$ vertices with a virtual graviton. These are the three orthogonal sectors that are matched with a similar decomposition of the correlators.

The transverse traceless 
part has been discussed in \cite{Bzowski:2013sza,Bzowski:2018fql} using a regularization of the conformal integrals (3K integrals) useful for the analysis of general conformal theories. The identification of the form factors appearing in the decomposition of this sector, the only relevant for the exchange of a spin-2 graviton in a BIM-like diagram can be conveniently done using a Gram decomposition. 
The $tt$ sector is decomposed, with a minor modification of the notation of \cite{Bzowski:2018fql}, as  

\begin{equation}
\langle t jj\rangle_{\mathrm{TT}}
=
\sum_{i=1}^{5}
\tilde{A}_i(s,s_1,s_2)\,
{T}_i^{\mu_1\nu_1\mu_2\mu_3}.
\label{eq:TTdecomp}
\end{equation}

\begin{align}
T_1^{\mu_1\nu_1\mu_2\mu_3}
&=
\Pi_{\mu_1\nu_1\alpha_1\beta_1}(p_1)\,
\pi^{\mu_2}_{\alpha_2}(p_2)\,
\pi^{\mu_3}_{\alpha_3}(p_3)\,
p_2^{\alpha_1} p_2^{\beta_1} p_3^{\alpha_2} p_1^{\alpha_3},
\\[2mm]
T_2^{\mu_1\nu_1\mu_2\mu_3}
&=
\Pi_{\mu_1\nu_1\alpha_1\beta_1}(p_1)\,
\pi^{\mu_2}_{\alpha_2}(p_2)\,
\pi^{\mu_3}_{\alpha_3}(p_3)\,
\delta^{\alpha_2\alpha_3}\,
p_2^{\alpha_1} p_2^{\beta_1},
\\[2mm]
T_3^{\mu_1\nu_1\mu_2\mu_3}
&=
\Pi_{\mu_1\nu_1\alpha_1\beta_1}(p_1)\,
\pi^{\mu_2}_{\alpha_2}(p_2)\,
\pi^{\mu_3}_{\alpha_3}(p_3)\,
\delta^{\alpha_1\alpha_2}\,
p_2^{\beta_1} p_1^{\alpha_3},
\\[2mm]
T_4^{\mu_1\nu_1\mu_2\mu_3}
&=
\Pi_{\mu_1\nu_1\alpha_1\beta_1}(p_1)\,
\pi^{\mu_2}_{\alpha_2}(p_2)\,
\pi^{\mu_3}_{\alpha_3}(p_3)\,
\delta^{\alpha_1\alpha_3}\,
p_2^{\beta_1} p_3^{\alpha_2},
\\[2mm]
T_5^{\mu_1\nu_1\mu_2\mu_3}
&=
\Pi_{\mu_1\nu_1\alpha_1\beta_1}(p_1)\,
\pi^{\mu_2}_{\alpha_2}(p_2)\,
\pi^{\mu_3}_{\alpha_3}(p_3)\,
\delta^{\alpha_1\alpha_3}\,
\delta^{\alpha_2\beta_1}.
\end{align}
Notice that $T_3$ and $T_4$ are related by the exchange of the momenta and indices of the two $J$ currents. The scalar functions $\tilde{A_i}$ are the $tt$ form factors. By construction, they are free from kinematic ambiguities and encode the genuinely dynamical content of the three--point function. 
We equip the $tt$ tensor space with the natural bilinear form
\begin{equation}
(X,Y)
\equiv
X^{\mu_1\nu_1\mu_2\mu_3}
Y_{\mu_1\nu_1\mu_2\mu_3}.
\label{eq:innerprod}
\end{equation}
Using this pairing, we define the Gram matrix associated with the basis \(\{{T}_i\}\) as
\begin{equation}
G_{ij}
\equiv
({T}_i,{T}_j)
=
{T}_i^{\mu_1\nu_1\mu_2\mu_3}
{T}_{j\,\mu_1\nu_1\mu_2\mu_3}.
\label{eq:Gram}
\end{equation}
For generic values of \((s,s_1,s_2)\), the tensors \({T}_i\) are linearly independent and the Gram matrix is non--degenerate. Its inverse \(G^{-1}\) therefore exists. Contracting the decomposition \eqref{eq:TTdecomp} with \({T}_j\) and using \eqref{eq:Gram}, one obtains the linear system
\begin{equation}
({T}_j,\langle TJJ\rangle)
=
\sum_{k=1}^{5}
G_{jk}\,\tilde{A}_k.
\label{eq:linearsystem}
\end{equation}
Multiplication by the inverse Gram matrix immediately yields the projection formula
\begin{equation}
\tilde A_i(s,s_1,s_2)
=
\sum_{j=1}^{5}
(G^{-1})_{ij}\,
{T}_j^{\mu_1\nu_1\mu_2\mu_3}
\,
\langle T_{\mu_1\nu_1} J_{\mu_2} J_{\mu_3} \rangle.
\label{eq:projection}
\end{equation}
This relation is purely algebraic and holds independently of any perturbative realization. In particular, the full correlator may be used on the right--hand side, since the $tt$ tensors are orthogonal to the non--$tt$ sector.
The regularization of the correlator follows the use of the counterterm action from which one extracts the corresponding vertex $\Delta_{ct} TJJ$ which is then projected onto all the components $\tilde{A}_i$ using the same Gram projection shown above, where the vertex extracted from the counterterm by functional differentiation is projected over all the $tt$ form factors

 \begin{equation}
A_{j }^{}= -\,  n_f\,   \frac{g_s^2}{16\pi^2}\, \bar{A}_j, \qquad j=1,2\ldots 4
 \end{equation}

\begin{eqnarray}
\label{Abar}
		\bar{A}_1  &=& \frac{1}{48 \left(p_1^2 p_2^2-(p_1 \cdot p_2)^2 \right)^4} \Big[ A_{10} + A _{11} B_0 (p_1^2) + A_{12} B_0(p_2^2) + A_{13} B_0(q^2) + A_{14} C_0 (p_1^2 , p_2^2 , q^2 ) \Big] \notag \\
		\bar{A}_2  &=& -\frac{1}{144 \left(p_1^2 p_2^2-(p_1 \cdot p_2)^2 \right)^3} \Big[ A_{20} + A_{21}  B_0 (p_1^2) + A_{22} B_0(p_2^2)  + A_{23} B_0(q^2) + A_{24} C_0 (p_1^2 , p_2^2 , q^2 ) \Big] \notag \\
		\bar{A}_3  &=& \frac{1}{72 \left(p_1^2 p_2^2-(p_1 \cdot p_2)^2 \right)^3} \Big[A_{31}  B_0 (p_1^2) + A_{32} B_0(p_2^2)  + A_{33} B_0(q^2) + A_{34} C_0 (p_1^2 , p_2^2 , q^2 ) \Big] \notag \\
		\bar{A}_5  &=& -\frac{1}{72 \left(p_1^2 p_2^2-(p_1 \cdot p_2)^2 \right)^2} \Big[A_{40} + A_{41}  B_0 (p_1^2) + A_{42} B_0(p_2^2) + A_{43} B_0(q^2) + A_{44} C_0 (p_1^2 , p_2^2 , q^2 ) \Big]. \notag \\
\end{eqnarray}
and $A_4$ obtained from $A_3$ with the exchange of $p_1$ with $p_2$.
We give here, as an example, the first polynomial appearing in $\bar{A}_1$, the other exhibiting a similar structure
\begin{eqnarray}
	A_{10} &=& 16\, (p_1\cdot p_2)^7+36 \, p_1^2 \, (p_1\cdot p_2)^6+36 \, p_2^2 \, (p_1\cdot p_2)^6+12 \, p_1^4 \, (p_1\cdot p_2)^5+12 \, p_2^4 \, (p_1\cdot p_2)^5+240 \, p_1^2 \, p_2^2 \, (p_1\cdot p_2)^5 \notag \\ &&
	+188 \, p_1^2 \, p_2^4 \, (p_1\cdot p_2)^4+188 \, p_1^4 \, p_2^2 \, (p_1\cdot p_2)^4+46 \, p_1^2 \, p_2^6 \, (p_1\cdot p_2)^3-108 \, p_1^4 \, p_2^4 \, (p_1\cdot p_2)^3+46 \, p_1^6 \, p_2^2 \, (p_1\cdot p_2)^3 \notag \\ &&
	-204 \, p_1^4 \, p_2^6 \, (p_1\cdot p_2)^2-204 \, p_1^6 \, p_2^4 \, (p_1\cdot p_2)^2-20 \, p_1^6 \, p_2^8 \,-20 \, p_1^8 \, p_2^6 \,-58 \, p_1^4 \, p_2^8 \, (p_1\cdot p_2) \notag \\ &&
	-148 \, p_1^6 \, p_2^6 \, (p_1\cdot p_2)-58 \, p_1^8 \, p_2^4 \, (p_1\cdot p_2) 
\end{eqnarray}
Their explicit expressions can be found in Appendix G of \cite{Coriano:2024qbr}, by sending the Casimir 
$C_A\to 0$.\\
The Gram--matrix method ensures that this extraction is unique and free from ambiguities related to trace or longitudinal terms. The tensors structures are identical to those introduced in 
 \cite{Bzowski:2018fql}, which are four,  since $\tilde{A_3}$ and $\tilde{A_4}$  and $T_3$ and $T_4$ differ just by the exchange of the momenta/polarization of the two photons.  
 As we have mentioned, this scheme can be perfectly matched by the perturbative analysis of two combined free field theory with multiplicities $n_s$ and $n_f$ 
describing the general structure allowed by the correlator in a CFT. Such correspondences have been investigated both within the abstract CFT approach and the perturbative realizations, and extended to parity odd correlators as well \cite{Coriano:2023hts,Coriano:2023gxa,Coriano:2023cvf}. 
\subsection{The general conformal solution and its on-shell limit} \label{sksubseq}
The general conformal solution is expressed in terms of an overall normalization constant $C_1$ and the normalization of the $JJ$ 2-point function $C_{JJ}$. Another coefficient $D_{JJ}$ is introduced by the renormalization scheme implemented in \cite{Bzowski:2018fql}.

\begin{equation}
\big\langle
J^{\mu}(p)\,J^{\nu}(-p)
\rangle
=
\pi^{\mu\nu}(p)\,
p^{2N-2}
\left[
C_{JJ}\,
\ln\!\left(\frac{p^{2}}{\mu^{2}}\right)
+
D_{JJ}
\right],
\label{eq:JJ2pt}
\end{equation}
One can check that the solution presented in free field theory or given in \cite{Bzowski:2018fql} 
 is infrared safe for on-shell spin-1. Indeed, in the on-shell photon limit the expression of the transverse traceless form factors  simplify drastically and take the form

\begin{align}
A_1 &= \frac{8 C_1}{p_1^2} \nn \\
A_2 &= 2 C_1 + 2 C_{JJ} \left(\frac{4}{3} - \log p_1^2\right) - 2 D_{JJ} \nn \\
A_3 &=A_4= -2 C_1 + 2 C_{JJ} \left(\frac{4}{3} - \log p_1^2\right) - 2 D_{JJ}\nn \\
A_5 &=  2p_1^2 \left[ C_1 - C_{JJ} \left(\frac{2}{3}  - \log p_1^2\right) + D_{JJ} \right].
\label{bmcs}
\end{align}
The complete expression of the correlator in the notations of \cite{Bzowski:2018fql} is given in Appendix 
\eqref{bms}. The finiteness of the result can be directly checked by an explicit computation and deserves few comments. 
In the on-shell limit of the two photons in which $p_2^2, p_3^2\to 0$, before using the transversality condition on the two external vector lines, the computation manifests a double pole $1/(d-4)^2$ as well 
as a single pole, which is attributed to the scalar $C_0$ integral. The double pole characterizes a soft and collinear singularity in the variable $q^2\equiv p_1^2$ describing the invariant mass of the off-shell graviton decaying into two on-shell spin 1. Once we use the transversality conditions on the photons, the amplitude for an off-shell graviton 
to decay into two on-shell photons is finite and nonzero.   

\subsection{The perturbative implementation of the $TJJ$ correlator}
Here we briefly review the perturbative computation of the correlator, in order to summarize its structure in the case of on-shell photons. \\

We consider the standard coupling in an Abelian theory, which for simplicity we take to be QED. The same analysis applies to a generic conformal sector. In that case, the result presented below is simply multiplied by a multiplicity factor \(n_f\), accounting for a generic number of fermions, while \(e\) denotes the Abelian gauge coupling. In QED, the energy--momentum tensor (EMT) of a fermion of mass \(m\) is
\begin{equation}
T_{\mu\nu}
=
\frac{i}{4}\,\bar{\psi}
\left(
\gamma_\mu \overleftrightarrow{D}_\nu
+
\gamma_\nu \overleftrightarrow{D}_\mu
\right)\psi
-
g_{\mu\nu}\mathcal{L},
\end{equation}
where \(\mathcal{L}\) is the QED Lagrangian. When computing the \(TJJ\) correlator
\begin{equation}
\langle T_{\mu\nu}(k) J_\alpha(p) J_\beta(q)\rangle,
\qquad k=p+q,
\end{equation}
one must include both the triangle diagrams and the contact terms generated by the covariant derivative
\begin{equation}
D_\mu=\partial_\mu+i e A_\mu.
\end{equation}
The fermionic part of the energy--momentum tensor is
\begin{equation}
T^{\mu\nu}_\psi
=
\frac{i}{4}
\bar\psi
\left(
\gamma^\mu \overleftrightarrow{\partial}^{\nu}
+
\gamma^\nu \overleftrightarrow{\partial}^{\mu}
\right)\psi
-\eta^{\mu\nu}\,
\bar\psi (i\gamma^\rho \partial_\rho - m)\psi,
\label{Tmunu_fermion}
\end{equation}
with
\begin{equation}
\bar\psi \overleftrightarrow{\partial}^\mu \psi
=
\bar\psi (\partial^\mu \psi)
-
(\partial^\mu \bar\psi)\psi.
\label{leftrightarrow_def}
\end{equation}

In momentum space, with incoming fermion momentum \(p\) and outgoing momentum \(p'\), the graviton--fermion--fermion vertex is
\begin{equation}
V^{\mu\nu}_{h\bar\psi\psi}(p',p)
=
-\,\frac{i\kappa}{8}
\left[
\gamma^\mu (p'+p)^\nu
+
\gamma^\nu (p'+p)^\mu
-2 \eta^{\mu\nu}\left(\gamma^\rho (p'+p)_\rho-2m\right)
\right].
\label{vertex_hpsipsi}
\end{equation}
The Maxwell energy--momentum tensor is
\begin{equation}
T^{\mu\nu}_{\text{Maxwell}}
=
F^{\mu\lambda}F^\nu_{\ \lambda}
-\frac14 \eta^{\mu\nu}F^{\rho\sigma}F_{\rho\sigma},
\label{Tmunu_Maxwell}
\end{equation}
and, for photon momenta \(p\) and \(q\), the \(hAA\) vertex reads
\begin{equation}
\begin{aligned}
V^{\mu\nu\alpha\beta}_{hAA}(p,q)
=
-\,i\kappa
\Big[
& \eta^{\alpha\beta} p^{(\mu} q^{\nu)}
-\eta^{\mu\nu} p^{(\alpha} q^{\beta)}
+ \eta^{\mu(\alpha} q^{\beta)} p^{\nu}
+ \eta^{\nu(\alpha} p^{\beta)} q^{\mu}
- \eta^{\mu(\alpha}\eta^{\beta)\nu} (p\!\cdot\! q)
\Big],
\end{aligned}
\label{vertex_hAA}
\end{equation}
where symmetrization is defined by
\begin{equation}
a^{(\mu} b^{\nu)}
=
\frac12 \left(a^\mu b^\nu + a^\nu b^\mu\right).
\end{equation}
Expanding the Dirac action in curved space also generates the mixed contact interaction
\begin{equation}
h_{\mu\nu} A_\alpha \bar\psi \psi,
\label{mixed_structure}
\end{equation}
whose momentum-space vertex is
\begin{equation}
V^{\mu\nu\alpha}_{hA\bar\psi\psi}
=
\frac{i\kappa e}{4}
\left[
\eta^{\mu\alpha}\gamma^\nu
+
\eta^{\nu\alpha}\gamma^\mu
-
\eta^{\mu\nu}\gamma^\alpha
\right].
\label{vertex_hApsipsi}
\end{equation}
The full Abelian \(TJJ\) amplitude is then
\begin{equation}
\Gamma^{\mu\nu\alpha\beta}
=
\Gamma^{\mu\nu\alpha\beta}_{\text{triangle}}(p,q)
+
\Gamma^{\mu\nu\beta\alpha}_{\text{triangle}}(q,p)
+
\Gamma^{\mu\nu\alpha\beta}_{\text{contact}}(p,q)
+
\Gamma^{\mu\nu\beta\alpha}_{\text{contact}}(q,p),
\end{equation}
where \(p\) and \(q\) are the photon momenta. The two contributions are given by the following loop integrals:
\begin{equation}
\Gamma^{\mu\nu\alpha\beta}_{\text{triangle}}(p,q)
=
-\int\frac{d^4 l}{(2\pi)^4}
\, {\rm tr}\left\{
V^{\mu\nu}_{h\bar\psi\psi}(l-q,l+p)
\frac{1}{\lsl-\qsl-m}
\gamma^{\beta}
\frac{1}{\lsl-m}
\gamma^{\alpha}
\frac{1}{\lsl+\psl-m}
\right\},
\end{equation}
and
\begin{equation}
\Gamma^{\mu\nu\alpha\beta}_{\text{contact}}(p,q)
=
-\int\frac{d^4 l}{(2\pi)^4}
\, {\rm tr}\left\{
V_{hA\bar\psi\psi}^{\mu\nu\alpha}
\frac{1}{\lsl-\qsl-m}
\gamma^{\beta}
\frac{1}{\lsl-m}
\right\}.
\label{loop}
\end{equation}
The full amplitude satisfies the Ward identities
\begin{equation}
p_\alpha \Gamma^{\mu\nu\alpha\beta}=0,
\qquad
q_\beta \Gamma^{\mu\nu\alpha\beta}=0.
\label{Ward_identities}
\end{equation}

\subsection{The anomaly}
\label{reff2}
At the operator level, the trace of the EMT is
\begin{equation}
T^\mu_{\ \mu}
=
\mathcal{A}(x),
\end{equation}
where \(\mathcal{A}(x)\) denotes the trace anomaly. For a single massless fermion, the electromagnetic contribution 
\begin{equation}
\mathcal{A}
=
\frac{\beta(e)}{2e}
F_{\mu\nu}F^{\mu\nu}
=
\frac{\alpha}{3\pi}
F_{\mu\nu}F^{\mu\nu}.
\end{equation}

\begin{figure}[t]
\begin{center}
\includegraphics[scale=0.7,angle=0]{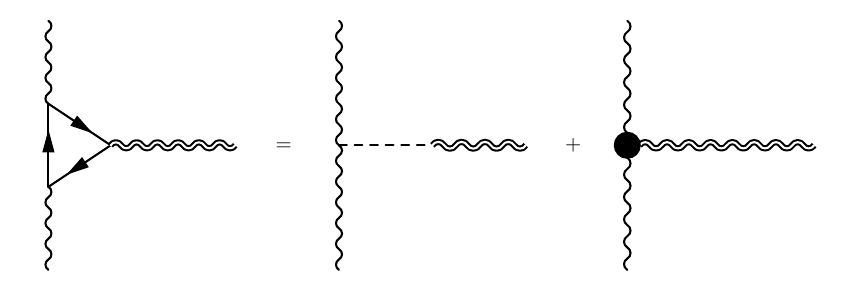}
\caption{Decomposition of the \(TJJ\) vertex into an anomalous contribution and a transverse--traceless sector (black blob).}
\label{expansionX}
\end{center}
\end{figure}

The Abelian correlator has been studied in the past in a basis of thirteen tensor structures \(t_1,\ldots,t_{13}\), reported in Appendix \ref{reff}, with corresponding form factors \(F_1,\ldots,F_{13}\). For on-shell photons and massless fermions, the nonvanishing form factors are ($s\equiv k^2$)
\bea
F_{1} (s, 0, 0, 0) &=& - \frac{e^2}{18 \pi^2  s}, \\
F_{3} (s, 0, 0, 0) &=&  F_{5} (s, 0, 0, 0) = - \frac{e^2}{144 \pi^2 \, s}, \\
F_{7} (s, 0, 0, 0) &=& -4 \, F_{3} (s, 0, 0, 0), \\
F_{13, R} (s, 0, 0, 0) &=& - \frac{e^2}{144 \pi^2} \, \left[ 12 \log \left(-\frac{s}{\mu^2}\right) - 35\right],
\qquad s <0 \eea
Accordingly, the correlator with two on-shell photons can be written as \cite{Armillis:2009pq}
\bea
\Gamma^{\mu\nu\alpha\beta}  (s; 0,0,0) &=&
F_{1} (s, 0, 0, 0) \, \widetilde{t}^{\, \mu \nu \alpha \beta}_{1}
+ F_{3} (s, 0, 0, 0) \,
\left(
\widetilde{t}^{\, \mu \nu \alpha \beta}_{3}
+ \widetilde{t}^{\, \mu \nu \alpha \beta}_{5}
- 4 \widetilde{t}^{\, \mu \nu \alpha \beta}_{7}
\right)
+ F_{13, R} \, \widetilde{t}^{\, \mu \nu \alpha \beta}_{13},
\label{gamma00}
\eea
with
\bea
 \widetilde t_1^{\, \mu \nu \a \b} &=&
 (s \, g^{\mu\nu} - k^{\mu}k^{\nu}) \, u^{\a \b} (p,q)
 =\hat{\pi}^{\mu\nu}(s) \, u^{\a \b} (p,q),
 \label{widetilde1}\\
 \widetilde t_3^{\, \mu \nu \a \b}
 + \widetilde t_5^{\, \mu \nu \a \b}
- 4 \widetilde t_7^{\, \mu \nu \a \b} &=&
- 2 \, u^{\a \b} (p,q)
\left(
s \, g^{\mu \nu}
+ 2 (p^\mu p^\nu + q^\mu q^\nu)
- 4 (p^\mu q^\nu + q^\mu p^\nu)
\right),
\eea
and
\bea
\widetilde{t}^{\, \mu \nu \alpha \beta}_{13} &=&
\big(p^{\mu} q^{\nu} + p^{\nu} q^{\mu}\big)g^{\alpha\beta}
+ \frac{s}{2} \big(g^{\alpha\nu} g^{\beta\mu} + g^{\alpha\mu} g^{\beta\nu}\big) \nn \\
&&
- g^{\mu\nu} \left(\frac{s}{2} g^{\alpha \beta}- q^{\alpha} p^{\beta}\right)
-\big(g^{\beta\nu} p^{\mu}
+ g^{\beta\mu} p^{\nu}\big)q^{\alpha}
-\big(g^{\alpha\nu} q^{\mu}
+ g^{\alpha\mu} q^{\nu }\big)p^{\beta}.
\label{widetilde13}
\eea
Each of these three tensor structures is separately conserved and gauge invariant on the photon lines in the massless limit. This solution can be mapped to the basis introduced in Section \ref{sksubseq} through
\bea
F_3(s,0,0,0)&=& -\frac{A_1 s^2-2 A_3 s-4 A_5}{24 s^2},\nn\\
F_{13,R}(s,0,0,0)&=& -\frac{A_3}{2}.
\eea
Two characteristic features of this expression should be emphasized. The first is associated with the tensor structure \(\tilde t_1\) and the corresponding form factor \(F_1\), which contains the anomaly pole and contributes to the traced part of the correlator (see Fig.~2). This term controls the exchange of the scalar component of the graviton and originates from the nonlocal nature of the effective interaction. A second pole appears in the traceless sector and is encoded in the form factor \(F_3\). This contribution has been pointed out in earlier analyses and interpreted as a possible ``plasmon'' mode \cite{Giannotti:2008cv}; in the present framework it is generated by exchange of the spin--2 component of the virtual graviton. While the scalar mode is interpolated by a  projector, as we are going to show in the next sections, the two plasmon modes couple through the spin--2 sector of the exchanged graviton.\\
The decomposition of this correlator derived in \cite{Armillis:2009pq} in the on-shell limit has been extended to the non-Abelian case in \cite{Coriano:2025fom}, in the analysis of dilaton sum rules in QCD. In particular, the residue at the \(k^2=0\) pole was computed in \cite{Coriano:2025fom}. The result contains two contributions: the first, proportional to \(\tilde\phi_1^{\mu\nu\alpha\beta}\), is associated with the conformal anomaly, while the second, proportional to \(\tilde\phi_2^{\mu\nu\alpha\beta}\), corresponds to an additional pole in a traceless tensor structure. The corresponding pole relation is
\beq\label{tjjpolelim}
\lim_{s \to 0} \, k^2 \braket{T^{\mu\nu}(k) J^{\alpha }(p) J^{\beta }(q)}
= -\frac{g_s^2}{48 \pi^2} \left(\frac{2}{3} n_f\right) \tilde\phi_1^{\mu\nu\alpha\beta}
- \frac{g_s^2}{288 \pi^2} n_f \, \tilde\phi_2^{\mu\nu\alpha\beta},
\eeq
which is in agreement with \eqref{gamma00}. Indeed, one finds
\begin{align}
\tilde\phi_1^{\, \mu \nu \alpha \beta} (p,q) &= \tilde{t}_1^{\mu\nu\a\b},
\label{widetilde1phi} \\
\tilde\phi_2^{\, \mu \nu \alpha \beta} (p,q) &=
\widetilde t_3^{\, \mu \nu \a \b}
+ \widetilde t_5^{\, \mu \nu \a \b}
- 4 \widetilde t_7^{\, \mu \nu \a \b}.
\label{widetilde2}
\end{align}
These identify massless exchanges in distinct channels. The anomaly pole contribution behaves as
\(\sim s^{-1}\tilde{\phi}_1^{\mu\nu \a\b}\),
while the additional pole in the traceless sector,
\(\sim s^{-1}\tilde{\phi}_2^{\mu\nu \a\b}\),
is reproduced by the nonlocal effective action
\begin{equation}
S_{\text{extra pole}}
=
\frac{g_s^2}{72\pi^2} n_f
\int d^4x \sqrt{-g}
\int d^4x' \sqrt{-g'}
\, h_{\mu\nu}(x)
\, \Box^{-1}_{x,x'}
\, \mathcal{O}^{\mu\nu}(x'),
\label{extra_pole_action}
\end{equation}
where
\begin{equation}
\mathcal{O}^{\mu\nu}
=
3(\partial^\mu\partial^\nu F_{\alpha\beta}^a)F^{\alpha\beta a}
+\frac14 (g^{\mu\nu}\Box-4\partial^\mu\partial^\nu)
F_{\alpha\beta}^aF^{\alpha\beta a}.
\label{operator_def}
\end{equation}
This is in agreement with the analysis of Giannotti and Mottola \cite{Giannotti:2008cv}. The second traceless structure therefore identifies an additional nonlocal contribution to the effective action, describing an interaction localized around the light cone,
\begin{align}\label{expoleac}
S_{\text{extra pole}}
=
\frac{g_s^2}{72 \pi^2} n_f
\int d^4x \sqrt{-g}
\int d^4x' \sqrt{-g'}\, h_{\mu\nu}(x) \Box^{-1}_{x,x'}
\left[
3 (\partial^\mu \partial^\nu F^a_{\alpha\beta}) F^{\alpha\beta a}
+ \frac{1}{4} \left( g^{\mu\nu} \Box - 4 \partial^\mu \partial^\nu \right) F^a_{\alpha\beta} F^{\alpha\beta a}
\right]_{x'},
\end{align}
whereas the first tensor structure is associated with the familiar anomaly action \cite{Armillis:2009pq}
\bea
S_{\text{anom}} &=&
\frac{1}{3} \, \frac{g^3}{16 \pi^2}
\left(  \frac{2}{3} \, n_f \right)
\int d^4 x \, d^4 y \,
R^{(1)}(x)\, \square^{-1}(x,y) \,
F^a_{\alpha \beta}F^{\alpha \beta a}.
\eea

In the next sections we are going to show how the nonlocal structure of the anomaly effective action with the inclusion of $S_{\text{anom}}$ is responsible for the activation of a scalar degree of freedom in the case of 4-point functions, discussing the gravitational scattering of photons. For this, we need to turn to 
the massless Fierz-Pauli Lagrangian, extracted from the EH action at quadratic level in the metric fluctuations and address the gauge dependence of the virtual graviton.

\section{Towards 4-point functions: spin projectors and the scalar sector in the de~Donder gauge}

In order to fix our conventions, we start at linearized level, with the metric in the  Einstein--Hilbert action in mostly minus signature
\begin{equation}
S_{\rm EH}=\frac{1}{2\kappa^{2}}\int d^{4}x\,\sqrt{-g}\,R,
\qquad
\kappa^{2}=16\pi G_N 
\label{EHaction}
\end{equation}
and
\begin{equation}
\kappa^2 = \frac{1}{M_P^2},
\end{equation}
where $M_P$ denotes the reduced Planck mass.
We linearize the metric around flat spacetime
\begin{equation}
g_{\mu\nu}=\eta_{\mu\nu}+\kappa h_{\mu\nu},
\qquad
|h_{\mu\nu}|\ll1 ,
\label{metricexpand}
\end{equation}
using $\eta_{\mu\nu}$ to raise and lower indices. The linearized Christoffel symbols are
\begin{equation}
\Gamma^{\rho}{}_{\mu\nu}
=\frac12\,\eta^{\rho\sigma}\left(
\partial_\mu h_{\nu\sigma}
+\partial_\nu h_{\mu\sigma}
-\partial_\sigma h_{\mu\nu}
\right),
\end{equation}
from which one obtains the linearized Ricci tensor and scalar
\begin{equation}
R^{(1)}_{\mu\nu}
=\frac12\left(
\partial_{\rho}\partial_{\mu}h^{\rho}{}_{\nu}
+\partial_{\rho}\partial_{\nu}h^{\rho}{}_{\mu}
-\Box h_{\mu\nu}
-\partial_{\mu}\partial_{\nu}h
\right),
\end{equation}
\begin{equation}
R^{(1)}=\partial_{\mu}\partial_{\nu}h^{\mu\nu}-\Box h ,
\end{equation}
with $\Box=\eta^{\mu\nu}\partial_\mu\partial_\nu$. The linearized Einstein tensor is
\begin{equation}
G^{(1)}_{\mu\nu}
=
\frac12\left(
\Box h_{\mu\nu}
-\partial_{\mu}\partial^{\rho}h_{\rho\nu}
-\partial_{\nu}\partial^{\rho}h_{\rho\mu}
+\partial_{\mu}\partial_{\nu}h
+\eta_{\mu\nu}\partial_{\rho}\partial_{\sigma}h^{\rho\sigma}
-\eta_{\mu\nu}\Box h
\right)
\label{FierzPauli}
\end{equation}
and the linearization of Einstein’s equations gives the Fierz--Pauli equations
\begin{equation}
G^{(1)}_{\mu\nu}=0 .
\label{FPeq}
\end{equation}
It is convenient to introduce the trace--reversed field
\begin{equation}
\bar h_{\mu\nu}=h_{\mu\nu}-\frac12\eta_{\mu\nu}h,
\qquad
\bar h=-h ,
\end{equation}
and impose the de~Donder (harmonic) gauge condition
\begin{equation}
\partial^\mu\bar h_{\mu\nu}=0 .
\label{dedonder}
\end{equation}
This condition implies
\begin{equation}
\partial^\mu h_{\mu\nu}=\frac12\,\partial_\nu h 
\end{equation}
and reduces the equations of motion to the diagonal form
\begin{equation}
\Box\,\bar h_{\mu\nu}=0 .
\label{waveeq}
\end{equation}
At the level of the action, the quadratic Einstein--Hilbert (massless Fierz-Pauli, FP) Lagrangian before gauge fixing is
\begin{equation}
\mathcal L^{(2)} = M_P^2\left(
\frac12\,\partial_\lambda h_{\mu\nu}\partial^\lambda h^{\mu\nu}
- \partial_\mu h^{\mu\nu}\partial^\lambda h_{\lambda\nu}
+ \partial_\mu h^{\mu\nu}\partial_\nu h
- \frac12\,\partial_\lambda h \partial^\lambda h
\right),
\label{FP}
\end{equation}
with $h=h^\mu{}_\mu$. Implementing the de~Donder gauge via the gauge--fixing function
\begin{equation}
\mathcal F_\nu=\partial^\mu h_{\mu\nu}-\frac12\partial_\nu h ,
\end{equation}
one adds the term
\begin{equation}
\mathcal L_{\rm GF}
=
-\frac{M_P^2}{2\xi_D}\,\mathcal F_\nu\mathcal F^\nu ,
\label{DD}
\end{equation}
where $\xi_D$ is a gauge parameter. Even after this gauge fixing, a residual gauge symmetry remains, corresponding to transformations with parameters $\xi_\mu$ satisfying
\begin{equation}
\Box\xi_\mu=0 .
\label{residual}
\end{equation}
This residual symmetry plays a crucial role in eliminating nonphysical components of \(h_{\mu\nu}\). To make the covariant structure of the field explicit, we introduce the decomposition
\begin{equation}
h_{\mu\nu}
=
h_{\mu\nu}^{TT}
+\partial_\mu V_\nu+\partial_\nu V_\mu
+\left(\partial_\mu\partial_\nu-\frac14\eta_{\mu\nu}\Box\right)w
+\frac14\eta_{\mu\nu}h,
\label{eq:cov_decomp_ped}
\end{equation}
with
\begin{equation}
\partial^\mu h_{\mu\nu}^{TT}=0,
\qquad
h^{TT\,\mu}{}_{\mu}=0,
\qquad
\partial^\mu V_\mu=0.
\label{eq:cov_constraints_ped}
\end{equation}

This decomposition is purely kinematical. The original symmetric tensor \(h_{\mu\nu}\) has \(10\) independent components, which are reorganized covariantly as
\begin{equation}
10 = 5 + 3 + 1 + 1,
\label{eq:10_5311_ped}
\end{equation}
namely the covariant transverse--traceless tensor \(h_{\mu\nu}^{TT}\), carrying \(5\) components;
 the transverse vector \(V_\mu\), carrying \(3\) components; the two scalars \(w\) and \(h\), carrying \(1+1\) components.\\
Under linearized diffeomorphisms
\begin{equation}
\delta h_{\mu\nu}=\partial_\mu \xi_\nu+\partial_\nu \xi_\mu,
\label{eq:lin_diff_ped}
\end{equation}
with
\begin{equation}
\xi_\mu=\xi_\mu^T+\partial_\mu \sigma,
\qquad
\partial^\mu \xi_\mu^T=0,
\label{eq:xi_split_ped}
\end{equation}
one finds
\begin{equation}
\delta h_{\mu\nu}^{TT}=0,
\qquad
\delta V_\mu=\xi_\mu^T,
\qquad
\delta w=2\sigma,
\qquad
\delta h=2\Box\sigma.
\label{eq:cov_gauge_transform_ped}
\end{equation}

Hence the \(3\) transverse components of \(\xi_\mu^T\) gauge away the \(3\) components of \(V_\mu\), while the scalar parameter \(\sigma\) removes one scalar combination of \(w\) and \(h\). Therefore the covariant gauge-invariant content is
\begin{equation}
6 = 5 + 1,
\label{eq:6_51_ped}
\end{equation}
where the unique scalar gauge-invariant combination is
\begin{equation}
\Phi_{cov}\equiv h-\Box w.
\label{eq:Phi_cov_ped}
\end{equation}

Neither \(w\) nor \(h\) is gauge invariant individually. However,
\begin{equation}
\delta(h-\Box w)=2\Box\sigma-\Box(2\sigma)=0,
\end{equation}
so \(\Phi_{cov}\) is gauge invariant. This is the component investigated by Mottola in \cite{Mottola:2016mpl}.
It is the unique gauge-invariant scalar built from \(h_{\mu\nu}\) at linear order in a covariant decomposition.\\
Notice that the reduction from 5 to 2 degrees propagating of freedom in $h_{\mu\nu}^{TT}$ is associated 
with the residual gauge freedom \eqref{residual}. Indeed, in the covariant de Donder gauge

At this point it is important to stress the distinction between gauge-invariant variables and propagating degrees of freedom. The counting \eqref{eq:6_51_ped} is kinematical: it tells us how many independent gauge-invariant covariant combinations can be built from \(h_{\mu\nu}\). It does \emph{not} mean that all these variables propagate.\\
Indeed, the linearized curvature scalar is related to \(\Phi_{cov}\) by
\begin{equation}
\delta R
=
\partial_\mu\partial_\nu h^{\mu\nu}
-\Box h
=
-\frac34\,\Box \Phi_{cov}.
\label{eq:R_Phicov_ped}
\end{equation}
Substituting the decomposition \eqref{eq:cov_decomp_ped} into the quadratic Einstein--Hilbert action, using the transversality conditions and integrating by parts, all contributions involving \(V_\mu\), \(w\), \(h\), and therefore \(\Phi_{cov}\), cancel identically up to boundary terms. The action reduces to
\begin{equation}
S_{EH}^{(2)}
=
\frac{1}{8}
\int d^4x\,
\partial_\lambda h_{\mu\nu}^{TT}\partial^\lambda h^{TT\,\mu\nu}.
\label{eq:SEH_TT_only_ped}
\end{equation}

Thus the gauge-invariant scalar \(\Phi_{cov}\) does not appear in the quadratic Einstein--Hilbert action. In particular,
\begin{equation}
\frac{\delta S_{EH}^{(2)}}{\delta \Phi_{cov}}=0,
\label{eq:no_Phicov_variation_ped}
\end{equation}
so there is no Euler--Lagrange equation associated with \(\Phi_{cov}\) and no independent scalar phase-space degree of freedom generated by the quadratic Einstein--Hilbert dynamics.\\
Therefore, covariantly, one can construct \(5+1\) gauge-invariant variables, namely \(h_{\mu\nu}^{TT}\) and \(\Phi_{cov}\). However, only the spin--2 sector contributes to the quadratic Einstein--Hilbert action, while the scalar invariant \(\Phi_{cov}\) is non-dynamical. Hence the covariant decomposition should be interpreted as a kinematical organization of the field variables, not as a direct counting of propagating modes. The physical massless graviton still carries only the two helicity-\(\pm 2\) propagating degrees of freedom.\\
In summary, while the de~Donder gauge gives rise to wave equations for the trace-reversed field and for its trace, the covariant decomposition of the quadratic Einstein--Hilbert action makes clear that the gauge-invariant scalar combination \(\Phi_{cov} = h - \Box w\) does not enter the action. The constraint structure of general relativity thus guarantees that only the two transverse--traceless tensor polarizations propagate as physical degrees of freedom. In the following sections, we revisit this issue in detail after including the anomaly contribution in the effective action.

\subsection{Spin decomposition of the graviton propagator}

\begin{figure}
	\centering
	\includegraphics[scale=1.2,angle=0]{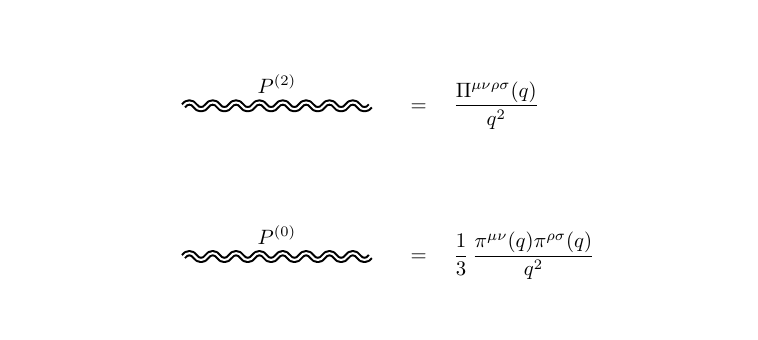}
	\caption{Spin-2 and spin-0 part of the graviton propagator.}
	\label{fig:graviton_propagator_decomposition}
\end{figure}

To make contact between virtual graviton exchange in Einstein--Hilbert gravity using the FP Lagrangian, it is useful to work in momentum space and decompose the graviton propagator into its spin components. This makes explicit which sectors are diffeomorphis invariant (gauge independent), which are gauge independent, and how the gauge-invariant scalar combination of the metric, introduced previously, propagate in this basis. \\For this reason, following Barnes and Rivers \cite{Barnes1965,Rivers1964}, define the transverse and longitudinal projectors
\begin{equation}
\pi_{\mu\nu}\equiv \eta_{\mu\nu}-\frac{k_\mu k_\nu}{k^2},
\qquad
\omega_{\mu\nu}\equiv \frac{k_\mu k_\nu}{k^2}.
\label{eq:piomega}
\end{equation}
Here \(k^2\neq 0\), and poles are treated in the propagator, not in the projector algebra. In four dimensions, the Barnes--Rivers projectors for symmetric rank-two tensors are
\begin{align}
P^{(2)}_{\mu\nu,\alpha\beta}
&=
\frac12\!\left(\pi_{\mu\alpha}\pi_{\nu\beta}+\pi_{\mu\beta}\pi_{\nu\alpha}\right)
-\frac13\,\pi_{\mu\nu}\pi_{\alpha\beta},
\label{eq:P2}
\\
P^{(1)}_{\mu\nu,\alpha\beta}
&=
\frac12\!\left(
\pi_{\mu\alpha}\omega_{\nu\beta}
+\pi_{\mu\beta}\omega_{\nu\alpha}
+\pi_{\nu\alpha}\omega_{\mu\beta}
+\pi_{\nu\beta}\omega_{\mu\alpha}
\right),
\label{eq:P1}
\\
P^{(0\text{-}s)}_{\mu\nu,\alpha\beta}
&=\frac13\,\pi_{\mu\nu}\pi_{\alpha\beta},
\qquad\qquad
P^{(0\text{-}w)}_{\mu\nu,\alpha\beta}
=\omega_{\mu\nu}\omega_{\alpha\beta}.
\label{eq:P0}
\end{align}
They satisfy
\begin{equation}
\left(P^{(2)}+P^{(1)}+P^{(0\text{-}s)}+P^{(0\text{-}w)}\right)_{\mu\nu,\alpha\beta}
=
I_{\mu\nu,\alpha\beta},
\qquad\qquad
I_{\mu\nu,\alpha\beta}\equiv
\frac12\!\left(\eta_{\mu\alpha}\eta_{\nu\beta}+\eta_{\mu\beta}\eta_{\nu\alpha}\right),
\label{eq:completeness}
\end{equation}
together with orthogonality and idempotency:
\(P^{(i)}P^{(j)}=\delta^{ij}P^{(i)}\). Contracting \(h_{\mu\nu}\) with \(\pi^{\mu\nu}\) isolates the transverse scalar component,
\begin{equation}
\pi^{\mu\nu}h_{\mu\nu}(k)=\Phi_{cov}(k),
\label{eq:PhiProj}
\end{equation}
which is the momentum-space representative of the gauge-invariant scalar combination discussed in the previous section. With a de Donder gauge-fixing term parametrized by \(\xi_D\) \eqref{DD}, the quadratic kinetic operator is
\begin{equation}
\mathcal O_{\mu\nu,\alpha\beta}(k)=
k^2\!\left[
P^{(2)}
-2\,P^{(0\text{-}s)}
+\frac{1}{\xi_D}\!\left(P^{(1)}+\frac12 P^{(0\text{-}w)}\right)
\right]_{\mu\nu,\alpha\beta}.
\label{eq:Ok}
\end{equation}
Using projector orthogonality, its inverse is decomposed as 
\begin{equation}
D_{\mu\nu,\alpha\beta}(k)=
\frac{1}{k^2}\!\left[
P^{(2)}
-\frac12 P^{(0\text{-}s)}
+\xi_D\!\left(P^{(1)}+\frac12 P^{(0\text{-}w)}\right)
\right]_{\mu\nu,\alpha\beta}.
\label{eq:Dk}
\end{equation}
We denote by $D^{(2)}$ and $D^{(0-s)}$ the spin-2 and spin-0 part of the propagator (Figure \ref{fig:graviton_propagator_decomposition}) 
\begin{equation}\label{eq:propagator_spin_decomposition}
	D^{(0-s)}_{\mu\nu,\alpha\beta}(k) = -\frac{1}{6}\frac{\pi^{\mu\nu}(k)\pi^{\alpha\beta}(k)}{k^2}, \qquad \qquad D^{(2)}_{\mu\nu,\alpha\beta}(k) = -\frac{1}{2}\frac{\Pi^{\mu\nu,\alpha\beta}(k)}{k^2}.
\end{equation}
All \(\xi_D\)-dependence is confined to \(P^{(1)}\) and \(P^{(0\text{-}w)}\), i.e. the pure-gauge sectors. The spin-2 part \(P^{(2)}\) and the scalar projector \(P^{(0\text{-}s)}\) are gauge-parameter independent (see Fig. 3). In Einstein gravity, this scalar sector does not represent an independent propagating physical degree of freedom. For \(\xi_D=1\) (de Donder in this convention),
\begin{equation}
D_{\mu\nu,\alpha\beta}(k)\big|_{\xi_D=1}
=
\frac{1}{k^2}
\left(
P^{(2)}-\frac12 P^{(0\text{-}s)}+P^{(1)}+\frac12 P^{(0\text{-}w)}
\right)_{\mu\nu,\alpha\beta}.
\label{eq:DkDeDonder}
\end{equation}
For conserved external sources (\(k_\mu T^{\mu\nu}=k_\alpha T'^{\alpha\beta}=0\)), the longitudinal projectors do not contribute, and one obtains the exact identity
\begin{equation}
T^{\mu\nu}D_{\mu\nu,\alpha\beta}(k)T'^{\alpha\beta}
=
\frac{1}{k^2}\,
T^{\mu\nu}
\left[
\frac12\!\left(
\eta_{\mu\alpha}\eta_{\nu\beta}
+\eta_{\mu\beta}\eta_{\nu\alpha}
-\eta_{\mu\nu}\eta_{\alpha\beta}
\right)
\right]
T'^{\alpha\beta}.
\label{eq:FPonConserved}
\end{equation}
Thus the exchange kernel, for conserved stres energy tensors, is exactly the Fierz--Pauli tensor structure on conserved sources.

\section{Tree--level graviton exchange and the spin--0 channel in the De~Donder gauge }
To characterize the role of the scalar exchange in the De Donder gauge, and extend \eqref{eq:FPonConserved}, we consider two independent matter sectors, redefined as $L$ and $R$, interacting only through gravity. A similar analysis was presented in \cite{Giannotti:2008cv}.
One can think of $L$ as a visible and of $R$ as a dark sector, with gravity mediating their interaction. At tree level, the exchange of a graviton is described by the amplitude
\begin{equation}
\mathcal A_t
=
\frac{\kappa^2}{4}\,
T^{\mu\nu}_L(k)\,
D_{\mu\nu,\rho\sigma}(k)\,
T^{\rho\sigma}_R(-k),
\label{Atree_def}
\end{equation}
where $k$ is the momentum transfer. Throughout this section the stress--energy tensors are assumed to be conserved
\begin{equation}
k_\mu T^{\mu\nu}_{L}(k)=0,
\qquad
k_\mu T^{\mu\nu}_{R}(k)=0,
\label{T_conservation}
\end{equation}
but not necessarily traceless
\begin{equation}
T^\mu{}_{\mu\,L}\neq0,
\qquad
T^\mu{}_{\mu\,R}\neq0 .
\label{T_trace}
\end{equation}
Using the covariant spin decomposition of the graviton propagator discussed in the previous section, the de~Donder propagator can be written as
\begin{equation}
D_{\mu\nu,\rho\sigma}(k)
=
\frac{i}{k^2+i\epsilon}
\left[
P^{(2)}_{\mu\nu,\rho\sigma}
-
\frac12\,P^{(0\text{-}s)}_{\mu\nu,\rho\sigma}
\right],
\label{prop_spin}
\end{equation}
where only the transverse spin--2 and transverse scalar projectors appear. Substituting \eqref{prop_spin} into \eqref{Atree_def}, the amplitude splits into a spin--2 and a spin--0 contribution,
\begin{equation}
\mathcal A_t
=
\frac{i\kappa^2}{4k^2}
\left[
T^{\mu\nu}_L
P^{(2)}_{\mu\nu,\rho\sigma}
T^{\rho\sigma}_R
-
\frac12\,
T^{\mu\nu}_L
P^{(0\text{-}s)}_{\mu\nu,\rho\sigma}
T^{\rho\sigma}_R
\right].
\label{Atree_split}
\end{equation}
Because of energy--momentum conservation, all longitudinal components drop out and the projectors act effectively as algebraic operators on the stress--energy tensors. The spin--2 contribution evaluates to
\begin{equation}
T^{\mu\nu}_L
P^{(2)}_{\mu\nu,\rho\sigma}
T^{\rho\sigma}_R
=
T^{\mu\nu}_L T_{R\,\mu\nu}
-
\frac13\,
T^\mu{}_{\mu\,L}\,
T^\nu{}_{\nu\,R},
\label{spin2}
\end{equation}
where the trace subtraction reflects the irreducibility of the spin--2 representation. The spin--0 contribution follows from
\begin{equation}
T^{\mu\nu}_L
P^{(0\text{-}s)}_{\mu\nu,\rho\sigma}
T^{\rho\sigma}_R
=
\frac13\,
\left(\pi_{\mu\nu}T_L^{\mu\nu}\right)
\left(\pi_{\rho\sigma}T_R^{\rho\sigma}\right).
\end{equation}
Using conservation once more, $\pi_{\mu\nu}T^{\mu\nu}_{L,R}=T^\mu{}_{\mu\,L,R}$, one finds
\begin{equation}
T^{\mu\nu}_L
P^{(0\text{-}s)}_{\mu\nu,\rho\sigma}
T^{\rho\sigma}_R
=
\frac13\,
T^\mu{}_{\mu\,L}\,
T^\nu{}_{\nu\,R}.
\label{spin0}
\end{equation}
Inserting \eqref{spin2} and \eqref{spin0} into \eqref{Atree_split}, the full tree--level amplitude becomes
\begin{equation}
\mathcal A_t
=
\frac{i\kappa^2}{4k^2}
\left[
T^{\mu\nu}_L T_{R\,\mu\nu}
-
\frac12\,
T^\mu{}_{\mu\,L}\,
T^\nu{}_{\nu\,R}
\right].
\label{Atree_final}
\end{equation}
Equation \eqref{Atree_final} shows that, whenever the stress--energy tensors are not traceless, an apparent scalar contribution proportional to the product of the traces remains alongside the spin--2 exchange. This scalar term is associated with the same massless pole, $1/k^2$, as the spin--2 part and matches precisely the scalar projector $P^{(0\text{-}s)}$ identified previously. Its interpretation, however, requires some care, since the mere appearance of a trace contribution does not necessarily imply that the trace contributes to a physical scattering amplitude through a genuinely propagating scalar degree of freedom. To make this point precise, it is useful to turn to the analysis of four--point functions and examine anomaly--mediated interactions in that framework, where the physical content of the exchange can be extracted more clearly. We shall first consider the case of explicit breaking and then move to anomalous breaking.

\section{Scalar gravitational scattering and explicit breaking:  the absence of scalar exchange in 4-point functions}
\label{sec:scalar_xi_dependence_summary}
As just mentioned, the appearance of a nonvanishing trace in the scalar-graviton vertex is not, by itself, sufficient to conclude that a scalar component of the metric is being exchanged. This statement is best understood by analyzing the same process in two complementary descriptions: first in the standard covariant formulation, where the full graviton propagator is used explicitly, and then in the transverse-traceless gauge, where only the propagating spin-2 part is manifest and the remaining metric components appear through constraint equations. The two derivations lead to the same physical amplitude. In particular, they show that in ordinary Einstein gravity the presence of explicit conformal-symmetry-breaking terms in the matter sector does not activate an independent propagating scalar mode.

The scalar theory considered below allows for two distinct sources of explicit breaking of conformal symmetry. The first is the non-minimal coupling \(\xi\neq 1/6\), which already breaks conformal invariance in the massless theory. The second is the scalar mass \(m\), which provides an additional and independent breaking. These two effects behave differently at the level of the amplitude. In the massless case with generic \(\xi\), the tree-level amplitude is \(\xi\)-dependent off-shell, but the \(\xi\)-dependence disappears on-shell. In the massive case, by contrast, the amplitude remains \(\xi\)-dependent even on-shell. One may therefore wonder whether the simultaneous presence of a trace contribution and of a mass term could signal the exchange of a scalar component of the metric. The answer is negative. The explicit computations below show that the trace of the vertex and the existence of a genuine scalar exchange are logically distinct issues.

We begin with the action in \(d=4\) for a real scalar field coupled non-minimally to gravity,
\begin{equation}
S_\phi=\int d^4x\,\sqrt{-g}\left[
\frac12 g^{\mu\nu}\nabla_\mu\phi\nabla_\nu\phi
-\frac12 m^2\phi^2
-\frac12 \xi R\phi^2
\right],
\end{equation}
using the mostly-minus convention
\begin{equation}
\eta_{\mu\nu}=\mathrm{diag}(+,-,-,-).
\end{equation}
At linear order in the metric fluctuation \(h_{\mu\nu}\), the scalar-graviton vertex is
\begin{equation}
V_{\mu\nu}(p_1,p_2)=
p_{1\mu}p_{2\nu}+p_{1\nu}p_{2\mu}
-\eta_{\mu\nu}(p_1\!\cdot\! p_2-m^2)
+2\xi\big(\eta_{\mu\nu}q^2-q_\mu q_\nu\big),
\qquad q=p_1+p_2.
\end{equation}
Its trace is
\begin{equation}
V^\mu{}_\mu=-2\,p_1\!\cdot\! p_2+4m^2+6\xi q^2.
\label{eq:traceV_standalone_rewritten}
\end{equation}
The fact that this trace does not vanish for generic \(\xi\) and \(m\) will be central in what follows. However, as we shall show, its presence does not imply the propagation of a scalar metric degree of freedom.

We consider the tree-level \(2\to2\) scattering of scalar particles without initially imposing any on-shell condition. The kinematics is conveniently described in terms of the Mandelstam invariants
\begin{equation}
s=s_{12}=(p_1+p_2)^2,
\qquad
t=s_{13}=(p_1-p_3)^2,
\qquad
u=s_{14}=(p_1-p_4)^2,
\end{equation}
which satisfy the off-shell identity
\begin{equation}
s+t+u=\sum_{i=1}^4 p_i^2.
\label{eq:stu_identity_standalone_rewritten}
\end{equation}
For each channel $s$, $t$, and $u$, the exchange amplitude can be written as
\begin{equation}
i\mathcal M_\mathrm{ch}=
-\frac{i\kappa^2}{4s_\mathrm{ch}}
\left[
V^{(\mathrm{ch})}_{\mu\nu}V^{(\mathrm{ch})\mu\nu}
-\frac12 V^{(\mathrm{ch})}V^{(\mathrm{ch})}
\right],
\label{eq:Mx_standalone_rewritten}
\end{equation}
where \(s_\mathrm{ch}\in\{s,t,u\}\) denotes the corresponding channel invariant.

To isolate the effect of the non-minimal coupling, it is convenient to split the vertex into a minimal part and a purely \(\xi\)-dependent contribution ($p$ and $q$ are generic momenta of the scalars in the graviton/scalar/scalar vertex)
\begin{equation}
V_{\mu\nu}^{(\mathrm{ch})}(p,q)=
V_{0\,\mu\nu}^{(\mathrm{ch})}(p,q)+2\xi\,X_{\mu\nu}^{(\mathrm{ch})}(p,q),
\end{equation}
with
\begin{equation}
X_{\mu\nu}^{(\mathrm{ch})}(p,q)=
\eta_{\mu\nu}(p+q)^2-(p+q)_\mu(p+q)_\nu,
\end{equation}
and
\begin{equation}
V_{0\,\mu\nu}^{(\mathrm{ch})}(p,q)=
p_\mu q_\nu+p_\nu q_\mu-\eta_{\mu\nu}(p\!\cdot\! q-m^2).
\end{equation}
Defining
\begin{equation}
N^{(\mathrm{ch})}\equiv
V^{(\mathrm{ch})}_{\mu\nu}V^{(\mathrm{ch})\mu\nu}
-\frac12 V^{(\mathrm{ch})}V^{(\mathrm{ch})},
\end{equation}
one finds that the \(\xi\)-dependence reorganizes in the simple form
\begin{equation}
N^{(\mathrm{ch})}
=
N_0^{(\mathrm{ch})}+(2\xi-6\xi^2)s_\mathrm{ch},
\label{eq:N_decomposition_rewritten}
\end{equation}
where \(N_0^{(\mathrm{ch})}\) denotes the $\xi$-independent contribution. Substituting this decomposition into \eqref{eq:Mx_standalone_rewritten}, the single-channel amplitude becomes
\begin{equation}
i\mathcal M_\mathrm{ch}
=
-\frac{i\kappa^2}{4s_\mathrm{ch}}N_0^{(\mathrm{ch})}
-\frac{i\kappa^2}{4}(2\xi-6\xi^2)s_\mathrm{ch}.
\end{equation}
Summing over the three channels gives
\begin{equation}
i\mathcal M(\xi,m)=
i\mathcal M_0(m)-\frac{i\kappa^2}{4}(2\xi-6\xi^2)(s+t+u),
\qquad
i\mathcal M_0(m)\equiv
-\frac{i\kappa^2}{4}
\left(
\frac{N_0^{(\mathrm{s})}}{s}+\frac{N_0^{(\mathrm{t})}}{t}+\frac{N_0^{(\mathrm{u})}}{u}
\right).
\label{eq:master_amp_standalone_rewritten}
\end{equation}
This is the basic formula from which the subsequent discussion follows. The minimal numerators can be written entirely in terms of the invariants \(s,t,u\) and the external virtualities \(p_i^2\):
\begin{align}
N_0^{(\mathrm{s})}&=
\frac12\!\left[
(t-p_1^2-p_3^2)(t-p_2^2-p_4^2)
+(u-p_1^2-p_4^2)(u-p_2^2-p_3^2)
-(s-p_1^2-p_2^2)(s-p_3^2-p_4^2)
\right] +\left(2sm^2-6m^4\right),\\[1mm]
N_0^{(\mathrm{t})}&=
\frac12\!\left[
(s-p_1^2-p_2^2)(s-p_3^2-p_4^2)
+(u-p_1^2-p_4^2)(u-p_2^2-p_3^2)
-(t-p_1^2-p_3^2)(t-p_2^2-p_4^2)
\right]
+\left(2tm^2-6m^4\right),\\[1mm]
N_0^{(\mathrm{u})}&=
\frac12\!\left[
(s-p_1^2-p_2^2)(s-p_3^2-p_4^2)
+(t-p_1^2-p_3^2)(t-p_2^2-p_4^2)
-(u-p_1^2-p_4^2)(u-p_2^2-p_3^2)
\right]
+\left(2um^2-6m^4\right).
\end{align}

\subsection{Off-shell and on-shell \texorpdfstring{$\xi$}{xi} dependence}

The structure of \eqref{eq:master_amp_standalone_rewritten} makes the dependence on \(\xi\) particularly transparent
\begin{equation}
i\mathcal M(\xi,m)=
i\mathcal M_0(m)-\frac{i\kappa^2}{4}(2\xi-6\xi^2)(s+t+u).
\label{eq:summary_master_rewritten}
\end{equation}
This immediately shows that the \(\xi\)-dependent part is polynomial and proportional to the sum of the channel invariants.

In the massless case, \(m=0\), the off-shell amplitude is
\begin{equation}
i\mathcal M^{\rm off}(\xi,0)=
i\mathcal M_0^{\rm off}(0)-\frac{i\kappa^2}{4}(2\xi-6\xi^2)(s+t+u),
\end{equation}
and is therefore \(\xi\)-dependent. If one goes on-shell, \(p_i^2=0\), then \(s+t+u=0\), so that
\begin{equation}
i\mathcal M^{\rm on}(\xi,0)=i\mathcal M_0^{\rm on}(0)
=-\frac{i\kappa^2}{4}
\left(
\frac{tu}{s}+\frac{su}{t}+\frac{st}{u}
\right).
\end{equation}
Thus the massless on-shell amplitude is independent of \(\xi\), even though the local vertex has a nonvanishing trace for generic \(\xi\).

The situation is different in the massive case. Off-shell, the amplitude is again \(\xi\)-dependent by \eqref{eq:summary_master_rewritten}. For equal external masses on-shell, \(p_i^2=m^2\), one has
\begin{equation}
s+t+u=4m^2,
\end{equation}
and therefore
\begin{equation}
i\mathcal M^{\rm on}(\xi,m)=
i\mathcal M_0^{\rm on}(m)-i\kappa^2m^2(2\xi-6\xi^2).
\label{eq:summary_massive_on_rewritten}
\end{equation}
Hence, unlike the massless case, the massive on-shell amplitude retains a nontrivial dependence on \(\xi\). In particular, at the conformal value \(\xi=1/6\),
\begin{equation}
i\mathcal M^{\rm on}(\xi=1/6,m)=
i\mathcal M_0^{\rm on}(m)-\frac{i\kappa^2m^2}{6}.
\end{equation}
The conclusion is therefore clear. In the massless theory, the amplitude is \(\xi\)-dependent off-shell but \(\xi\)-independent on-shell. In the massive theory, by contrast, the amplitude depends on \(\xi\) both off-shell and on-shell. This difference reflects the distinct role played by the two explicit breakings of conformal symmetry, namely the non-minimal coupling and the scalar mass. However, this distinction should not be confused with the propagation of an independent scalar gravitational degree of freedom. Even when the trace of the vertex is nonzero, and even when the amplitude retains a nontrivial dependence on \(\xi\), the full exchange still need not correspond to scalar propagation.

\section{The activation of the scalar sector in the Fierz-Pauli Lagrangian with the anomaly}
\label{closing}
To illustrate the unusual behaviour of the scalar interaction mediated by virtual gravitational exchanges 
we start from the quadratic Einstein--Hilbert action in Fierz--Pauli form, and isolate the scalar longitudinal sector in \(d=4\) from \eqref{eq:cov_decomp_ped}  
\begin{equation}
h_{\mu\nu}^{(S)}
=
\left(\partial_\mu\partial_\nu-\frac14\eta_{\mu\nu}\Box\right)w
+\frac14\eta_{\mu\nu}h,
\qquad
\Phi_{cov}\equiv h-\Box w,
\label{eq:scalar_decomp}
\end{equation}
having used the covariant decomposition.
Introducing \(U\equiv \Box w\), one has
\begin{equation}
\partial_\mu h^{\mu\nu}=\frac14\,\partial^\nu(h+3U),\qquad
h=\eta^{\mu\nu}h_{\mu\nu}^{(S)}.
\label{eq:divh}
\end{equation}
Substituting \eqref{eq:scalar_decomp} into the Fierz-Pauli Lagrangian \eqref{FP} we obtain
\begin{equation}
\mathcal L_{S}^{(2)}
=
-\frac38\Big[(\partial h)^2-2\,\partial h\!\cdot\!\partial U+(\partial U)^2\Big]
=
-\frac38\big(\partial(h-U)\big)^2
=
-\frac38(\partial\Phi_{cov})^2.
\label{eq:Lsc}
\end{equation}
Therefore the scalar contribution generates the action
\begin{equation}
S_{EH,S}^{(2)}
=
-\frac{3}{32\kappa^2}\int d^4x\,(\partial\Phi_{cov})^2
=
\frac{3}{32\kappa^2}\int d^4x\,\Phi_{cov}\,\Box\Phi_{cov},
\label{eq:SEHscalar}
\end{equation}
up to boundary terms, which is the scalar kinetic term quoted in the main text. Using the 
linearized expression of the Ricci scalar 
\begin{equation}
R^{(1)}=\partial_\mu\partial_\nu h^{\mu\nu}-\Box h
\label{eq:R1def}
\end{equation}
from \eqref{eq:scalar_decomp},
\begin{equation}
\partial_\mu\partial_\nu h^{\mu\nu}
=\frac14\Box h+\frac34\Box^2 w,
\end{equation}
one derives the relation
\begin{equation}
R^{(1)}
=
-\frac34\Box(h-\Box w)
=
-\frac34\Box\Phi_{cov}.
\label{eq:R1Phi}
\end{equation}
At this point, if we include the anomaly-pole interaction from the $TJJ$ vertex we rewrite the pole action as
\begin{equation}
S_{\mathrm{pole}}
=
\frac{b}{8\pi^2}\int d^4x\,
R^{(1)}\frac1\Box F^2
=
\frac{b}{8\pi^2}\int d^4x
\left(-\frac34\Box\Phi_{cov}\right)\frac1\Box F^2
=
-\frac{3b}{32\pi^2}\int d^4x\,\Phi_{cov} F^2.
\label{scalar1}
\end{equation}
Combining \eqref{eq:SEHscalar} and \eqref{scalar1} the action accounting for the anomaly becomes
\begin{equation}
S_\Phi
=
\int d^4x\left[
\frac{3}{32\kappa^2}\Phi_{cov}\Box\Phi_{cov}
-\frac{3b}{32\pi^2}\Phi_{cov} F^2
\right].
\label{eq:SphiTotal}
\end{equation}
that once varies in the presence of an anomaly, eq. \eqref{eq:no_Phicov_variation_ped}, gives the
modified equation
\begin{equation}
\label{cov}
\frac{3}{32\kappa^2}\Box\Phi_{cov}-\frac{3b}{32\pi^2}F^2=0
\quad\Longrightarrow\quad
\Box\Phi_{cov}=\frac{\kappa^2 b}{\pi^2}F^2.
\end{equation}
with $b$ denoting the multiplicity factor (the number of massless conformal states) of the conformal states which have been integrated over in the quantum corrections (in the $TJJ$).
Clearly, $\Phi_{cov}$ is sourced by the anomaly.\\
At this point it is useful to distinguish between two related but conceptually different effective descriptions of the anomaly-induced interaction.

If one integrates out the matter fields while keeping both the metric and the gauge field as explicit variables, the anomaly appears in the mixed 1PI effective action \(\Gamma[g,A]\) through the nonlocal \(TJJ\) vertex. To linear order in the metric fluctuation and quadratic order in the gauge field, this gives a contribution of the schematic form
\begin{equation}
\Gamma_{\rm anom}[g,A]\supset \int d^4x\, R^{(1)}\,\Box^{-1}F^2,
\label{eq:mixed_anom_action}
\end{equation}
or, equivalently, in terms of the covariant scalar sector,
\begin{equation}
\Gamma_{\rm anom}[g,A]\supset \int d^4x\, \Phi_{cov}\,F^2,
\label{eq:phicov_f2_action}
\end{equation}
using \eqref{eq:R1Phi}. In this representation the anomaly-induced term is part of the nonlocal effective action of the coupled gravity-matter system; it is not an external source inserted by hand. Most importantly, the anomaly pole selects a unique scalar direction in the metric sector, namely \(\Phi_{cov}\), when the latter is described in the covariant decomposition, as emphasized in \cite{Mottola:2016mpl}.

A different effective description is obtained if one eliminates the graviton as well. In that case one sews two \(TJJ\) vertices through an internal graviton line and obtains a reduced effective action for the gauge field alone. As we will show below, this generates schematically a nonlocal four-point interaction of the form, setting \(\kappa=1\),
\begin{equation}
\Gamma_{\rm red}[A]\supset \int d^4x\, d^4y\,
F^2(x)\,G(x-y)\,F^2(y),
\label{eq:reduced_gauge_action}
\end{equation}
where \(G\) denotes the kernel induced by the exchanged gravitational channel and is schematically of \(\Box^{-1}\) type. Restoring the Planck scale, the corresponding momentum-space structure is
\begin{equation}
\Gamma_{\rm red}[A]\sim \frac{1}{M_P^2}\int d^4 q\,
F^2(q)\,\frac{1}{q^2}\,F^2(-q)
\label{eq:F2boxF2}
\end{equation}
up to numerical coefficients, with 
\begin{equation}
F^2(q)=
2\int \frac{d^4p}{(2\pi)^4}
\left[
p_\mu (q-p)_\nu-\eta_{\mu\nu}\,p\!\cdot\!(q-p)
\right]
A^\mu(p)\,A^\nu(q-p).
\end{equation}

It is important to stress that \eqref{eq:F2boxF2} is not a new anomaly structure independent of the original \(TJJ\) vertex. Rather, it is the four-point exchange amplitude generated by the same anomaly-selected scalar channel. For this reason, the appearance of an \(F^2\Box^{-1}F^2\)-type term should be interpreted as the manifestation of a single exchanged scalar contribution in the effective interaction, not as evidence for two independent scalar gravitational degrees of freedom. In covariant language this exchanged channel is again the unique invariant \(\Phi_{cov}\); in the non-covariant decomposition it corresponds to one specific linear combination of two gauge-invariant variables, while the orthogonal scalar combination remains non-propagating. The non covariant decomposition is discussed in Appendix B.\\
A final remark is in order. Since this four-point amplitude is obtained by combining two conformal \(TJJ\) vertices with an intermediate gravitational propagator governed by the Einstein--Hilbert action, which is not conformally invariant, the pole structure of the full sewn amplitude is not guaranteed to remain manifest in a completely factorized form once one passes to momentum space and includes the explicit expressions of the two $F^2$ invariant in \eqref{eq:F2boxF2}. Thus the existence of the anomaly pole is clearest at the level of the mixed \(TJJ\) effective action, while its realization in the reduced four-point amplitude requires a separate analysis.
\section{The anomaly interaction in the conformal limit and the photon 4-point function}
\label{section9}
The structure of the $TJJ$ correlator, as we have seen, is quite simple in the on-shell limit, being characterized by one pole in the trace sector, and one extra pole and a log 
in the traceless sector. The two poles are interpolated by the virtual graviton, coupled to its spin-2 and spin-0 components. Both components can be discussed separately. In each $L/R$ sector the exchange of such poles is interpreted as a light-cone dominated process. This is the kinematical situation in which the residue of the pole at each of the two TJJ vertices is nonvanishing. This point is rather technical and has been discussed  in detail in \cite{Coriano:2025fom}. In this section we are going to 
discuss the structure of this decomposition, extending our former analysis and isolating the contributions related to the spin-2 and spin-0 gravitational components in a rigorous way. We consider the scattering of two on-shell photons mediated by a graviton.  
In this specific kinematic configuration each of the vertices
is characterized by an anomaly form factor satisfying a
sum rule.
We have the amplitudes in the $t$, $s$ and $u$ channels.

We introduce the effective coupling
\begin{equation}
\frac{\bar C_0}{M_P^2}
\equiv
\kappa^2 \frac{3}{2}
\left(\frac{e^2}{48\pi^2}\right)^2 ,
\end{equation}
Hence
\begin{equation}
\label{C0}
\bar C_0 = \frac{3}{2}
\left(\frac{e^2}{48\pi^2}\right)^2 .
\end{equation}
Notice that the effect of the inclusion of an intermediate scalar projector, once we carry out the contractions 
in the $s , t$ and $u$ channels  is just to produce a factor $9/q^2$ in the exchange.   
Since we have the contractions
\begin{align}
\pi_{\mu_2\nu_2}(q)\,
{P^{(0)\,\mu_2\nu_2}{}_{\rho_2\sigma_2}(q)}\frac{1}{q^2}\,
\pi^{\rho_2\sigma_2}(q) 
=
\left(
\pi_{\mu_2\nu_2}(q)\,
\pi^{\mu_2\nu_2}(q)
\right)
\left(
\pi_{\rho_2\sigma_2}(q)\,
\pi^{\rho_2\sigma_2}(q)
\right)
\frac{1}{q^2} 
= \frac{9}{q^2}
\end{align}
on each internal scalar line.  The result is that the interaction between the two sectors is mediated by a single scalar exchange, with the two anomaly component of each vertex appearing symmetrically. 
A similar amplitude can be computed in the other $s$ and $u$ channels. The scalar–exchange amplitudes then take the compact form

\begin{equation}
\mathcal{A}_s^{(0)}
=
-\,\frac{\bar C_0}{M_P^2}
\, u^{\alpha\gamma}(p_1,p_2)
\frac{1}{s}
u^{\beta\delta}(p_3,p_4) ,
\end{equation}

\begin{equation}
\mathcal{A}_t^{(0)}
=
-\,\frac{\bar C_0}{M_P^2}
\, u^{\alpha\beta}(-p_1,p_3)
\frac{1}{t}
u^{\gamma\delta}(-p_2,p_4) ,
\end{equation}

\begin{equation}
\mathcal{A}_u^{(0)}
=
-\,\frac{\bar C_0}{M_P^2}
\, u^{\alpha\delta}(-p_1,p_4)
\frac{1}{u}
u^{\gamma\beta}(-p_2,p_3) .
\end{equation}

It is useful to stress the meaning of this scalar-exchange representation. The
spin--0 contribution isolated in this way should be understood as an effective
anomaly-induced scalar channel, equivalently described covariantly by
\(\Phi_{cov}\), rather than as a new elementary asymptotic graviton degree of
freedom. In other words, the scalar exchange appearing in the factorized
amplitudes is a convenient parametrization of the nonlocal pole structure of the
full \(TJJ\)-mediated interaction. It reproduces the same \(1/\Box\) behaviour of
the complete amplitude, but it should not be interpreted as the LSZ exchange of
an independent fundamental scalar particle. This interpretation is consistent
with the analysis of Section~\ref{app:TT_anomaly_reorganized}, where the scalar
sector of the Einstein--Hilbert/Fierz--Pauli action remains constrained and the
anomaly activates a distinguished nonlocal scalar response channel without
introducing an additional local propagating graviscalar in the spectrum.

\begin{table}[t]
\centering
\small
\renewcommand{\arraystretch}{1.12}
\setlength{\tabcolsep}{4.5pt}
\begin{tabular}{cc cc c @{\hspace{1em} \vrule width 0.75pt\hspace{1em} } cc cc c}
\toprule
$p_i$ & $\epsilon_i$ & $p_j$ & $\epsilon_j$ & $\epsilon(p_i)_\alpha \epsilon(p_j)_\beta u^{\alpha\beta}(p_i,p_j)$ &
$p_i$ & $\epsilon_i$ & $p_j$ & $\epsilon_j$ & $\epsilon(p_i)_\alpha \epsilon(p_j)_\beta u^{\alpha\beta}(p_i,p_j)$ \\
\midrule
$p_1$ & $+$ & $p_1$ & $+$ & $0$ &
$p_2$ & $+$ & $p_2$ & $+$ & $0$ \\

$p_1$ & $+$ & $p_1$ & $-$ & $0$ &
$p_2$ & $+$ & $p_2$ & $-$ & $0$ \\

$p_1$ & $+$ & $p_2$ & $+$ & $-\frac{s}{2}$ &
$p_2$ & $+$ & $p_3$ & $+$ & $\frac{u(t-u)}{2s}$ \\

$p_1$ & $+$ & $p_2$ & $-$ & $0$ &
$p_2$ & $+$ & $p_3$ & $-$ & $\frac{tu}{s}$ \\

$p_1$ & $+$ & $p_3$ & $+$ & $\frac{t(s+2t)}{2s}$ &
$p_2$ & $+$ & $p_4$ & $+$ & $\frac{t(u-t)}{2s}$ \\

$p_1$ & $+$ & $p_3$ & $-$ & $-\frac{tu}{s}$ &
$p_2$ & $+$ & $p_4$ & $-$ & $\frac{tu}{s}$ \\

$p_1$ & $+$ & $p_4$ & $+$ & $\frac{u(u-t)}{2s}$ &
$p_2$ & $-$ & $p_2$ & $-$ & $0$ \\

$p_1$ & $+$ & $p_4$ & $-$ & $-\frac{tu}{s}$ &
$p_2$ & $-$ & $p_3$ & $+$ & $\frac{tu}{s}$ \\

$p_1$ & $-$ & $p_1$ & $-$ & $0$ &
$p_2$ & $-$ & $p_3$ & $-$ & $\frac{u(t-u)}{2s}$ \\

$p_1$ & $-$ & $p_2$ & $+$ & $0$ &
$p_2$ & $-$ & $p_4$ & $+$ & $\frac{tu}{s}$ \\

$p_1$ & $-$ & $p_2$ & $-$ & $-\frac{s}{2}$ &
$p_2$ & $-$ & $p_4$ & $-$ & $\frac{t(u-t)}{2s}$ \\

$p_1$ & $-$ & $p_3$ & $+$ & $-\frac{tu}{s}$ &
$p_3$ & $+$ & $p_3$ & $+$ & $0$ \\

$p_1$ & $-$ & $p_3$ & $-$ & $\frac{t(s+2t)}{2s}$ &
$p_3$ & $+$ & $p_3$ & $-$ & $0$ \\

$p_1$ & $-$ & $p_4$ & $+$ & $-\frac{tu}{s}$ &
$p_3$ & $+$ & $p_4$ & $+$ & $\frac{s}{2}$ \\

$p_1$ & $-$ & $p_4$ & $-$ & $\frac{u(u-t)}{2s}$ &
$p_3$ & $+$ & $p_4$ & $-$ & $0$ \\

$p_3$ & $-$ & $p_3$ & $-$ & $0$ &
$p_4$ & $+$ & $p_4$ & $+$ & $0$ \\

$p_3$ & $-$ & $p_4$ & $+$ & $0$ &
$p_4$ & $+$ & $p_4$ & $-$ & $0$ \\

$p_3$ & $-$ & $p_4$ & $-$ & $\frac{s}{2}$ &
$p_4$ & $-$ & $p_4$ & $-$ & $0$ \\
\bottomrule
\end{tabular}
\caption{All possible contractions of polarization vectors with $u^{\alpha\beta}$.}
\label{tab:u_eps_eps}
\end{table}
Introducing the polarization four-vectors in the center of mass frame of the scattering photons
\begin{align}
\varepsilon^\mu(p_1,\pm)
&= \frac{1}{\sqrt{2}}(0,1,\pm i,0),
&
\varepsilon^\mu(p_2,\pm)
&= \frac{1}{\sqrt{2}}(0,1,\mp i,0),
\\[6pt]
\varepsilon^\mu(p_3,\pm)
&= \frac{1}{\sqrt{2}}(0,\cos\theta,\pm i,-\sin\theta),
&
\varepsilon^\mu(p_4,\pm)
&= \frac{1}{\sqrt{2}}(0,-\cos\theta,\pm i,\sin\theta).
\end{align}
we compute the contraction
\begin{equation}
\mathcal U_{p,q}(\lambda_1,\lambda_2)
=
\varepsilon_\alpha(p,\lambda_1)\,
\varepsilon_\beta(q,\lambda_2)\,
u^{\alpha\beta}(p,q).
\end{equation}
for the various helicity configurations. The results are collected in Table \ref{tab:u_eps_eps}. Notice that $\mathcal{U}_{p,q}$ satisfies the following property
\begin{equation}\label{eq:U_property}
\mathcal U_{-p,q}(\lambda_1,\lambda_2)
=
-\,\mathcal U_{p,q}(\lambda_1,\lambda_2),
\end{equation}
indeed, using transversality, the contraction on  the left hand side reduces to
\begin{equation}
\mathcal U_{p,q}
=
(p\!\cdot\!q)\,
\varepsilon(p)\!\cdot\!\varepsilon(q);
\end{equation}
and using the identity
\begin{equation}
u^{\alpha\beta}(-p,q) = -\,u^{\alpha\beta}(p,q),
\end{equation}
eq. \eqref{eq:U_property} follows immediately.

Upon contraction with the polaritazion vectors the amplitudes for the three channels read:
\begin{align}\label{eq:channels_TJJ}
	\mathcal{M}_s(\lambda_1,\lambda_2,\lambda_3,\lambda_4) &= \frac{\bar C_0}{M_P^2}\mathcal{U}_{p_1,p_2}(\lambda_1,\lambda_2)\frac{1}{s}\mathcal{U}_{p_1,p_2}(\lambda_3,\lambda_4),\\
	\mathcal{M}_t(\lambda_1,\lambda_2,\lambda_3,\lambda_4) &= \frac{\bar C_0}{M_P^2}\mathcal{U}_{-p_1,p_3}(\lambda_1,\lambda_3)\frac{1}{t}\mathcal{U}_{-p_2,p_4}(\lambda_2,\lambda_4),\\
	\mathcal{M}_u(\lambda_1,\lambda_2,\lambda_3,\lambda_4) &= \frac{\bar C_0}{M_P^2}\mathcal{U}_{-p_1,p_4}(\lambda_1,\lambda_4)\frac{1}{u}\mathcal{U}_{-p_2,p_3}(\lambda_2,\lambda_3),
\end{align}
\begin{table}[t]
\centering
\small
\renewcommand{\arraystretch}{1.12}
\setlength{\tabcolsep}{4.2pt}
\resizebox{\textwidth}{!}{%
\begin{tabular}{cccc c c c c  @{\hspace{1em} \vrule width 0.75pt\hspace{1em} }cccc c c c c}
\toprule
$p_1$&$p_2$&$p_3$&$p_4$&
$\bar{ \mathcal{M}}_s$&$\bar{ \mathcal{M}}_t$&$\bar{ \mathcal{M}}_u$&$\bar{ \mathcal{M}}$&
$p_1$&$p_2$&$p_3$&$p_4$&
$\bar{ \mathcal{M}}_s$&$\bar{ \mathcal{M}}_t$&$\bar{ \mathcal{M}}_u$&$\bar{ \mathcal{M}}$
 \\
\midrule
$+$&$+$&$+$&$+$&$-\frac{s}{4}$&$-\frac{t (t-u)^2}{4 s^2}$&$-\frac{u (t-u)^2}{4 s^2}$&$-\frac{t u}{s}$&
$-$&$+$&$+$&$+$&$0$&$\frac{t u (t-u)}{2 s^2}$&$\frac{t u (u-t)}{2 s^2}$&$0$ \\

$+$&$+$&$+$&$-$&$0$&$\frac{t u (t-u)}{2 s^2}$&$\frac{t u (u-t)}{2 s^2}$&$0$&
$-$&$+$&$+$&$-$&$0$&$-\frac{t u^2}{s^2}$&$-\frac{u (t-u)^2}{4 s^2}$&$-\frac{u}{4}$ \\

$+$&$+$&$-$&$+$&$0$&$\frac{t u (t-u)}{2 s^2}$&$\frac{t u (u-t)}{2 s^2}$&$0$&
$-$&$+$&$-$&$+$&$0$&$-\frac{t (t-u)^2}{4 s^2}$&$-\frac{t^2 u}{s^2}$&$-\frac{t}{4}$ \\

$+$&$+$&$-$&$-$&$-\frac{s}{4}$&$-\frac{t u^2}{s^2}$&$-\frac{t^2 u}{s^2}$&$-\frac{(t-u)^2}{4 s}$&
$-$&$+$&$-$&$-$&$0$&$\frac{t u (t-u)}{2 s^2}$&$\frac{t u (u-t)}{2 s^2}$&$0$ \\

$+$&$-$&$+$&$+$&$0$&$\frac{t u (t-u)}{2 s^2}$&$\frac{t u (u-t)}{2 s^2}$&$0$&
$-$&$-$&$+$&$+$&$-\frac{s}{4}$&$-\frac{t u^2}{s^2}$&$-\frac{t^2 u}{s^2}$&$-\frac{(t-u)^2}{4 s}$ \\

$+$&$-$&$+$&$-$&$0$&$-\frac{t (t-u)^2}{4 s^2}$&$-\frac{t^2 u}{s^2}$&$-\frac{t}{4}$&
$-$&$-$&$+$&$-$&$0$&$\frac{t u (t-u)}{2 s^2}$&$\frac{t u (u-t)}{2 s^2}$&$0$ \\

$+$&$-$&$-$&$+$&$0$&$-\frac{t u^2}{s^2}$&$-\frac{u (t-u)^2}{4 s^2}$&$-\frac{u}{4}$&
$-$&$-$&$-$&$+$&$0$&$\frac{t u (t-u)}{2 s^2}$&$\frac{t u (u-t)}{2 s^2}$&$0$ \\

$+$&$-$&$-$&$-$&$0$&$\frac{t u (t-u)}{2 s^2}$&$\frac{t u (u-t)}{2 s^2}$&$0$&
$-$&$-$&$-$&$-$&$-\frac{s}{4}$&$-\frac{t (t-u)^2}{4 s^2}$&$-\frac{u (t-u)^2}{4 s^2}$&$-\frac{t u}{s}$ \\
\bottomrule
\end{tabular}}
\caption{Four-particle configurations and corresponding values. Here $\bar{ \mathcal{M}}_s=\mathcal U(p_1,p_2)\frac1s\mathcal U(p_3,p_4)$, $\bar{\mathcal M}_t=\mathcal U(-p_1,p_3)\frac1t\mathcal U(-p_2,p_4)$, $\bar{\mathcal M}_u=\mathcal U(-p_1,p_4)\frac1u\mathcal U(-p_2,p_3)$, and $\bar{\mathcal M}$ is the total amplitude. In $\bar{ \mathcal{M}}$ we have omitted an overall constant $C/M_P^2$.}
\label{tab:TJJ-TTJ}
\end{table}
The amplitudes can be generated from the action 
\begin{equation}
S_{\text{eff}}
=
\frac{\bar C_0}{M_P^2}
\int d^4x \, d^4y \,
\big(F_{\mu\nu}F^{\mu\nu}\big)(x)\,
G(x-y)\,
\big(F_{\rho\sigma}F^{\rho\sigma}\big)(y),
\tag{6.81}
\end{equation}
where the Green's function satisfies
\begin{equation}
\Box_x G(x-y)=\delta^{(4)}(x-y).
\end{equation}
Equivalently,
\begin{equation}
\mathcal S_{\rm eff}
=
\frac{\bar C_0^2}{M_P^2}
\int d^4x \,
\big(F_{\mu\nu}F^{\mu\nu}\big)\,
\frac{1}{\Box}\,
\big(F_{\rho\sigma}F^{\rho\sigma}\big).
\end{equation}
Although the overall mass dimension of the Lagrangian density is four,
\begin{equation}
\left[
\frac{1}{M_P^2}
(F^2)\frac{1}{\Box}(F^2)
\right]=4,
\end{equation}
the operator is intrinsically nonlocal because
\begin{equation}
\left(\frac{1}{\Box} f\right)(x)
=
\int d^4y\, G(x-y)\, f(y),
\end{equation}
so the action couples field strengths at different spacetime points. 
This nonlocality, however, does not imply that the corresponding momentum-space amplitude exhibits a genuine propagating massless scalar exchange. As we show below, in the present kinematic setting it reduces instead to a contact contribution, namely to a term polynomial in the external momenta. This should not be surprising: contact interactions commonly arise when an underlying interaction is probed at energies much smaller than the characteristic mass scale of its mediator, although in the anomaly-induced case the interpretation is more subtle.\\
For instance, in Fermi theory of weak interactions one encounters the four-fermion contact term
\begin{equation}
\mathcal L_F
=
-\frac{G_F}{\sqrt{2}}
\left(\bar\psi \gamma^\mu (1-\gamma^5)\psi\right)
\left(\bar\chi \gamma_\mu (1-\gamma^5)\chi\right),
\label{fermitheory}
\end{equation}
which is a local dimension-six operator suppressed by a heavy mass scale,
\begin{equation}
G_F \sim \frac{1}{M_{W/Z}^2}.
\end{equation}
In this case the heavy mediators have been integrated out, producing a genuine local contact interaction valid at energies \(E \ll M_{W/Z}\). By contrast, the term
\begin{equation}
\frac{1}{M_P^2}\,F^2\,\frac{1}{\Box}\,F^2
\end{equation}
does not originate from integrating out a heavy field. Rather, it is associated with the exchange of a massless channel encoded in a nonlocal effective interaction, with the overall strength suppressed by the Planck scale. For this reason, the inverse d'Alembertian cannot be expanded into a finite tower of local operators, and the effective action remains intrinsically nonlocal even at low energies.\\
`
\subsection{Contact interactions mediated by the EH action}
It is interesting to investigate the implications of this result more closely. 
In momentum space the amplitude is of the form 
\begin{equation}
\mathcal M_\Phi(q)\propto \frac{V_{F^2}(q)\,V'_{F^2}(q)}{q^2},
\qquad
\operatorname{Res}_{q^2=0}\mathcal M_\Phi
\propto
\lim_{q^2\to0}V_{F^2}(q)V'_{F^2}(q),
\label{eq:ResidueCriterion}
\end{equation}
more details will we given in the next sections.
For photon scattering, the anomaly-induced spin--0 exchange can be represented
in terms of an effective vertex using \eqref{uab},
\begin{equation}
iV^{\alpha\beta}(p_i,p_j)
=
ig_\Phi\,u^{\alpha\beta}(p_i,p_j),
\qquad
u^{\alpha\beta}(p_i,p_j)
=
(p_i\!\cdot p_j)\eta^{\alpha\beta}-p_j^\alpha p_i^\beta.
\label{eq:Vertexu}
\end{equation}
where \(\Phi\) denotes the effective scalar channel associated with the
anomaly-induced spin--0 exchange. \\
A detailed analysis of these 2-to-2 amplitudes will be discussed in the next sections, but here we anticipate a result in order to characterize their specific features. \\
Recalling that by definition 
\begin{equation}
\mathcal{U}_{ij} = 
u^{\alpha\beta}(p_i,p_j)\,\varepsilon_{i\alpha}\varepsilon_{j\beta}
=
(p_i\!\cdot p_j)(\varepsilon_i\!\cdot\varepsilon_j)
-(p_j\!\cdot\varepsilon_i)(p_i\!\cdot\varepsilon_j),
\label{eq:Uij}
\end{equation}
with photon polarization 4-vectors $\varepsilon_j$ and scalar propagator \(i/(q^2+i0)\), the three channels give \begin{equation}
i\mathcal M_s=i g_\Phi^2\frac{\mathcal{U}_{12}\,\mathcal{U}_{34}}{s+i0},\qquad
i\mathcal M_t=i g_\Phi^2\frac{\mathcal{U}_{13}\,\mathcal{U}_{24}}{t+i0},\qquad
i\mathcal M_u=i g_\Phi^2\frac{\mathcal{U}_{14}\,U_{23}}{u+i0},
\end{equation}
thus
\begin{equation}
i\mathcal M_{\gamma\gamma\to\gamma\gamma}^{(\Phi)}
=
i g_\Phi^2\left(
\frac{\mathcal{U}_{12}\,\mathcal{U}_{34}}{s+i0}
+\frac{\mathcal{U}_{13}\,\mathcal{U}_{24}}{t+i0}
+\frac{\mathcal{U}_{14}\,\mathcal{U}_{23}}{u+i0}
\right).
\label{eq:FullAmpPhi}
\end{equation}
and for on-shell photons (with $\kappa\to 1$)
\begin{equation}
\frac{\mathcal{U}\,\mathcal{U}'}{s_{\mathrm{ch}}}\sim s_{\mathrm{ch}},
\end{equation}
where $s_{\mathrm{ch}}\in\{s,t,u\}$,
i.e. each channel is polynomial as \(s_{\mathrm{ch}}\to0\) and the pole residue vanishes. Correcting by the Planck mass, the amplitude takes the form of a contact term 
\beq
\mathcal M_{\mathrm{ch}}\sim \frac{s_{\mathrm{ch}}}{M_P^2}.
\label{eq:OffshellPole}
\end{equation}
Therefore the \(1/\Box\) structure is present at the level of the effective action \eqref{eq:ResidueCriterion}, while the visibility of a physical scalar pole in a given observable is controlled by the channel-dependent residue criterion in the same equation.  

One should not be surprised of the unusual behaviour associated with anomaly-mediated interactions. One can show by a careful analysis of the interaction, that the coupling of massless poles and their residues are strictly connected with the off-shell/on-shell kinematics of each 3-point vertex.\\
 For example, one can show that the actual residue of a single pole in each anomaly vertex is nonzero only if  the graviton line is off-shell and the two spin-1 are on-shell. The interaction has to take place strictly on the light cone in order for each residue in each 3-point vertex to be nonzero \cite{Giannotti:2008cv,Coriano:2025fom}. \\
Notice that the existence of a massless pole in the \(TJJ\) correlator must be tested through the residue
\begin{equation}
\mathcal R^{\mu\nu\alpha\beta}(p_1,p_2)
\equiv
\lim_{s\to 0}
s\,\langle T^{\mu\nu}(q)\,J^\alpha(p_1)\,J^\beta(p_2)\rangle,
\qquad s\equiv q^2,\quad q=p_1+p_2.
\label{eq:Rdef_lightcone}
\end{equation}
A genuine particle-like massless exchange requires \(\mathcal R^{\mu\nu\alpha\beta}\neq0\). The key result  is that this condition is highly restrictive: the residue survives only in the massless, on-shell light-cone configuration.
To make this explicit, one compares two limiting prescriptions for \(q^2\to0\): (i) a light-cone scaling at fixed large \(q^+\), and (ii) a soft limit \(q^\lambda\to0\), as discussed in \cite{Coriano:2025fom}. In both cases, the same selection rule emerges. If at least one external current is off-shell, or if the loop fermion is massive, the residue vanishes:
\begin{equation}
\lim_{s\to0}
s\,\langle T^{\mu\nu}(q)\,J^\alpha(p_1)\,J^\beta(p_2)\rangle
=0,
\qquad
(s_1\neq0\ \text{or}\ s_2\neq0\ \text{or}\ m\neq0),
\label{eq:residue_zero_generic}
\end{equation}
where \(s_1=p_1^2\), \(s_2=p_2^2\). Conversely, in the strict on-shell massless configuration,
\begin{equation}
\lim_{s\to0}
s\,\langle T^{\mu\nu}(q)\,J^\alpha(p_1)\,J^\beta(p_2)\rangle
\neq0,
\qquad
(s_1=s_2=0,\ m=0).
\label{eq:residue_nonzero_strict}
\end{equation}
Therefore, the anomaly-coupled pole is not a generic off-shell feature of the correlator; it is activated only on a specific kinematic support.
As discussed in Section \ref{reff2}, in this limit the residue contains two tensor structures,
\begin{equation}
\mathcal R^{\mu\nu\alpha\beta}
=
c_1\,\tilde\phi_1^{\mu\nu\alpha\beta}
+
c_2\,\tilde\phi_2^{\mu\nu\alpha\beta},
\label{eq:residue_two_structures}
\end{equation}
defined in \eqref{widetilde1} and \eqref{widetilde2}. The first term is the anomaly-related trace contribution, while the second is traceless and corresponds to the additional pole term identified in \eqref{expoleac}.\\
Thus, even when the residue is nonvanishing, the conformal case is more restrictive than the naive picture of an ordinary particle pole mediating the interaction.\\
This clarifies why anomaly mediation need not generate an observable pole in a given four-point process and provides the appropriate conceptual bridge: anomaly poles couple only under sharply constrained kinematical conditions, and a nonzero residue is a light-cone/on-shell property rather than a generic off-shell feature of an anomaly vertex. This explains why one may consistently find no anomaly-induced scalar pole in a specific four-point amplitude, despite the existence of a nonlocal anomaly action. In such situations, the surviving contributions should instead be interpreted as contact terms, which are a genuine and characteristic manifestation of anomaly-mediated interactions.

\begin{figure}[t]
\begin{center}
\includegraphics[scale=0.7,angle=0]{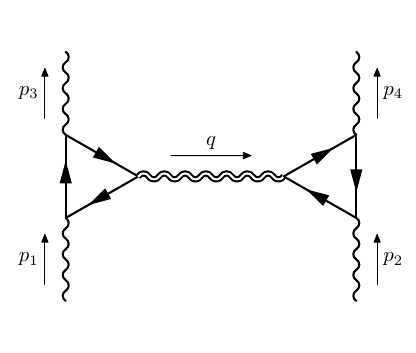}
\includegraphics[scale=0.7,angle=0]{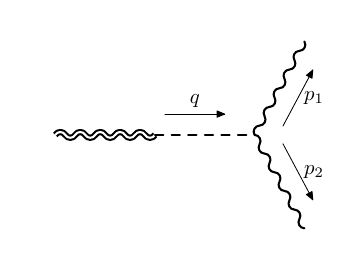}
\caption{The free field theory realization of the $TJJ$ vertex, here represented for an Abelian theory such as QED. The $J$ are coupled to photon lines (left). The mixing interaction induced by the graviton and the, corresponding in momentum space of a projector $\pi^{\mu\nu}$ on the external graviton line of any anomaly vertex (right).} 
\label{expansionX1}
\end{center}
\end{figure}

\section{Anomaly mediation with a massive fermion}

We now turn to the on-shell regime and analyze the spectral properties of the conformal-anomaly form factor \(\Phi_{TJJ}\) in the presence of $n_F$ massive fermions. The spectral analysis of this form factor has been presented in \cite{Coriano:2025fom} for QCD, but it can be easily extrapolated to the Abelian 
case. We are going to briefly summarize it.  \\
The anomaly contribution 
\beq
\frac{1}{ 3 \, q^2} \hat\pi^{\mu \nu}(q) \mathcal{A}^{\alpha \beta}
\eeq
appearing in \eqref{eq:TJJreconstructed2}, with the anomaly $\mathcal{A}^{\alpha \beta}$ given by\eqref{anoms} is replaced by 

\begin{equation}\label{PhiTJJ_onshell}
\Phi_{0}(s,m)
=
\frac{n_F g^2}{72\pi^2 s^2}
\Bigl[3m^2(s-4m^2)\,C_0(s,m^2)-6m^2+s\Bigr],
\end{equation}
where \(s=q^2\) and \(C_0(s,m^2)\equiv C_0(s,0,0,m^2)\) denotes the standard scalar three-point function with two onshell external lines
\begin{equation}\label{C0_onshell}
C_0(s,m^2)
=
\frac{1}{2s}
\log^2\!\left(
\frac{\sqrt{s(s-4m^2)}+2m^2-s}{2m^2}
\right).
\end{equation}

The spectral content of \(\Phi_0\) is extracted from the discontinuity across the physical cut along the real axis, \(s\ge 0\). We define
\begin{equation}
\text{disc}\,F(s)\equiv F(s+i0)-F(s-i0),
\end{equation}
for any function admitting boundary values on the cut. 
Introducing
\begin{equation}
\tau(s,m^2)=1-\frac{4m^2}{s},
\qquad
A(s)=C_0(s+i0,m^2)+C_0(s-i0,m^2),
\end{equation}
one finds
\begin{align}\label{disc_C0_over_s_clean}
\text{disc}\!\left(\frac{C_0(s,m^2)}{s}\right)
&=
-2\pi i\,\frac{1}{s^2}
\log\!\left(
\frac{1+\sqrt{\tau(s,m^2)}}{1-\sqrt{\tau(s,m^2)}}
\right)\theta(s-4m^2)
-\pi i\,A(0)\,\delta(s),
\\[2mm]
\text{disc}\!\left(\frac{C_0(s,m^2)}{s^2}\right)
&=
-2\pi i\,\frac{1}{s^3}
\log\!\left(
\frac{1+\sqrt{\tau(s,m^2)}}{1-\sqrt{\tau(s,m^2)}}
\right)\theta(s-4m^2)
+\pi i\,\delta'(s)\,A(s).
\end{align}
The term proportional to \(\delta'(s)\) is understood as
\begin{equation}\label{disc_powers_clean}
\delta'(s)\,A(s)=\delta'(s)\,A(0)-\delta(s)\,A'(0),
\qquad
A(0)=-\frac{1}{m^2},
\qquad
A'(0)=-\frac{1}{12m^4}.
\end{equation}

Using these relations, the discontinuity of the fermionic contribution to \(\Phi_{TJJ}\) can be organized as
\begin{equation}
\text{disc}\bigl[\Phi_0(s,m)\bigr]_{\text{fermion}}
=
\frac{n_F g^2}{72\pi^2}
\text{disc}\!\left(
\frac{3m^2(s-4m^2)C_0(s,m^2)-6m^2+s}{s^2}
\right).
\end{equation}
Substituting \eqref{disc_powers_clean} and \eqref{disc_C0_over_s_clean}, one finds that all terms proportional to \(\delta'(s)\) and \(\delta(s)\) cancel identically for \(m\neq0\). Therefore the fermionic sector does not generate any pole contribution to the spectral density; only the continuous cut above threshold survives:
\begin{equation}\label{disc_Phi_fermion_final_clean}
\text{disc}\bigl[\Phi_{0}(s,m)\bigr]_{\text{fermion}}
=
-2\pi i\,
\frac{n_F e^2\,\delta^{ab}}{72\pi}
\frac{3m^2(s-4m^2)}{s^3}
\log\!\left(
\frac{1+\sqrt{\tau(s,m^2)}}{1-\sqrt{\tau(s,m^2)}}
\right)
\theta(s-4m^2).
\end{equation}
This explicitly shows that the anomaly pole is absent in the fermionic sector at finite mass: the massless pole dissolves into a branch cut. In the conformal limit \(m\to0\), however, the cut collapses to the origin,
\begin{equation}
\text{disc}\bigl[\Phi_0(s)\bigr]
=
-\frac{(2n_F)\,i e^2}{72\pi}\,\delta(s),
\end{equation}
Introducing the standard factor \((2i)^{-1}\) in the definition of the spectral density, one recovers
\begin{equation}
\Delta\Phi_0(s)
\equiv\frac{1}{2 i}\text{disc}\bigl[\Phi_0(s)\bigr]=
-\frac{1}{3}\,\frac{\beta(e)}{e}\,\delta(s),
\end{equation}
which reproduces the anomaly sum rule upon integration over \(s\ge0\),
\beq
\int_{4m^2}^{\infty}\Delta\Phi_0(s,m^2)\,ds
=
\frac{g_s^2}{144\pi^2}\,(-2n_f)
=
-\frac{1}{3}\frac{\beta(e)}{e}.
\eeq
In this case the anomaly interaction is characterized by the presence of branch cuts rather than poles on each vertex. The scalar amplitude is modified to
\begin{equation}
\widetilde{\mathcal{A}}^{(0)}
=
\widetilde{\mathcal{A}}_s^{(0)}
+
\widetilde{\mathcal{A}}_t^{(0)}
+
\widetilde{\mathcal{A}}_u^{(0)}.
\end{equation}
In components this takes the form
\begin{equation}
\begin{aligned}
\widetilde{\mathcal{A}}^{(0)\alpha\beta\gamma\delta}
=
\frac{\bar C_0}{M_P^2}
\Bigg[&
\Phi_0^2(s,m)\,
u^{\alpha\gamma}(p_1,p_2)\,
\frac{1}{s}\,
u^{\beta\delta}(p_3,p_4)
\\
&\quad
+
\Phi_0^2(t,m)\,
u^{\alpha\beta}(p_1,p_3)\,
\frac{1}{t}\,
u^{\gamma\delta}(p_2,p_4)
+
\Phi_0^2(u,m)\,
u^{\alpha\delta}(p_1,p_4)\,
\frac{1}{u}\,
u^{\gamma\beta}(p_2,p_3)
\Bigg].
\end{aligned}
\end{equation}
A direct computation gives
\begin{equation}
  \widetilde{\mathcal{A}}_{\lambda_1\lambda_2\lambda_3\lambda_4}
  =
  Y
  \begin{dcases}
  -\frac{4 (t+u)^6}{s^3} \Phi_0(s,m)^2 - t (s-t+3 u)^2 \Phi_0(t,m)^2 - u (s+3 t-u)^2 \Phi_0(u,m)^2   \hfill\text{for } \pm\pm\pm\pm \\
  -\frac{4 (t+u)^6}{s^3} \Phi_0(s,m)^2 - \frac{t^2 (s+t-3 u)^2}{u} \Phi_0(u,m)^2 - \frac{u^2 (s-3 t+u)^2}{t} \Phi_0(t,m)^2 \quad \text{for } \pm\pm\mp\mp \\
  -t (s-t+3 u)^2 \Phi_0(t,m)^2 - \frac{t^2 (s+t-3 u)^2}{u} \Phi_0(u,m)^2 \hfill \text{for } \pm\mp\pm\mp \\
  -\frac{u^2 (s-3 t+u)^2}{t} \Phi_0(t,m)^2 - u (s+3 t-u)^2 \Phi_0(u,m)^2
  \hfill  \text{for } \pm\mp\mp\pm \\
  u (s-3 t+u) (s-t+3 u) \Phi_0(t,m)^2 + t (s+3 t-u) (s+t-3 u) \Phi_0(u,m)^2
  \hfill \text{for odd flips}
  \end{dcases}
\end{equation}
with
\beq
Y\equiv \frac{\bar C_0}{16M_P^2(t+u)^2},
\eeq
and \(\bar C_0\) given by \eqref{C0}.

\begin{figure}[t]
\begin{center}
\includegraphics[scale=0.6,angle=0]{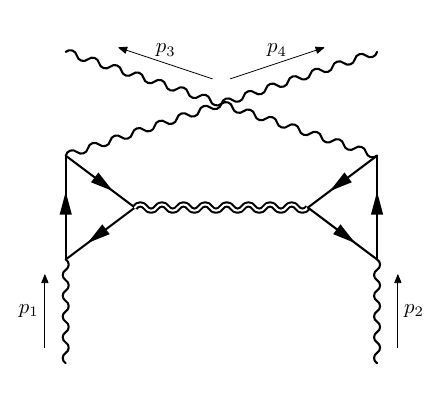}
\includegraphics[scale=0.6,angle=0]{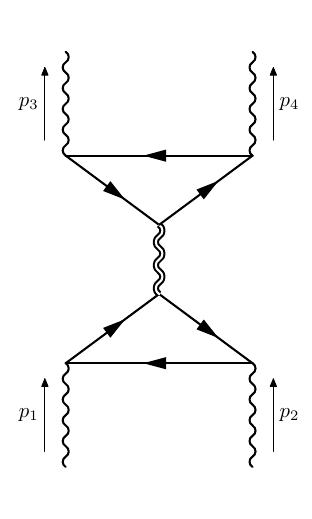}
\caption{$t$ and $s$ channel diagrams in the scattering of spin-1 with anomaly mediation. } 
\label{expansionX2}
\end{center}
\end{figure}

\section{ Graviton-Graviton scattering at tree level and the anomaly corrections}
The extension of our previous analysis to graviton--graviton scattering can be carried out using the explicit expression of the \(TTT\) correlator. This correlator has been analysed within the general framework based on the solution of the conformal Ward identities developed in \cite{Bzowski:2013sza}, and subsequently re-investigated perturbatively in \cite{Coriano:2018bsy}. In the latter work, the integration constants appearing in the general conformal solution of \cite{Osborn:1993cr}, reconstructed in momentum space in \cite{Bzowski:2013sza}, were matched to three independent multiplicity factors in arbitrary spacetime dimensions.\\
The analysis of \cite{Coriano:2018bsy} is, to our knowledge, the only one that yields infrared-safe results for the vertex when two of the three external momenta are taken on shell, a feature that we have already shown to occur in the case of the \(TJJ\) correlator. The difference with respect to the dimensional-regularization scheme adopted in \cite{Bzowski:2013sza} lies in the specific subtraction prescription used in the renormalization of the triple-\(K\) integrals, which generates additional logarithmic terms depending on the renormalization scale and which fail to remain finite in the double on-shell limit.\\
While this issue requires a dedicated discussion that we defer to future work, our present interest is focused on the anomaly contribution to the vertex, for which the approaches of \cite{Bzowski:2013sza} and \cite{Coriano:2018bsy} are in agreement. The same anomaly-induced effective vertex has also been shown to coincide with the result obtained from the nonlocal anomaly action, a method that we now briefly review.\\
It becomes possible to recast the generally covariant non-local effective action (\ref{Snonlsq}) in local form by the introduction of  
only a single new scalar field $\vf$, the conformalon field \cite{Mottola:2016mpl}. 
Thus, if the anomaly effective action is recast in local form
\bea
&&\hspace{-1.5cm} \cS_{\rm anom}[g;\vf] \equiv -\sdfrac{b'}{2} \int d^4x\,\sqrt{-g}\, \Big[ (\sq \vf)^2 - 2 \big(R^{\m\n} - \tfrac{1}{3} R g^{\m\n}\big)
(\na_\m\vf)(\na_\n \vf)\Big]\nn \\
&& \hspace{1.5cm} +\, \sdfrac{1}{2}\,\int d^4x\,\sqrt{-g}\  \Big[b'\big(E - \tfrac{2}{3}\sq R\big) + b\,C^2 \Big]\,\vf
\label{Sanom}
\eea
and varied with respect to $\varphi$, the linear eq. of motion takes the form 
\be
\sqrt{-g}\,\D_4\, \vf = \sqrt{-g}\left[\sdfrac{E}{2}- \sdfrac{\!\sq R\!}{3} + \sdfrac{b}{2b'}\, C^2\right] = \frac{1}{2b'} \,\cA
\label{phieom}
\ee

\begin{figure}[t]
\centering
\includegraphics[scale=0.7]{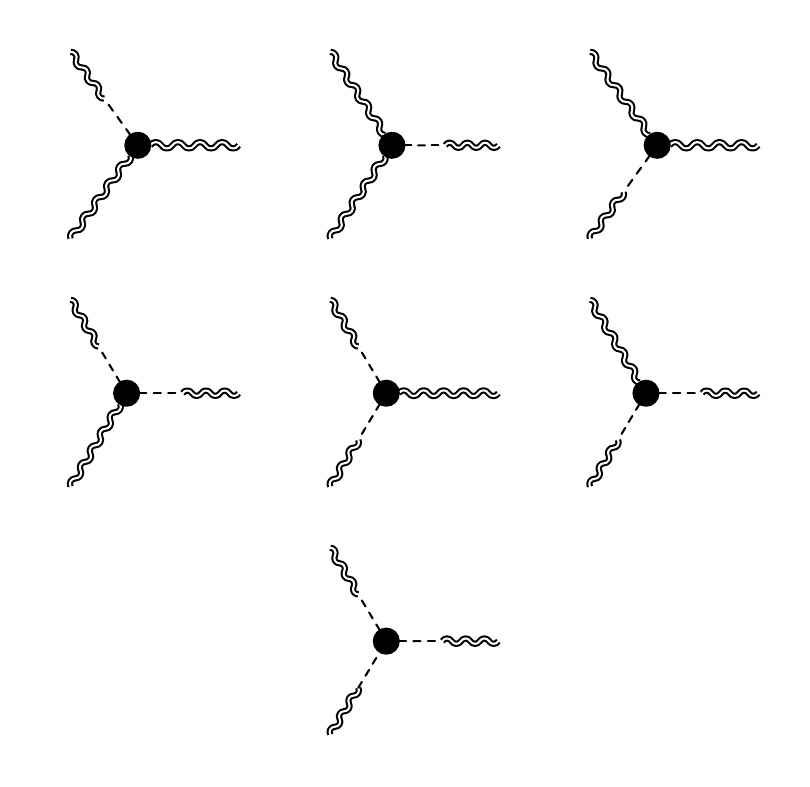}
\caption{Anomaly interactions mediated by the exchange of one, two or three poles. The poles are generated by the renormalization of the longitudinal sector of the $TTT$.}
\label{dec}
\end{figure}

In order to obtain the contributions of the anomaly effective action (\ref{Sanom}) to the three-point function
$\cS_3^{\m_1\n_1\m_2\n_2\m_3\n_3}$ we require the expansion of $\cS_{\rm anom}[g;\vf]$ to third order
in deviations from flat space. The consistent expansion  of $\cS_{\rm anom}[g;\vf]$ around flat space is 
defined by the simultaneous expansion of the metric $g_{\m\n}$ and conformalon scalar $\vf$
\bes
\bea
g_{\m\n} &=& g_{\m\n}^{(0)} + g_{\m\n}^{(1)} + g_{\m\n}^{(2)} + \dots \equiv \eta_{\m\n} + h_{\m\n} + h_{\m\n}^{(2)} + \dots\\
\vf &=& \vf^{(0)} +  \vf^{(1)} +  \vf^{(2)}  + \dots
\eea
\ees
substitution of these expansions into (\ref{phieom}) and identification of terms of the same order. Thus to the lowest three orders of the expansion we have
\bes
\bea
&&\hspace{4cm}\sqb^2 \vf^{(0)} = 0 \label{eom0}\\
&&\hspace{-1.5cm}(\sqrt{-g} \D_4)^{(1)} \vf^{(0)} + \sqb^2 \vf^{(1)} = \left[\sqrt{-g}
\left( \sdfrac{E}{2}- \sdfrac{\!\sq R\!}{3} + \sdfrac{b}{2b'}\, C^2 \right)\right]^{(1)}
= - \sdfrac{\!1\!}{3}\, \sqb R^{(1)} \label{eom1}\\
&&\hspace{-2cm}(\sqrt{-g} \D_4)^{(2)} \vf^{(0)} + (\sqrt{-g} \D_4)^{(1)} \vf^{(1)} + \sqb^2 \vf^{(2)} =
\left[\sqrt{-g}\left(\sdfrac{E}{2}- \sdfrac{\!\sq R\!}{3} + \sdfrac{b}{2b'}\, C^2 \right)\right]^{(2)} \nn \\
&&\hspace{2cm}= \sdfrac{1}{2}E^{(2)} - \sdfrac{1}{3}\, [\sqrt{-g}\sq R]^{(2)} + \sdfrac{b}{2b'}\, [C^2]^{(2)} \label{eom2}
\eea
\ees
where $\sqb$ is the d'Alembert wave operator in flat Minkowski spacetime, and we have used the fact that $E$ and $C^2$ are second order in curvature invariants while the Ricci scalar $R$ starts at first order. 

\be
\vf^{(1)} = - \sdfrac{\!1 \!}{3\sqb} \, R^{(1)}
\label{vf1}
\ee
and hence the solution of (\ref{eom2}) is
\be
\vf^{(2)} = \sdfrac{1}{\sqb^2} \left\{ (\sqrt{-g} \D_4)^{(1)}\sdfrac{\!1 \!}{3\sqb} \, R^{(1)} 
+  \sdfrac{1}{2}E^{(2)} - \sdfrac{1}{3}\, [\sqrt{-g}\sq R]^{(2)} + \sdfrac{b}{2b'}\, [C^2]^{(2)}\right\}
\label{vf2}
\ee
giving at cubic order the effective action
\bea
&&\hspace{-5mm} \cS_{\rm anom}^{(3)} =
 \sdfrac{b'}{9} \int\! d^4x \int\!d^4x'\!\int\!d^4x''\!\left\{\big(\pa_{\m} R^{(1)})_x\left(\sdfrac{1}{\sqb}\right)_{\!xx'}  
 \!\left(R^{(1)\m\n}\! - \!\sdfrac{1}{3} \eta^{\m\n} R^{(1)}\right)_{x'}\!
\left(\sdfrac{1}{\sqb}\right)_{\!x'x''}\!\big(\pa_{\n} R^{(1)})_{x''}\right\}\nn \\
&&\hspace{-6mm}- \sdfrac{1}{6}\! \int\! d^4x\! \int\!d^4x'\! \left(b'\, E^{\!(2)} + b\,  [C^2]^{(2)}\right)_{\!x}\! \left(\sdfrac{1}{\sqb}\right)_{\!xx'} \!R^{(1)}_{x'}
 + \sdfrac{b'}{18} \! \int\! d^4x\, R^{(1)}\left(2\, R^{\!(2)} + (\sqrt{-g})^{(1)} R^{(1)}\right)
\label{S3anom3}
\eea
which in momentum space takes the form 
\begin{align}
^{(\cA)}S_3&^{\m_1\n_1\m_2\n_2\m_3\n_3} = \sdfrac{1}{3}\, \pi^{\m_1\n_1}(p_1)\,\eta_{\a_1\b_1}\,^{(\cA)}S_3^{\a_1\b_1\m_2\n_2\m_3\n_3}
+\sdfrac{1}{3}\, \pi^{\m_2\n_2}(p_2)\,\eta_{\a_2\b_2}\,^{(\cA)}S_3^{\m_1\n_1\a_2\b_2\m_3\n_3} \nonumber\\
& +\sdfrac{1}{3}\, \pi^{\m_3\n_3}(p_3)\,\eta_{\a_3\b_3}\,^{(\cA)}S_3^{\m_1\n_1\m_2\n_2\a_3\b_3} 
-\sdfrac{1}{9}\,\pi^{\m_1\n_1}(p_1)\,\pi^{\m_3\n_3}(p_3)\,\eta_{\a_1\b_1}\eta_{\a_3\b_3}\,^{(\cA)}S_3^{\a_1\b_1\m_2\n_2\a_3\b_3} \nonumber\\
&-\sdfrac{1}{9}\,\pi^{\m_2\n_2}(p_2)\pi^{\m_3\n_3}(p_3)\,\eta_{\a_2\b_2}\eta_{\a_3\b_3}\,^{(\cA)}S_3^{\m_1\n_1\a_2\b_2\a_3\b_3}
-\sdfrac{1}{9}\,\pi^{\m_1\n_1}(p_1)\pi^{\m_2\n_2}(p_2)\eta_{\a_1\b_1}\, \eta_{\a_2\b_2}\,^{(\cA)}S_3^{\a_1\b_1\a_2\b_2\m_3\n_3} \nonumber\\
& +\sdfrac{1}{27}\,\pi^{\m_1\n_1}(p_1)\pi_2^{\m_2\n_2} (p_2)\pi^{\m_3\n_3}(p_3)\,\eta_{\a_1\b_1}\eta_{\a_2\b_2}\eta_{\a_3\b_3}\,^{(\cA)}S_3^{\a_1\b_1\a_2\b_2\a_3\b_3} \,. \label{fin1}
\end{align}
where 
\bea
&&\hspace{-1.5cm}\eta_{\a_1\b_1}\,^{\!(\cA)}\cS_3^{\a_1\b_1\m_2\n_2\m_3\n_3}(p_1,p_2,p_3)\Big\vert_{p_3 = -(p_1 + p_2)} =\nn \\
&& 8 b\, \big[(C^2)^{(2)}\big]^{\m_2\n_2\m_3\n_3} (p_2, p_3) 
+ \,8b'\, \big[E^{(2)}\big]^{\m_2\n_2\m_3\n_3} (p_2, p_3) = 4 \, \cA_2^{\m_2\n_2\m_3\n_3} (p_2, p_3)
\label{traceS3}
\eea
denotes the trace of the third functional derivative of anomaly functional around flat space. the other structures, describing contact interactions in coordinate space are obtained by double tracing

\bea
&&\eta_{\a_1\b_1}\eta_{\a_3\b_3}\,^{\!(\cA)}\cS_3^{\a_1\b_1\m_2\n_2\a_3\b_3}(p_1,p_2,p_3)\big\vert_{p_3 = -(p_1 + p_2)} 
= 8b' \, \eta_{\a_3\b_3}\big[E^{(2)}\big]^{\m_2\n_2\a_3\b_3} (p_2, p_3)  \nn \\
&& \qquad = 16b'\, Q^{\m_2\n_2}(p_1,p_2,p_3)\big\vert_{p_3 = -(p_1 + p_2)} \, + \ 8b'\,p_2^2\, \left( p_1^2  + p_1\cdot p_2\right)\pi^{\m_2\n_2}(p_2) 
\label{dbltrace3}
\eea
with 
\bea
&&Q^{\m_2\nu_2}(p_1,p_2,p_3) \equiv p_{1\m}\, [R^{\m\nu}]^{\m_2\nu_2}(p_2)\,p_{3\n} \nn \\
&&= \sdfrac{1}{2} \,\Big\{(p_1\cdot p_2)(p_2\cdot p_3)\, \eta^{\m_2\n_2} 
+ p_2^2 \ p_1^{(\m_2}\, p_3^{\n_2)} - (p_2\cdot p_3) \, p_1^{(\m_2}\, p_2^{\n_2)} - (p_1\cdot p_2) \, p_2^{(\m_2}\, p_3^{\n_2)}\Big\}
\eea

and triple traces of the same vertex
\be
\eta_{\a_1\b_1}\eta_{\a_2\b_2}\eta_{\a_3\b_3}\,^{\!(\cA)}\cS_3^{\a_1\b_1\a_2\b_2\a_3\b_3}(p_1,p_2,p_3)\big\vert_{p_3 = -(p_1 + p_2)} =
16b' \left[ p_1^2\,p_2^2 - (p_1\cdot p_2)^2\right].
\label{triptrace3}
\ee

Second functional derivatives of the anomaly functional in flat space have a specific local structure 

\bea
&&\hspace{-1.5cm}\big[E^{(2)}\big]^{\m_i\nu_i\m_j\nu_j} =\big[R_{\m\a\n\b}^{(1)}R^{(1)\m\a \n\b}\big]^{\m_i\nu_i\m_j\nu_j}
-4\,\big[R_{\m\n}^{(1)}R^{(1)\m\n}\big]^{\m_i\nu_i\m_j\nu_j}
+\big[ \big(R^{(1)}\big)^2\big]^{\m_i\nu_i\m_j\nu_j}\\
&&\hspace{-1.5cm} \big[(C^2)^{(2)}\big]^{\m_i\nu_i\m_j\nu_j}= \big[R_{\m\a\n\b}^{(1)}R^{(1)\m\a \n\b}\big]^{\m_i\nu_i\m_j\nu_j}
-2\,\big[R_{\m\n}^{(1)}R^{(1)\m\n}\big]^{\m_i\nu_i\m_j\nu_j}
+ \sdfrac{1}{3}\,\big[(R^{(1)})^2\big]^{\m_i\nu_i\m_j\nu_j}
\eea
Notice that $\big[(C^2)^{(2)}\big]^{\m_i\nu_i\m_j\nu_j}$ is proportional to the transverse traceless projector 
$P^{(2)}$ in flat space.

\begin{figure}[t]
\begin{center}
\includegraphics[scale=0.5,angle=0]{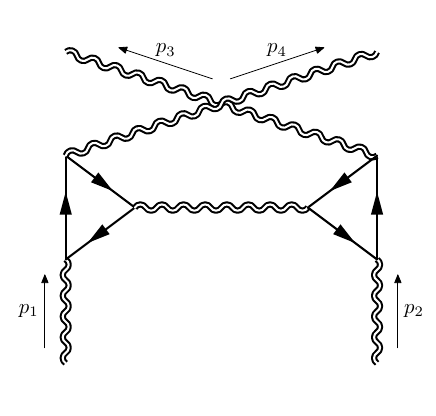}
\includegraphics[scale=0.5,angle=0]{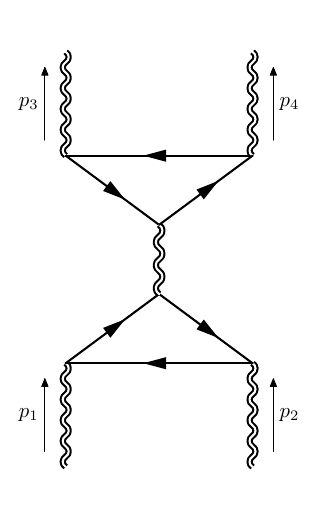}
\includegraphics[scale=0.5,angle=0]{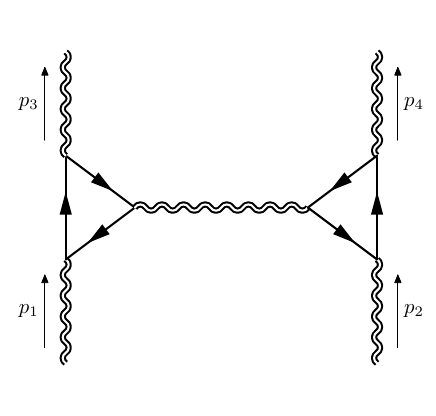}
\caption{Anomaly mediation of graviton-graviton scattering. Shown are the $s, t$ and $u$ channels. } 
\label{expansionX3}
\end{center}
\end{figure}

\subsection{The scalar contribution to a 2-to-2  scattering and the double copy}
In order to discuss the impact of anomaly mediation we can send on-shell two of the three graviton lines and consider only the interpolating graviton in its scalar component, in order to build the scalar exchange shown in Fig. 1 (right). The amplitude is built by sewing together two $TTT$ 3-point functions using the scalar component of the De Donder propagator (see Fig. 7). 
Since we are only interested in this specific component of the exchange, together with small momentum transfers which 
are relevant for the study of the eikonal limit, we need just to select a single term from \eqref{fin1}, which in this case is given by (for the left vertex)
\beq
\Gamma^{\mu_1\nu_1,\mu_2\nu_2,\mu_3\nu_3}(p_1,p_2,p_3)=\sdfrac{1}{3}\, \pi^{\mu_2\nu_2}(p_2)\,\eta_{\sigma_1\sigma_2}\,^{(\cA)}S_3^{\mu_1\nu_1\sigma_1\sigma_2\mu_3\nu_3}(p_1,p_2,p_3)
\eeq
The anomaly appears via a single trace contraction of $S_3$ which is proportional to  $\mathcal{A}_2(p_1,q)$ as specified by \eqref{traceS3}
\begin{equation}
\Gamma^{\mu_1\nu_1,\mu_2\nu_2,\mu_3\nu_3}(p_1,p_2,p_3)
=
\frac{1}{3}\mathcal{A}_2^{\mu_1\nu_1,\mu_3\nu_3}(p_1,p_3)\,
\pi^{\mu_2\nu_2}(p_2),
\end{equation}
where
\begin{equation}
\pi^{\mu_2\nu_2}(p_2)
\equiv
\left(
{R}^{(1)}(p_2)\,\Box^{-1}
\right)^{\mu_2\nu_2}
\end{equation}
is the scalar projector appearing on the exchanged gravitational leg (see Fig. 1).
The intermediate propagator is given by the scalar component of the De Donder propagator \eqref{eq:propagator_spin_decomposition}, which is proportional to the spin-0 projector in flat space
\beq
D^{(0-s)}(k)_{\rho\sigma\alpha\beta}  = \frac{1}{2} \frac{P^{(0){}}_{\rho\sigma\alpha\beta}(k)}{k^2} = -\frac{1}{6}\frac{\pi_{\rho\sigma}\pi_{\alpha\beta}}{k^2} .
\end{equation}
We contract the two vertices to obtain for the channel $t$
\begin{align}
  \Gamma_4^{\mu_1\nu_1,\mu_2\nu_2,\mu_3\nu_3,\mu_4\nu_4}(p_1,p_2,p_3,p_4) & \equiv \Gamma^{\mu_1\nu_1,\rho\sigma,\mu_2\nu_2}(p_1,-q,p_3)\,D^{(0-s)}_{\rho\sigma\alpha\beta}(q)\,\Gamma^{\mu_3\nu_3,\alpha\beta,\mu_4\nu_4}(p_3,q,p_4) \nonumber\\
  &= -\frac{1}{6}\,\tilde{\mathcal{A}}_2^{\mu_1\nu_1,\mu_3\nu_3}(p_1,p_3)\,
\frac{1}{q^2}\tilde{\mathcal{A}}_2^{\mu_2\nu_2,\mu_4\nu_4}(p_2,p_4)\,
,
\end{align}
where $q$ is the exchanged momentum. Adding the contributions of all the three channels we obtain 
\begin{equation}\label{eq:tot_ampl_gg-D0s-gg}
\begin{aligned}
\mathcal{M}^{(0)}_{\mu_1\nu_1\,\mu_2\nu_2\,\mu_3\nu_3\,\mu_4\nu_4}
=
-\frac{1}{6M_P^6}
\Big[
&\mathcal{A}_2^{\mu_1\nu_1\,\mu_2\nu_2}(p_1,p_2)
\frac{1}{s}
\mathcal{A}_2^{\mu_3\nu_3\,\mu_4\nu_4}(p_3,p_4) + 
\mathcal{A}_2^{\mu_1\nu_1\,\mu_3\nu_3}(-p_1,p_3)
\frac{1}{t}
\mathcal{A}_2^{\mu_2\nu_2\,\mu_4\nu_4}(-p_2,p_4)
\\
&+
\mathcal{A}_2^{\mu_1\nu_1\,\mu_4\nu_4}(-p_1,p_4)
\frac{1}{u}
\mathcal{A}_2^{\mu_2\nu_2\,\mu_3\nu_3}(-p_2,p_3)
\Big] .
\end{aligned}
\end{equation}
with $\cA_2^{\mu_1\nu_1,\mu_2,\nu_2}$ given by \eqref{traceS3}. For the generic channel we have 
\begin{equation}\label{eq:A2_simplified}
\begin{aligned}
\mathcal{A}^{\mu_1\nu_1,\mu_2,\nu_2}_2(p_i,p_j)&=\frac{1}{4}\,(b+b')\Big(
-2s_\mathrm{ch}\,p_j^{\mu_1}p_i^{\mu_2}g^{\nu_1\nu_2}
-2s\,p_i^{\mu_2}p_j^{\nu_1}g^{\mu_1\nu_2}
-2s\,p_j^{\mu_1}p_i^{\nu_2}g^{\mu_2\nu_1}
-2s\,p_j^{\nu_1}p_i^{\nu_2}g^{\mu_1\mu_2}
\\
&\qquad\qquad
+s^{2}g^{\mu_1\nu_2}g^{\mu_2\nu_1}
+s^{2}g^{\mu_1\mu_2}g^{\nu_1\nu_2}
+8\,p_j^{\mu_1}p_i^{\mu_2}p_j^{\nu_1}p_i^{\nu_2}
\Big).
\end{aligned}
\end{equation}
with $s_\mathrm{ch} \in\{ s, t, u\}$. Observe that on-shell $\mathcal{A}_2^{\mu_1\nu_1\mu_2\nu_2}$ can be rewritten in the following way:
\begin{align}\label{eq:A2_u_decomposition}
\mathcal{A}^{\alpha\beta\rho\sigma}_2(p,q)&\equiv \prescript{(\mathcal{A})}{}{S_3}^{\alpha\beta\rho\sigma\mu\nu}(p,q,-p-q)\eta_{\mu\nu}\nonumber \\
&= 4(b+b')\left[u^{\alpha\rho}(p,q)u^{\beta\sigma}(p,q)+u^{\alpha\sigma}(p,q)u^{\beta\rho}(p,q)\right],
\end{align}
where $u^{\alpha\beta}(p,q)$ defined in \eqref{uab} represent exactly the anomaly part of the $TJJ$. Notice that as one should expect the rhs of \eqref{eq:A2_u_decomposition} is invariant under the exchange of the two gravitons and is symmetric under $\{\alpha, \beta\} \leftrightarrow \{\rho, \sigma\}$. The relationship:
\begin{equation}
	\mathcal{A}_2^{\alpha\beta\rho\sigma}=4(b+b')\left[u^{\alpha\rho}u^{\beta\sigma}+u^{\alpha\sigma}u^{\beta\rho}\right]	
\end{equation}
can be interpreted as a manifestation of the double-copy.

The polarization tensor of the graviton can be written as product of two polarization vectors:
\beq
\varepsilon_{\mu\nu}(p,\lambda) = \varepsilon_{\mu}(p,\lambda)\varepsilon_{\nu}(p,\lambda)
\eeq
If we require it to be transverse-traceless then:
\begin{align}
\varepsilon^{TT}_{\mu\nu}(p,\lambda)& = \Pi_{\mu\nu}^{\;\;\;\;\alpha\beta}(p)\varepsilon_{\alpha\beta}(p,\lambda)\nonumber\\
&= \left[\frac{1}{2}\Big(\pi_{\mu}^{\;\;\alpha}(p) \pi_{\nu}^{\;\;\beta}(p)+ \pi_{\nu}^{\;\;\alpha}(p) \pi_{\mu}^{\;\;\beta}(p)\Big) - \frac{1}{3} \pi_{\mu\nu}(p) \pi^{\alpha\beta}(p)\right]\varepsilon_{\alpha}(p)\varepsilon_{\beta}(p).
\end{align}
If $\varepsilon_{\alpha}(p)$ and $\varepsilon_{\beta}(p)$ are transverse this expression simplifies:
\begin{align}
	\varepsilon^{TT}_{\mu\nu}(p,\lambda)& = \varepsilon^T_\mu(p) \varepsilon^T_\nu(p)- \frac{1}{3}\pi_{\mu\nu}\, \varepsilon^T(p)\cdot\varepsilon^T(p).
\end{align}
Notice that helicity projectors are both transverse and lightlike $p\cdot \varepsilon(p,\pm) = \varepsilon(p,\pm)\cdot\varepsilon(p,\pm)$ = 0, so that:
\begin{equation}
	\varepsilon^{TT}_{\mu\nu}(p,\pm) = \varepsilon_{\mu}(p,\pm)\varepsilon_{\nu}(p,\pm).
\end{equation}
The contraction of the $\mathcal{A}_2$ vertex with the polarization tensors is given by:
\begin{align}\label{eq:mathsfA2}
\varepsilon_{\mu_1\nu_1}(p_1,\lambda_1) \varepsilon_{\mu_2\nu_2}(p_2,\lambda_2) \mathcal{A}^{\mu_1\nu_1\mu_2\nu_2}_2(p_1,p_2) =8(b+b')\mathcal{U}^2_{p_1,p_2}(\lambda_1,\lambda_2). 
\end{align}
This relation can be inserted into \eqref{eq:tot_ampl_gg-D0s-gg} to compute the total scattering amplitude. Here we report the results for the three channels separately:
\begin{align}
	\mathcal{M}_s(\lambda_1,\lambda_2,\lambda_3,\lambda_4) &= -\frac{32\,(b+b')^2}{3 M_P^6} \mathcal{U}^2_{p_1,p_2}(\lambda_1,\lambda_2)^2\frac{1}{s}\,\mathcal{U}^2_{p_3,p_4}(\lambda_3,\lambda_4),\\
	\mathcal{M}_t (\lambda_1,\lambda_2,\lambda_3,\lambda_4) &= -\frac{32\,(b+b')^2}{3M_P^6} \mathcal{U}^2_{-p_1,p_3}(\lambda_1,\lambda_3)\frac{1}{t}\,\mathcal{U}^2_{-p_2,p_4}(\lambda_2,\lambda_4),\\
  \mathcal{M}_u (\lambda_1,\lambda_2,\lambda_3,\lambda_4) &=-\frac{32\,(b+b')^2}{3 M_P^6}  \mathcal{U}^2_{-p_1,p_4}(\lambda_1,\lambda_4)\frac{1}{u}\,\mathcal{U}^2_{-p_2,p_3}(\lambda_2,\lambda_3).
\end{align}
A comparison with \eqref{eq:channels_TJJ} shows the presence of a "double copy relation" for these amplitudes. The explicit results for the scattering amplitudes corresponding to various helicity configurations are collected in Table \ref{table:graviton_DC1}.

\begin{table}[t]
	\centering
	\small
	\renewcommand{\arraystretch}{1.12}
	\setlength{\tabcolsep}{4.2pt}
	\resizebox{\textwidth}{!}{%
		\begin{tabular}{cccc c c c c @{\hspace{1.0em}  \vrule width 0.75pt\hspace{1em} } cccc c c c c}
			\toprule
			$\lambda_1$&$\lambda_2$&$\lambda_3$&$\lambda_4$&
			$\mathcal{M}_s$&$\mathcal{M}_t$&$\mathcal{M}_u$&$\mathcal{M}_{\text{tot}}$&
			$\lambda_1$&$\lambda_2$&$\lambda_3$&$\lambda_4$&
			$\mathcal{M}_s$&$\mathcal{M}_t$&$\mathcal{M}_u$&$\mathcal{M}_{\text{tot}}$
			\\
			\midrule
			$+$&$+$&$+$&$+$&$-\frac{s}{4}$&$-\frac{t (t-u)^2}{4 s^2}$&$-\frac{u (t-u)^2}{4 s^2}$&$\frac{s^7+(t-u)^4 \left(t^3+u^3\right)}{16 s^4}$&
			$-$&$+$&$+$&$+$&$0$&$\frac{t u (t-u)}{2 s^2}$&$\frac{t u (u-t)}{2 s^2}$&$-\frac{t^2 u^2 (t-u)^2}{4 s^3}$ \\[0.3em]
			
			$+$&$+$&$+$&$-$&$0$&$\frac{t u (t-u)}{2 s^2}$&$\frac{t u (u-t)}{2 s^2}$&$-\frac{t^2 u^2 (t-u)^2}{4 s^3}$&
			$-$&$+$&$+$&$-$&$0$&$-\frac{t u^2}{s^2}$&$-\frac{u (t-u)^2}{4 s^2}$&$\frac{u^3 \left(16 t^3 u+(t-u)^4\right)}{16 s^4}$ \\[0.3em]
			
			$+$&$+$&$-$&$+$&$0$&$\frac{t u (t-u)}{2 s^2}$&$\frac{t u (u-t)}{2 s^2}$&$-\frac{t^2 u^2 (t-u)^2}{4 s^3}$&
			$-$&$+$&$-$&$+$&$0$&$-\frac{t (t-u)^2}{4 s^2}$&$-\frac{t^2 u}{s^2}$&$\frac{t^3 \left(16 t u^3+(t-u)^4\right)}{16 s^4}$ \\[0.3em]
			
			$+$&$+$&$-$&$-$&$-\frac{s}{4}$&$-\frac{t u^2}{s^2}$&$-\frac{t^2 u}{s^2}$&$\frac{s^3}{16}-\frac{t^3 u^3}{s^3}$&
			$-$&$+$&$-$&$-$&$0$&$\frac{t u (t-u)}{2 s^2}$&$\frac{t u (u-t)}{2 s^2}$&$-\frac{t^2 u^2 (t-u)^2}{4 s^3}$ \\[0.3em]
			
			$+$&$-$&$+$&$+$&$0$&$\frac{t u (t-u)}{2 s^2}$&$\frac{t u (u-t)}{2 s^2}$&$-\frac{t^2 u^2 (t-u)^2}{4 s^3}$&
			$-$&$-$&$+$&$+$&$-\frac{s}{4}$&$-\frac{t u^2}{s^2}$&$-\frac{t^2 u}{s^2}$&$\frac{s^3}{16}-\frac{t^3 u^3}{s^3}$ \\[0.3em]
			
			$+$&$-$&$+$&$-$&$0$&$-\frac{t (t-u)^2}{4 s^2}$&$-\frac{t^2 u}{s^2}$&$\frac{t^3 \left(16 t u^3+(t-u)^4\right)}{16 s^4}$&
			$-$&$-$&$+$&$-$&$0$&$\frac{t u (t-u)}{2 s^2}$&$\frac{t u (u-t)}{2 s^2}$&$-\frac{t^2 u^2 (t-u)^2}{4 s^3}$ \\[0.3em]
			
			$+$&$-$&$-$&$+$&$0$&$-\frac{t u^2}{s^2}$&$-\frac{u (t-u)^2}{4 s^2}$&$\frac{u^3 \left(16 t^3 u+(t-u)^4\right)}{16 s^4}$&
			$-$&$-$&$-$&$+$&$0$&$\frac{t u (t-u)}{2 s^2}$&$\frac{t u (u-t)}{2 s^2}$&$-\frac{t^2 u^2 (t-u)^2}{4 s^3}$ \\[0.3em]
			
			$+$&$-$&$-$&$-$&$0$&$\frac{t u (t-u)}{2 s^2}$&$\frac{t u (u-t)}{2 s^2}$&$-\frac{t^2 u^2 (t-u)^2}{4 s^3}$&
			$-$&$-$&$-$&$-$&$-\frac{s}{4}$&$-\frac{t (t-u)^2}{4 s^2}$&$-\frac{u (t-u)^2}{4 s^2}$&$\frac{s^7+(t-u)^4 \left(t^3+u^3\right)}{16 s^4}$ \\[0.3em]
			\bottomrule
	\end{tabular}}
	\caption{Four-particle configurations and corresponding scattering amplitude values for four gravitons. Here $\bar{\mathcal{M}}_s$, $\bar{\mathcal{M}}_t$, and $\bar{\mathcal{M}}_u$ represent the respective channels, and $\bar{\mathcal{M}}_{\text{tot}}$ is the total amplitude. As in Table \ref{tab:TJJ-TTJ} we have omitted the common factor $-32\,(b+b')^2/(3 M_P^6)$ for simplicity.}
	\label{table:graviton_DC1}
\end{table}

\section{Scalar emission in graviton-graviton scattering}

\begin{figure}[t]
	\begin{center}
		\includegraphics[scale=0.4,angle=0]{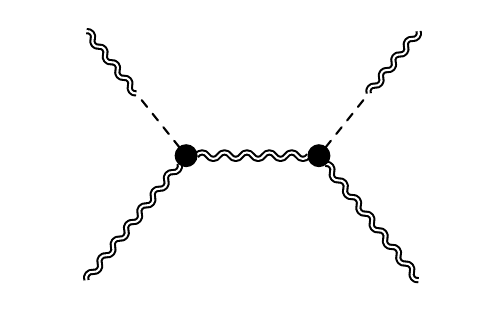}
		\includegraphics[scale=0.4,angle=0]{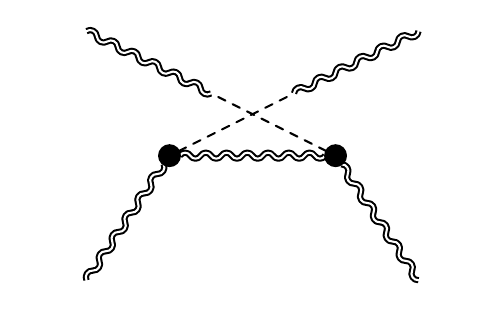}
		\includegraphics[scale=0.4,angle=0]{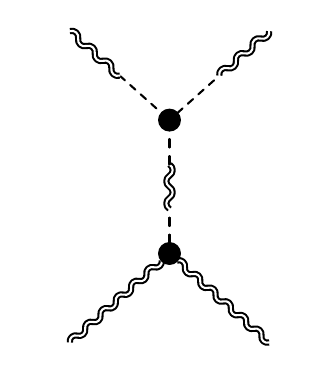}
		\includegraphics[scale=0.4,angle=0]{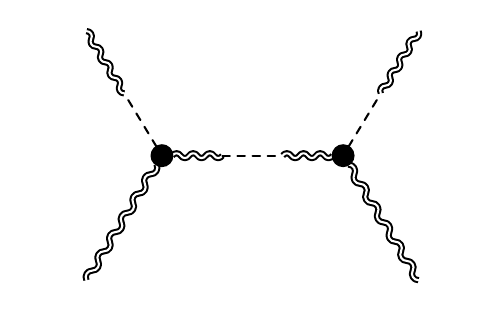}
		\includegraphics[scale=0.4,angle=0]{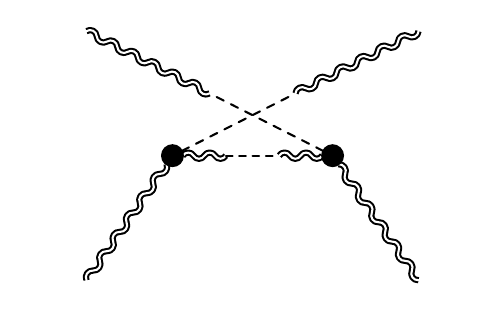}
		\caption{ Scalar emission in graviton-graviton scattering with anomaly mediation.} 
		\label{fig:scalar_emission}
	\end{center}
\end{figure}

We now study the three channels for the process
\[
g_{\lambda_1}(p_1)\,g_{\lambda_2}(p_2)\to \phi(p_3)\,\phi(p_4),
\]
where \(\phi\) is the spin--0 component of the graviton (Figure \ref{fig:scalar_emission}). We define
\begin{equation}
	\mathcal{A}_0(p_1,p_2) \equiv 4 \eta_{\mu_1\nu_1}\eta_{\mu_2\nu_2}\mathcal{A}_2^{\mu_1\nu_1\mu_2\nu_2}(p_1,p_2) = 16\,b'(p_1^2 p_2^2-p_1\cdot p_2)
\end{equation}
The polarization tensor can be decomposed in a spin-2 and spin-0 part as:
\begin{equation}
	\varepsilon_{\mu\nu}(p) = \varepsilon^{(0)}_{\mu\nu}(p) + \varepsilon^{(2)})_{\mu\nu}(p)
\end{equation}
where:
\begin{align}
	\varepsilon^{(2)}_{\mu\nu}(p) &= \Pi_{\mu\nu}^{\;\;\;\alpha\beta}(p)\epsilon_{\alpha\beta}(p) = \varepsilon_{\mu\nu}(p,\pm), \label{eq:epsilon2_def}\\
	\varepsilon^{(0)}_{\mu\nu}(p) &= \frac{1}{3}\pi_{\mu\nu}(p).\label{eq:epsilon0_def}
\end{align}
We will use \eqref{eq:epsilon2_def} for the gravitons of helicities $\pm 2$ and \eqref{eq:epsilon0_def} for the scalars. Notice also that $\varepsilon^{(0)}$ satisfies:
\begin{equation} \label{eq:epsilon0_times_pi}
	\pi^{\mu\nu}(p)\varepsilon^{(0)}_{\mu\nu}(p) = 1.
\end{equation}
The $s$-channel open indices amplitude for spin-0 polarization vector reads:
\begin{align}
	\mathcal{M}_s^{\mu_1\nu_1\mu_2\nu_2\mu_3\nu_3\mu_4\nu_4} (p_1,p_2,p_3,p_4) = & \,\frac{1}{M_P^6}\left[\frac{1}{3}\,\mathcal{A}_2^{\mu_1\nu_1\mu_2\nu_2}(p_1,p_2)\pi^{\rho \sigma}(p_1+p_2)\right]D^{(0-s)}_{\rho\sigma\alpha\beta}(p)\nonumber\\
	& \times \left[\frac{1}{27}\,\pi^{\alpha \beta}(p)\mathcal{A}_0(-p_3,-p_4)\pi^{\mu_3\nu_3}(-p_3)\pi^{\mu_4\nu_4}(-p_4)\right]
\end{align}
where the expression for the verteces contained inside the square brakets can be extrapolated from \eqref{fin1}. Notice that only the $D^{(0-s)}$ part of the propagator contributes, since
\begin{equation}
	\pi^{\rho\sigma}(p)D^{(2)}_{\rho\sigma\alpha\beta}(p)=0.
\end{equation}
Using \eqref{eq:epsilon0_times_pi} and 
\begin{equation}
	\pi^{\rho\sigma}(p)D^{(0-s)}_{\rho\sigma\alpha\beta}(p)\pi_{\alpha\beta}(p) = -\frac{3}{2}\frac{1}{p^2} 
\end{equation}
we can compute the scattering amplitude:
\begin{equation}
	\mathcal{M}_s(\lambda_1,\lambda_2) \equiv \varepsilon^{(2)}_{\mu_1\nu_1}(p_1,\lambda_1) \varepsilon^{(2)}_{\mu_2\nu_2}(p_2,\lambda_2) \varepsilon^{(0)}_{\mu_3\nu_3}(p_3) \varepsilon^{(0)}_{\mu_4\nu_4}(p_4) \,\mathcal{M}_s^{\mu_1\nu_1\mu_2\nu_2\mu_3\nu_3\mu_4\nu_4} (p_1,p_2,p_3,p_4);
\end{equation}
for different polarizations. The result is:
\begin{equation}
	\mathcal{M}_s(\lambda_1,\lambda_2)= \begin{dcases}
		\;\frac{1}{27 M_P^6}\, b'(b'+b) \, s^3 \qquad &\text{if} \qquad \lambda_1 = \lambda_2,
		\\
		\;0 \qquad &\text{if} \qquad \lambda_1 \neq \lambda_2.
	\end{dcases}\label{eq:amp_ggss_sChannel}
\end{equation}
In the $t$-channel both $D^{(0-s)}$ and $D^{(2)}$ contribute: 
\begin{align}
	\mathcal{M}_t^{\mu_1\nu_1\mu_2\nu_2\mu_3\nu_3\mu_4\nu_4} (p_1,p_2,p_3,p_4) = & \frac{1}{M_P^6}\left[\frac{1}{3}\,\mathcal{A}_2^{\mu_1\nu_1\rho\sigma}(p_1,p_3-p_1)\pi^{\mu_3\nu_3}(p_3)\right]\Big(D^{(0-s)}_{\rho\sigma\alpha\beta}+D^{(2)}_{\rho\sigma\alpha\beta}\Big)(p_1-p_3)\nonumber\\
	& \times \left[\frac{1}{3}\,\mathcal{A}_2^{\mu_2\nu_2\alpha\beta}(p_2,p_4-p_2)\pi^{\mu_4\nu_4}(-p_4)\right]
\end{align}
hence given the amplitude
\begin{align}
	\mathcal{M}_t(\lambda_1,\lambda_2) \equiv \varepsilon^{(2)}_{\mu_1\nu_1}(p_1,\lambda_1) \varepsilon^{(1)}_{\mu_2\nu_2}(p_2,\lambda_2) \varepsilon^{(0)}_{\mu_3\nu_3}(p_3) \varepsilon^{(0)}_{\mu_4\nu_4}(p_4) \mathcal{M}_t^{\mu_1\nu_1\mu_2\nu_2\mu_3\nu_3\mu_4\nu_4} (p_1,p_2,p_3,p_4) \nonumber 
\end{align}
we can separate it into two components: the first one where the scattering is mediated by a spin-0 propagator and the second one where it is mediated by a spin-2 propagator \eqref{expansionX2}
\begin{equation}
	\mathcal{M}_t = \mathcal{M}_t^{(0-s)} + \mathcal{M}_t^{(2)}.
\end{equation}
On shell one finds:
\begin{align}
	\mathcal{M}_t^{(0-s)} &= 0\\
	\mathcal{M}_t^{(2)}(\lambda_1,\lambda_2) &= \begin{dcases}
		-\frac{(b+b')^2}{36} \, \left[t^3\left(1 + \frac{4\, u^2}{s^2}\right)^2\right]\qquad &\text{if} \qquad \lambda_1 = \lambda_2,\\
		-\frac{4(b+b')^2}{9M_P^6}\, \left(t^3\,\frac{u^4}{s^4}\right)\qquad &\text{if} \qquad \lambda_1 \neq \lambda_2.
	\end{dcases}\label{eq:amp_ggss_tChannel}
\end{align}
Following the same line of reasoning for the $u$-channel:
\begin{align}
	\mathcal{M}_u^{\mu_1\nu_1\mu_2\nu_2\mu_3\nu_3\mu_4\nu_4} (p_1,p_2,p_3,p_4) = & \frac{1}{M_P^6}\left[\frac{1}{3}\mathcal{A}_2^{\mu_1\nu_1\rho\sigma}(p_1,p_4-p_1)\pi^{\mu_4\nu_4}(p_4)\right]\Big(D^{(0-s)}_{\rho\sigma\alpha\beta}+D^{(2)}_{\rho\sigma\alpha\beta}\Big)(p_1-p_4)\nonumber\\
	& \times \left[\mathcal{A}_2^{\mu_2\nu_2\alpha\beta}(p_2,p_3-p_2)\pi^{\mu_4\nu_4}(-p_3)\right]
\end{align}
and separating the amplitude based on the spin of the propagator:
\begin{equation}
	\mathcal{M}_u = \mathcal{M}_u^{(0-s)} + \mathcal{M}_u^{(2)}
\end{equation}
going on shell one finds:
\begin{align}
	\mathcal{M}_u^{(0-s)} &= 0\\
	\mathcal{M}_u^{(2)}(\lambda_1,\lambda_2) &= \begin{dcases}
		- \frac{(b+b')^2}{36M_P^6} \, \left[u^3\left(1 + \frac{4\,t^2}{s^2}\right)^2\right]\qquad &\text{if} \qquad \lambda_1 = \lambda_2,\\
		-\frac{4(b+b')^2}{9M_P^6}\, \left(u^3\,\frac{t^4}{s^4}\right)\qquad &\text{if} \qquad \lambda_1 \neq \lambda_2.
	\end{dcases}\label{eq:amp_ggss_uChannel}
\end{align}
The result of this analysis is that the interaction manifests only as contact terms.
\subsection{Total amplitude cross section}
Using equations \eqref{eq:amp_ggss_sChannel},\eqref{eq:amp_ggss_tChannel}and \eqref{eq:amp_ggss_uChannel} we can compute the total scattering amplitude for equal and opposite polarizations
\begin{align}
	\mathcal{M}_\text{eq} &\equiv  \mathcal{M}^{(2)}_t(\pm,\pm) + \mathcal{M}^{(2)}_u(\pm,\pm)= + \frac{4}{9 M_P^6}   \left[\frac{b'(b'+b)}{108}\,s^3 + (b+b')^2\left(\frac{t^3 u^3}{s^3} + \frac{t^2 u^2}{2s} -\frac{t^3 + u^3}{16}\right) \right], \\
	\mathcal{M}_\text{opp} &\equiv \mathcal{M}^{(2)}_t(\pm,\mp) + \mathcal{M}^{(2)}_u(\pm,\mp) =  \frac{4(b+b')^2}{9M_P^6} \,\left(\frac{t^3 u^3}{s^3} \right),
\end{align}
and derive the differential cross section which is given by
\begin{equation}
  \frac{d\sigma}{d\Omega} = \frac{1}{2}\frac{1}{64\pi^2 s}\,|\mathcal M|^2 = \frac{1}{128\pi^2 s}\left[|\mathcal M_\text{eq}|^2 + |\mathcal M_\text{opp}|^2 + 2\,\text{Re}(\mathcal M_\text{eq} \mathcal M_\text{opp}^*)\right]
\end{equation}
where the factor of $1/2$ accounts for the identical final states. Integrating over the solid angle we find the total cross sections
\begin{equation}
  \sigma_\text{eq} = \frac{1}{64\pi s}\int_{-1}^{1} dx\,|\mathcal M_\text{eq}|^2 
  \qquad \qquad 
  \sigma_\text{opp} = \frac{1}{64\pi s}\int_{-1}^{1} dx\,|\mathcal M_\text{opp}|^2 \qquad \qquad 
  \sigma_\text{int} = \frac{1}{64\pi s}\int_{-1}^{1} dx\,\text{Re}(\mathcal M_\text{eq} \mathcal M_\text{opp}^*),
\end{equation}
where we have defined $x \equiv \cos\theta$. Explicitly it is given by 
\begin{align}
\sigma_\text{eq} &= \frac{\tilde{C}^2 }{M_P^{12}}\,s^5 (b+b')^2 \left[c_{\text{eq}_1}b^2 + c_{\text{eq}_2} b\,b' + c_{\text{eq}_3} b'^2\right], \label{eq:tot_sigma_eq}\\
\sigma_\text{opp} &= \frac{\tilde{C}^2 }{M_P^{12}}\,s^5 (b+b')^4 c_{\text{opp}}, \label{eq:tot_sigma_opp}\\
\sigma_\text{int} &= \frac{\tilde{C}^2 }{M_P^{12}}\,s^5 (b+b')^3 \left[c_{\text{int}_1}b+c_{\text{int}_2}b'\right],\label{eq:tot_sigma_int}
\end{align}
where the numerical coefficients are
\begin{gather}
  c_{\text{eq}_1} = \frac{2609}{138378240\pi}, \qquad
  c_{\text{eq}_2} = \frac{739807}{16812956160 \pi}, \qquad
  c_{\text{eq}_3} = \frac{7770803}{302633210880\pi}, \nonumber\\
  c_{\text{opp}_1} = \frac{1}{1945944\pi},  \nonumber \\
  c_{\text{int}_1} = \frac{1189}{467026560\pi}, \qquad
  c_{\text{int}_2} = \frac{4139}{1401079680\pi}.
\end{gather}
Classical GR conventionally employs a linear representation of the excitations. The two are related by a simple change of basis
\beq
|+\rangle_{\rm lin}=\frac{|+2\rangle+|-2\rangle}{\sqrt2},
\qquad
|\times\rangle_{\rm lin}=\frac{|+2\rangle-|-2\rangle}{\sqrt2},
\eeq
and express the scattering amplitudes in the linear basis as
\beq
\mathcal M_{[+][+]}=\mathcal M_{\rm eq}+\mathcal M_{\rm opp},
\qquad
\mathcal M_{[\times][\times]}=\mathcal M_{\rm eq}-\mathcal M_{\rm opp},
\qquad
\mathcal M_{[+][\times]}=\mathcal M_{[\times][+]}=0.
\eeq
The corresponding cross sections are given by
\begin{align}\label{eq:sigma_plus_cross_def}
  \sigma_{[+][+]}=\sigma_{\text{eq}}+\sigma_{\text{opp}}+\sigma_{\text{int}},\qquad \qquad \sigma_{[\times][\times]}=\sigma_{\text{eq}}+\sigma_{\text{opp}}-\sigma_{\text{int}}.
\end{align}
Substituting the expressions for the three total cross sections from eq.~\eqref{eq:tot_sigma_eq}-\eqref{eq:tot_sigma_int} we find
\begin{align}
\sigma_{[+][+]}&=\frac{\tilde{C}^2 }{M_P^{12}}\,s^5 (b+b')^2 \left[(c_{\text{eq}_1} +c_{\text{opp}} + c_{\text{int}_1})b^2 + (c_{\text{eq}_2} + 2c_{\text{opp}} + c_{\text{int}_1} + c_{\text{int}_2})bb' + (c_{\text{eq}_3} +c_{\text{opp}}+c_{\text{int}_2}){b'}^2\right],\\
\sigma_{[\times][\times]}&=\frac{\tilde{C}^2 }{M_P^{12}}\,s^5 (b+b')^2 \left[(c_{\text{eq}_1} +c_{\text{opp}} - c_{\text{int}_1})b^2 + (c_{\text{eq}_2} + 2c_{\text{opp}} - c_{\text{int}_1} - c_{\text{int}_2})bb' + (c_{\text{eq}_3} +c_{\text{opp}} - c_{\text{int}_2}){b'}^2\right],\\
\sigma_{[+][\times]}&=\sigma_{[\times][+]}=0.
\end{align}
The unpolarized cross section is obtained by averaging over all possible initial helicity configurations, and can be expressed in terms of the helicity cross sections as
\beq
\sigma_{\rm unpol}
=
\frac14\left(
\sigma_{++}+\sigma_{+-}+\sigma_{-+}+\sigma_{--}
\right)
=
\frac12(\sigma_{\rm eq}+\sigma_{\rm opp}).
\eeq
Using \eqref{eq:sigma_plus_cross_def} we can also express the result in terms of the linear polarization cross sections as
\begin{equation}
\sigma_{\rm unpol} = \frac{1}{4}\left(\sigma_{[+][+]}+\sigma_{[\times][\times]}\right).
\end{equation}
Substituting the explicit expressions either for \(\sigma_{\rm eq}\) and \(\sigma_{\rm opp}\) or for \(\sigma_{[+][+]}\) and \(\sigma_{[\times][\times]}\),  the unpolarized cross section takes the form
\begin{equation}
\sigma_{\rm unpol} = \frac{\tilde{C}^2 }{2 M_P^{12}}\,s^5 (b+b')^2 \left[(c_{\text{eq}_1} +c_{\text{opp}})b^2 + (c_{\text{eq}_2} + 2c_{\text{opp}})bb' + (c_{\text{eq}_3} +c_{\text{opp}}){b'}^2\right].
\end{equation}
The cross section has a behaviour of $\mathcal{O}(s^5/{M_P}^{12})$ and it is quartic in the multiplicites of the conformal states appearing in the virtual corrections.

\section{Anomaly correction to the eikonal phase}

In this section we analyze how anomaly-mediated corrections modify high-energy photon scattering in impact-parameter space. Our goal is to separate long-distance effects, controlled by non-analytic \(t\)-channel singularities, from short-distance contributions, which are analytic in \(t\) and therefore local in impact-parameter space. We focus on a helicity-preserving channel and suppress explicit helicity labels whenever no confusion can arise. We use the standard eikonal representation
\begin{equation}
\mathcal{M}(s,t)
=
2s\int d^2\mathbf{b}\;e^{-i\mathbf{q}\cdot\mathbf{b}}
\left(e^{i\chi(s,\mathbf{b})}-1\right),
\qquad
t=-\mathbf{q}^2,
\label{eq:eikonal-convention}
\end{equation}
which at leading eikonal order reduces to
\begin{equation}
e^{i\chi}-1\simeq i\chi,
\qquad
\chi(s,\mathbf{b})
=
\frac{1}{2si}\int\frac{d^2\mathbf{q}}{(2\pi)^2}\,
e^{i\mathbf{q}\cdot\mathbf{b}}\,
\mathcal{M}(s,-\mathbf{q}^2).
\label{eq:chi-linear}
\end{equation}
At tree level, the Einstein--Hilbert contribution is
\begin{equation}
\mathcal{M}^{(0)}(s,t,u)
=
\mathcal{M}^{(0)}_{s}
+\mathcal{M}^{(0)}_{t}
+\mathcal{M}^{(0)}_{u}.
\label{eq:tree-full-stu-clean}
\end{equation}
In the Regge/eikonal limit,
\begin{equation}
s\gg |t|,
\qquad
t=-\mathbf{q}^{2},
\label{eq:regge-limit}
\end{equation}
the only term that generates a long-range phase is the \(t\)-channel pole,
\begin{equation}
\mathcal{M}^{(0)}_{t}(s,t)\simeq i\,\frac{\kappa^2 s^2}{4t},
\label{eq:t-pole}
\end{equation}
while \(\mathcal{M}^{(0)}_{s}\), \(\mathcal{M}^{(0)}_{u}\), and the regular part of \(\mathcal{M}^{(0)}_{t}\) are analytic in \(t\), and therefore correspond to contact terms in \(\mathbf b\)-space. For \(b\neq 0\), one then finds
\begin{equation}
\chi^{(0)}(s,b)
=
\frac{1}{2si}\int\frac{d^2\mathbf{q}}{(2\pi)^2}\,
e^{i\mathbf{q}\cdot\mathbf{b}}
\left(i\,\frac{\kappa^2 s^2}{4t}\right)
=
-\frac{\kappa^2 s}{8}
\int\frac{d^2\mathbf{q}}{(2\pi)^2}\,
\frac{e^{i\mathbf{q}\cdot\mathbf{b}}}{\mathbf{q}^2}.
\label{eq:chi0-pre}
\end{equation}
Using
\begin{equation}
\int\frac{d^2\mathbf{q}}{(2\pi)^2}\,
\frac{e^{i\mathbf{q}\cdot\mathbf{b}}}{\mathbf{q}^2}
=
\frac{1}{2\pi}\log\frac{L}{b},
\label{eq:2d-log-ft}
\end{equation}
we obtain
\begin{equation}
\chi^{(0)}(s,b)
=
-\frac{\kappa^2 s}{16\pi}\log\frac{L^2}{b^2},
\qquad (b\neq0),
\label{eq:chi-tree-final-clean}
\end{equation}
up to the overall sign convention for \(\mathcal M\). The physical meaning of \(\chi(s,b)\) follows from the stationary-phase approximation applied to \eqref{eq:eikonal-convention}. The phase of the Fourier transform is
\begin{equation}
\Phi(\mathbf b)=\mathbf q\cdot \mathbf b+\chi(s,b),
\label{eq:phase-def}
\end{equation}
and the saddle-point condition reads
\begin{equation}
\nabla_b \Phi(\mathbf b)=0
\qquad\Longrightarrow\qquad
\mathbf q_\perp=-\nabla_b \chi(s,b).
\label{eq:saddle-condition}
\end{equation}
Thus the transverse momentum transfer is determined directly by the gradient of the eikonal phase. For two massless particles in the center-of-mass frame,
\begin{equation}
p=\frac{\sqrt{s}}{2},
\qquad
t=-2p^2(1-\cos\theta),
\label{eq:t-angle}
\end{equation}
and for small-angle scattering,
\begin{equation}
q_\perp^2=-t\simeq p^2\theta^2.
\label{eq:small-angle-relation}
\end{equation}
Therefore the deflection angle is
\begin{equation}
\theta(s,b)
=
-\frac{1}{p}\,\partial_b \delta(s,b),
\qquad
\delta(s,b)\equiv \Re \chi(s,b),
\label{eq:theta-from-eikonal}
\end{equation}
or, equivalently, in absolute value,
\begin{equation}
|\theta(s,b)|=\frac{1}{p}\,|\partial_b\delta(s,b)|.
\label{eq:theta-abs}
\end{equation}
Since in the present tree-level channel \(\chi^{(0)}(s,b)\) is real up to conventions, we may identify \(\delta=\chi^{(0)}\), and \eqref{eq:chi-tree-final-clean} gives
\begin{equation}
\theta^{(0)}(s,b)
=
-\frac{1}{p}\,\partial_b \chi^{(0)}(s,b)
=
\frac{\kappa^2\sqrt{s}}{4\pi b},
\label{eq:theta-tree}
\end{equation}
again up to the overall sign convention, while the magnitude of the deflection is unambiguous.

\subsection{Anomaly correction}

From the channel decomposition obtained in the previous section,
\begin{equation}
\mathcal{M}^{\mathrm{anom}}_{s}
=
-\frac{s}{4},
\qquad
\mathcal{M}^{\mathrm{anom}}_{t}
=
-\frac{t(t-u)^2}{4s^2},
\qquad
\mathcal{M}^{\mathrm{anom}}_{u}
=
-\frac{u(t-u)^2}{4s^2},
\end{equation}
hence
\begin{equation}
\mathcal{M}^{\mathrm{anom}}(s,t,u)
=
-\frac{tu}{s}.
\label{eq:anom-total-clean}
\end{equation}
Using \(u=-s-t\), this becomes
\begin{equation}
-\frac{tu}{s}
=
-\frac{t(-s-t)}{s}
=
t+\frac{t^2}{s}.
\label{eq:anom-expanded}
\end{equation}
With \(t=-\mathbf q^2\), we obtain
\begin{equation}
\mathcal{M}^{\mathrm{anom}}(s,-\mathbf q^2)
=
-\mathbf q^2+\frac{\mathbf q^4}{s}.
\label{eq:anom-qspace-clean}
\end{equation}
This expression is analytic at \(t=0\): in particular, it contains no \(1/t\) pole. Therefore it cannot generate the logarithmic long-distance eikonal phase associated with massless exchange. Let \(C_{\mathrm{anom}}\) denote the overall normalization of the anomaly amplitude. Then
\begin{equation}
\delta\chi^{\mathrm{anom}}(s,\mathbf b)
=
\frac{C_{\mathrm{anom}}}{2si}
\int\frac{d^2\mathbf q}{(2\pi)^2}\,
e^{i\mathbf q\cdot\mathbf b}
\left(-\mathbf q^2+\frac{\mathbf q^4}{s}\right).
\label{eq:delta-chi-anom-int-clean}
\end{equation}
Using
\begin{equation}
\int\frac{d^2\mathbf q}{(2\pi)^2}
e^{i\mathbf q\cdot\mathbf b}\,\mathbf q^2
=
-\nabla_b^2\delta^{(2)}(\mathbf b),
\qquad
\int\frac{d^2\mathbf q}{(2\pi)^2}
e^{i\mathbf q\cdot\mathbf b}\,\mathbf q^4
=
\nabla_b^4\delta^{(2)}(\mathbf b),
\label{eq:delta-identities}
\end{equation}
we find
\begin{equation}
\delta\chi^{\mathrm{anom}}(s,\mathbf b)
=
\frac{C_{\mathrm{anom}}}{2si}
\left[
\nabla_b^2\delta^{(2)}(\mathbf b)
+\frac{1}{s}\nabla_b^4\delta^{(2)}(\mathbf b)
\right].
\label{eq:delta-chi-anom-clean}
\end{equation}
Hence
\begin{equation}
\delta\chi^{\mathrm{anom}}(s,b)=0,
\qquad b\neq0.
\label{eq:delta-chi-zero}
\end{equation}
The total eikonal phase is therefore
\begin{equation}
\chi(s,\mathbf b)
=
\chi^{(0)}(s,\mathbf b)
+
\delta\chi^{\mathrm{anom}}(s,\mathbf b),
\end{equation}
and for nonzero impact parameter
\begin{equation}
\chi(s,b)
=
-\frac{\kappa^2 s}{16\pi}\log\frac{L^2}{b^2},
\qquad b\neq0.
\label{eq:chi-total-final-clean}
\end{equation}
It follows that the anomaly contribution does not modify the long-distance scattering angle:
\begin{equation}
\theta(s,b)=\theta^{(0)}(s,b),
\qquad b\neq0,
\label{eq:theta-no-anom}
\end{equation}
up to contact terms localized at \(\mathbf b=0\). In other words, the anomaly affects the eikonal phase only through ultra-local terms in impact-parameter space, while the classical deflection is entirely controlled by the non-analytic graviton pole.

\section{Extension to dark sectors}
In this section we comment on anomaly--mediated interactions between a visible and a dark conformal sector, extending the formal framework developed in the previous sections and emphasizing its dynamical implications. The technical structure of the relevant correlators and effective vertices has already been established; here we focus instead on how these ingredients operate when two otherwise decoupled sectors communicate through gravity and, more specifically, through the trace anomaly.

As a concrete example, consider a visible spin--1 conformal sector interacting with a virtual spin--1 field belonging to a corresponding dark one. Within the Einstein--Hilbert description, the interaction proceeds through metric exchange and mirrors structurally the photon--photon scattering process analyzed earlier. The dominant contribution arises from the $t$--channel diagram, leading schematically to
\begin{equation}
\mathcal{M}(s,t) \sim \kappa^2 \, T^{\mu\nu}_{\text{vis}} \,
\frac{P_{\mu\nu\rho\sigma}}{t} \,
T^{\rho\sigma}_{\text{dark}} \, ,
\label{eq:visible-dark-amplitude}
\end{equation}
where $P_{\mu\nu\rho\sigma}$ denotes the graviton propagator and $T^{\mu\nu}_{\text{vis}}$, $T^{\rho\sigma}_{\text{dark}}$ are the corresponding energy--momentum tensors. The coupling is thus entirely controlled by the stress--energy content of the two sectors and inherits the tensorial structure already discussed in the purely visible case.\\
The novelty emerges once quantum conformal breaking is included. Even if both sectors are classically scale invariant, the trace anomaly induces a nonvanishing trace of the energy--momentum tensor, with separate contributions from visible and dark sectors. Assuming free field theory realizations of the anomaly in both sectors, say $L$ and $R$, where $L$ is visible an $R$ is dark 
\be
\mathcal{A}_{L R}(x)=   b_{ L R}\, C^2 +
 b_{L R}' \big(E - \tfrac{2}{3}\sq R\big) + b_{L R}'' \sq R .
\label{trace2}
\ee
the virtual corrections can be integrated out in both sectors, generating a nonlocal action which is, however, unique for both, since the total conformal anomaly is simply the sum of the anomalies of the two $L$ and $R$ sectors. 
\beq
\mathcal{A}(x)= \mathcal{A}_{L }(x) + \mathcal{A}_{R}(x).
 \eeq
As a consequence, alongside the transverse--traceless spin--2 exchange, an effective scalar component of the metric contributes to the interaction. From the perspective of the visible sector, this scalar exchange activates channels that are absent at the classical level and whose structure is dictated by the anomalous form factors derived previously. The dark sector enters explicitly through its anomaly coefficients: if it contains multiple conformal species, their multiplicities enhance the trace anomaly and therefore the strength of the effective coupling. The hidden sector thus imprints its internal structure directly onto the amplitude through the anomaly.\\
In the high--energy limit, the eikonal treatment applies to this mixed visible--dark process as well. The resulting phase receives contributions from both the spin--2 and effective spin--0 exchanges, while contact terms ensure consistency with the Ward identities and with energy--momentum conservation. In this sense, conformal symmetry breaking is communicated between sectors by the scalar component of the gravitational field, which acts as the carrier of the anomalous trace.\\
\subsection{ IR effects from the $\Delta_4^{-1}$ interactions}
Within the minimal Einstein--Hilbert framework, however, these effects remain parametrically suppressed by the Planck scale. The interaction becomes relevant only in the regime where gravitational effects are unsuppressed, whereas at lower energies the exchanges are negligible.\\
A different situation arises when the interaction is described by the full anomaly--induced effective action rather than by the minimal Einstein--Hilbert term, as done in this work. In that case the dynamics is governed by a nonlocal functional which can be cast in local form through the introduction of the conformalon field \cite{Mottola:2016mpl}. This scalar resums the infrared contributions associated with the trace anomaly and encodes conformal symmetry breaking in a scale--independent manner.\\
In such a scenario, the breaking of conformal symmetry in the visible sector can be described in terms of the Riegert action, with coefficients determined entirely by the multiplicities, separately, of the hidden and the visible sectors. The resulting scalar exchange is no longer directly tied to Planck--suppressed couplings but instead controlled by the anomaly coefficients themselves. The interaction between visible and dark sectors thus acquires an intrinsically infrared character, driven by quantum conformal breaking rather than by ultraviolet gravitational exchange.\\
It should be noted, however, that the anomaly--induced action has known limitations. As emphasized in~\cite{Coriano:2022jkn} and \cite{Coriano:2021nvn}, for four--point functions the inclusion of Weyl--invariant terms is necessary in order to construct an effective action fully consistent with the hierarchical conformal Ward identities formulated around Minkowski space. \\ 
Differently from the Einstein--Hilbert analysis performed here, the contributions emerging from the infrared structure of the Riegert action can become dominant in regimes of large curvature and are not suppressed by the Planck scale. 

For this reason, it is natural to ask whether such effects may become relevant in strongly curved backgrounds, for instance in astrophysical processes such as neutron--star mergers, where curvature scales are large and anomaly--induced contributions could, in principle, compete with or even dominate over the minimal gravitational exchange.

\section{Conclusions}

In this work we have analyzed the structure of anomaly--mediated gravitational interactions between conformal sectors, focusing on the scalar contribution associated with the trace anomaly and its interplay with the Einstein--Hilbert (EH) action. The analysis has been performed at the level of scattering amplitudes around flat spacetime, allowing a direct comparison between the standard perturbative description of gravity and the effective nonlocal dynamics induced by the anomaly.

The EH action provides the gravitational kernel connecting the two matter sectors. Expanding the metric in the de Donder gauge, the quadratic EH action determines the graviton propagator, whose covariant spin decomposition separates the transverse--traceless spin--2 component from the transverse scalar projector $P^{(0\text{-}s)}$. Although this scalar component does not correspond to an independent propagating mode in pure Einstein gravity, it becomes dynamically relevant when coupled to anomalous three--point correlators. In the perturbative framework considered here the corresponding interaction behaves effectively as a contact term, since the propagator structure $\sim 1/(q^2 M_P^2)$ leads to amplitudes scaling as $s/M_P^2$, in close analogy with Fermi-type effective interactions.

The four--point amplitudes are obtained by sewing two three--point correlators, such as $TJJ$ or $TTT$, through single--graviton exchange. In this construction the EH propagator interpolates between the two sectors, while the anomaly enters through the trace part of the vertices encoded in the transverse projector $\pi_{\mu\nu}$. The resulting amplitudes exhibit a clear separation between the standard spin--2 exchange and an anomaly--induced scalar contribution. This scalar channel is not introduced as an additional field but emerges from the covariant structure of the graviton propagator once the trace anomaly is present.

In the absence of conformal symmetry breaking such scalar contributions cancel or can be removed by improvement terms, but once the anomaly is included the trace component becomes unavoidable and the scalar channel acquires physical significance. In the eikonal regime this anomaly--mediated interaction modifies the gravitational phase through contact--type contributions in impact--parameter space, while the long--distance behaviour remains dominated by the spin--2 sector. \\
The EH action therefore plays a dual role: it governs the standard gravitational dynamics and provides the propagator structure through which the anomaly pole contained in conformal correlators manifests itself as an effective scalar interaction. In this way scalar gravitational effects arise within a purely metric framework, without introducing fundamental scalar fields. Unlike scenarios such as the Starobinsky model, where $R^2$ corrections can be rewritten as Einstein gravity plus a scalar degree of freedom, here the scalar dynamics originates from the quantum breaking of conformal symmetry.

A natural extension of this analysis consists in promoting the gravitational sector to the full nonlocal anomaly action, namely the Riegert effective action,
\begin{equation}
S_{\mathrm{g}} = S_{\mathrm{EH}} + S_{\mathrm{anom, vis + dark}} ,
\end{equation}
where $S_{\mathrm{anom, vis + dark}}$ is scale independent and resums the infrared effects associated with the trace anomaly. In this framework the scalar degree of freedom, often referred to as the conformalon, acquires its own nonlocal kinetic structure and the amplitudes incorporate the fully resummed infrared dynamics of conformal symmetry breaking.

In summary, anomaly--mediated scalar gravitational interactions can be consistently described within a perturbative Einstein--Hilbert framework, where graviton exchange interpolates between conformal sectors and the scalar channel emerges from the trace anomaly. Extending the analysis to the full nonlocal anomaly action will allow one to capture the complete infrared structure of anomaly--induced gravity and explore the phenomenological implications of such scalar gravitational modes.

\centerline{\bf Acknowledgements} 
We thank Emil Mottola for discussions. This work is partially supported by INFN, inziativa specifica {\em QG-sky}, by the the grant PRIN 2022BP52A MUR "The Holographic Universe for all Lambdas" Lecce-Naples, and by the European Union, Next Generation EU, PNRR project "National Centre for HPC, Big Data and Quantum Computing", project code CN00000013.

\newpage

\appendix

\section{ Basis of the Abelian $TJJ$ and the on-shell limit of the spin-1}
\label{reff}
The correlator can be built using the  set of $13$ tensors catalogued in Table \ref{genbasis}. They are linearly independent for generic $k^2, p^2, q^2$
different from zero. Five of the $13$ tensors are Bose symmetric, namely,
\bea
t_i^{\mu\nu\alpha\beta}(p,q) = t_i^{\mu\nu\beta\alpha}(q,p)\,,\qquad i=1,2,7,8,13\,,
\eea
while the remaining eight tensors form four pairs which are overall related by Bose symmetry
\bea
&&t_3^{\mu\nu\alpha\beta}(p,q) = t_5^{\mu\nu\beta\alpha}(q,p)\,,\\
&&t_4^{\mu\nu\alpha\beta}(p,q) = t_6^{\mu\nu\beta\alpha}(q,p)\,,\\
&&t_9^{\mu\nu\alpha\beta}(p,q) = t_{10}^{\mu\nu\beta\alpha}(q,p)\,,\\
&&t_{11}^{\mu\nu\alpha\beta}(p,q) = t_{12}^{\mu\nu\beta\alpha}(q,p)\,.
\label{tpairs}
\eea
\bea
 \widetilde t_1^{\, \mu \nu \a \b} &=& \lim_{s_1,s_2 \rightarrow 0} \,  t_1^{\, \mu \nu \a \b} =
 (s \, g^{\mu\nu} - k^{\mu}k^{\nu}) \, u^{\a \b} (p,q)  
 \label{widetilde1x}\\
 \widetilde t_3^{\, \mu \nu \a \b} \, + \,  \widetilde t_5^{\, \mu \nu \a \b}
- \, 4 \, \widetilde t_7^{\, \mu \nu \a \b} &=&  \lim_{s_1,s_2 \rightarrow 0}  \, (t_3^{\, \mu \nu \a \b} \, + \,  t_5^{\, \mu \nu \a \b}
- \, 4 \,  t_7^{\, \mu \nu \a \b}) = \nn  \\
&& \hspace{-0.5cm} - 2 \, u^{\a \b} (p,q) \left( s \, g^{\mu \nu} + 2 (p^\mu \, p^\nu + q^\mu \, q^\nu )
- 4 \, (p^\mu \, q^\nu + q^\mu \, p^\nu) \right)  \\
\widetilde{t}^{\, \mu \nu \alpha \beta}_{13} &=&  \lim_{s_1,s_2 \rightarrow 0} \,  t_{13}^{\, \mu \nu \a \b} =
\big(p^{\mu} q^{\nu} + p^{\nu} q^{\mu}\big)g^{\alpha\beta}
+ \frac{s}{2} \big(g^{\alpha\nu} g^{\beta\mu} + g^{\alpha\mu} g^{\beta\nu}\big) \nn \\
&&  \hspace{-0.5cm} - g^{\mu\nu} (\frac{s}{2} g^{\alpha \beta}- q^{\alpha} p^{\beta})
-\big(g^{\beta\nu} p^{\mu}
+ g^{\beta\mu} p^{\nu}\big)q^{\alpha}
 - \big (g^{\alpha\nu} q^{\mu}
+ g^{\alpha\mu} q^{\nu }\big)p^{\beta}, 
\label{widetilde13}
\nn \\
\eea

\subsection{Evaluations of the $TJJ$}
The ordinary massless scalar master integral is given by
\begin{equation}
\label{lco}
		C_0\left(q^2,p_1^2,p_2^2\right)= \frac{1}{i \pi^2}\int d^n k\frac{1}{k^2 (k- q)^2 (k - p_1)^2},\,
	\end{equation}
In the explicit evaluation we can use the relations  

\bea
 &  C_0 ( p_1^2,p_2^2,q^2) = \frac{ 1}{q^2} \Phi (x,y),
 \eea
where the function $\Phi (x, y)$  is defined as
\cite{Usyukina:1993ch}
\bea
\Phi( x, y) &=& \frac{1}{\lambda} \biggl\{ 2 [Li_2(-\rho  x) + Li_2(- \rho y)]  +
\ln \frac{y}{ x}\ln \frac{1+ \rho y }{1 + \rho x}+ \ln (\rho x) \ln (\rho  y) + \frac{\pi^2}{3} \biggr\},
\label{Phi}
\eea
with
\bea
 \lambda(x,y) = \sqrt {\Delta},
 \qquad  \qquad \Delta=(1-  x- y)^2 - 4  x  y,
\label{lambda} \\
\rho( x,y) = 2 (1-  x-  y+\lambda)^{-1},
  \qquad  \qquad x=\frac{p_1^2}{q^2} \, ,\qquad \qquad y= \frac {p_2^2}{q^2}\, .
\eea
and 
\begin{equation}
{B}_0(p^2)=\sdfrac{1}{i\p^\frac{d}{2}}\int\,d^d l\ \frac{1}{l^2(l-p_1)^2}=\frac{ \left[\G\left(\frac{d}{2}-1\right)\right]^2\G\left(2-\frac{d}{2}\right)}{\G\left(d-2\right)(p^2)^{2-\frac{d}{2}}}\label{B0ex}.
\end{equation}
with 
\beq
B_0(p^2)=\frac{1}{\varepsilon}+\bar{B}_0(p^2)
\eeq
and
\beq
B_0^R(p^2,0,0)\equiv \bar{B}_0(p^2)= 2 + \log(\mu^2/p^2) 
\label{BB}
\eeq
is the finite part in $d=4$ of the scalar integral in the $\overline{MS}$ scheme. 
\subsection{The conformal solution}
\label{bms}
The general solution has been given in \cite{Bzowski:2018fql} that here we report  for convenience,

\begin{table}
$$
\begin{array}{|c|c|}\hline
i & t_i^{\mu\nu\alpha\beta}(p,q)\\ \hline\hline
1 &
\left(k^2 g^{\mu\nu} - k^{\mu } k^{\nu}\right) u^{\alpha\beta}(p.q)\\ \hline
2 &
\left(k^2g^{\mu\nu} - k^{\mu} k^{\nu}\right) w^{\alpha\beta}(p.q)  \\ \hline
3 & \left(p^2 g^{\mu\nu} - 4 p^{\mu}  p^{\nu}\right)
u^{\alpha\beta}(p.q)\\ \hline
4 & \left(p^2 g^{\mu\nu} - 4 p^{\mu} p^{\nu}\right)
w^{\alpha\beta}(p.q)\\ \hline
5 & \left(q^2 g^{\mu\nu} - 4 q^{\mu} q^{\nu}\right)
u^{\alpha\beta}(p.q)\\ \hline
6 & \left(q^2 g^{\mu\nu} - 4 q^{\mu} q^{\nu}\right)
w^{\alpha\beta}(p.q) \\ \hline
7 & \left[p\cdot q\, g^{\mu\nu}
-2 (q^{\mu} p^{\nu} + p^{\mu} q^{\nu})\right] u^{\alpha\beta}(p.q) \\ \hline
8 & \left[p\cdot q\, g^{\mu\nu}
-2 (q^{\mu} p^{\nu} + p^{\mu} q^{\nu})\right] w^{\alpha\beta}(p.q)\\ \hline
9 & \left(p\cdot q \,p^{\alpha}  - p^2 q^{\alpha}\right)
\big[p^{\beta} \left(q^{\mu} p^{\nu} + p^{\mu} q^{\nu} \right) - p\cdot q\,
(g^{\beta\nu} p^{\mu} + g^{\beta\mu} p^{\nu})\big]  \\ \hline
10 & \big(p\cdot q \,q^{\beta} - q^2 p^{\beta}\big)\,
\big[q^{\alpha} \left(q^{\mu} p^{\nu} + p^{\mu} q^{\nu} \right) - p\cdot q\,
(g^{\alpha\nu} q^{\mu} + g^{\alpha\mu} q^{\nu})\big]  \\ \hline
11 & \left(p\cdot q \,p^{\alpha} - p^2 q^{\alpha}\right)
\big[2\, q^{\beta} q^{\mu} q^{\nu} - q^2 (g^{\beta\nu} q^ {\mu}
+ g^{\beta\mu} q^{\nu})\big]  \\ \hline
12 & \big(p\cdot q \,q^{\beta} - q^2 p^{\beta}\big)\,
\big[2 \, p^{\alpha} p^{\mu} p^{\nu} - p^2 (g^{\alpha\nu} p^ {\mu}
+ g^{\alpha\mu} p^{\nu})\big] \\ \hline
13 & \big(p^{\mu} q^{\nu} + p^{\nu} q^{\mu}\big)g^{\alpha\beta}
+ p\cdot q\, \big(g^{\alpha\nu} g^{\beta\mu}
+ g^{\alpha\mu} g^{\beta\nu}\big) - g^{\mu\nu} u^{\alpha\beta} \\
& -\big(g^{\beta\nu} p^{\mu}
+ g^{\beta\mu} p^{\nu}\big)q^{\alpha}
- \big (g^{\alpha\nu} q^{\mu}
+ g^{\alpha\mu} q^{\nu }\big)p^{\beta}  \\ \hline
\end{array}
$$
\caption{Basis of 13 fourth rank tensors satisfying the vector current conservation on the external lines with momenta $p$ and $q$. \label{genbasis}}
\end{table}

\begin{align}
A_1
&=
C_1\, I_{5\{211\}}
\nonumber\\
&=
C_1\, p_1 p_2 p_3
\left(
p_1 \frac{\partial}{\partial p_1}-1
\right)
\frac{\partial^3}{\partial p_1\,\partial p_2\,\partial p_3}
I_{1\{000\}},
\label{eq:A1}
\\
A_2
&=
2 C_{JJ}
\left(
2- p_1 \frac{\partial}{\partial p_1}
\right)
I^{(\mathrm{fin})}_{2\{111\}}
-
C_1\, p_1^{3} p_2 p_3
\frac{\partial^3}{\partial p_1\,\partial p_2\,\partial p_3}
I_{1\{000\}}
\nonumber\\
&\quad
-\frac{2}{3} C_{JJ}
\left[
\ln\!\frac{p_1^{2}}{\mu^{2}}
+
\ln\!\frac{p_2^{2}}{\mu^{2}}
+
\ln\!\frac{p_3^{2}}{\mu^{2}}
\right]
+\frac{2}{3} C_{JJ}
-2 D_{JJ},
\label{eq:A2}
\\
A_3
&=
2 C_{JJ}
\left(
2- p_1 \frac{\partial}{\partial p_1}
\right)
I^{(\mathrm{fin})}_{2\{111\}}
+
2 C_1\, p_1 p_2 p_3^{2}
\left(
1- p_1 \frac{\partial}{\partial p_1}
\right)
\frac{\partial^{2}}{\partial p_1\,\partial p_2}
I_{1\{000\}}
\nonumber\\
&\quad
-\frac{2}{3} C_{JJ}
\left[
\ln\!\frac{p_1^{2}}{\mu^{2}}
+
\ln\!\frac{p_2^{2}}{\mu^{2}}
+
\ln\!\frac{p_3^{2}}{\mu^{2}}
\right]
-2\left(C_1 + D_{JJ}\right)
+\frac{2}{3} C_{JJ},
\label{eq:A3}
\\
A_4
&=
-2 C_{JJ}\, p_1^{2}\, I^{(\mathrm{fin})}_{2\{111\}}
-
2 C_1\, p_2^{2} p_3^{2} p_1
\left(
1- p_1 \frac{\partial}{\partial p_1}
\right)
\frac{\partial}{\partial p_1}
I_{1\{000\}}
\nonumber\\
&\quad
+\frac{1}{3} C_{JJ}\, p_1^{2}
\left[
\ln\!\frac{p_1^{2}}{\mu^{2}}
+
\ln\!\frac{p_2^{2}}{\mu^{2}}
+
\ln\!\frac{p_3^{2}}{\mu^{2}}
\right]
-
C_{JJ}
\left[
p_2^{2}\ln\!\frac{p_2^{2}}{\mu^{2}}
+
p_3^{2}\ln\!\frac{p_3^{2}}{\mu^{2}}
\right]
+
\left(C_1 + D_{JJ}\right)
\left(
p_1^{2}-p_2^{2}-p_3^{2}
\right)
-
C_{JJ}\, p_1^{2}.
\label{eq:A4}
\end{align}
shares a finite IR limit, shown in \eqref{bmcs} with the perturbative solution presented in \cite{Armillis:2009pq}. 
The integrals are given by

\begin{align}
I_{2\{111\}}^{(\mathrm{fin})}
=
-\frac{4 p_1^2 p_2^2 p_3^2}{J^2} \, I_{1\{000\}}
-\frac{1}{6 J^2}
\Bigg[&
p_1^2 \left(p_2^2 + p_3^2 - p_1^2\right)
\ln\!\left(\frac{p_1^4}{p_2^2 p_3^2}\right)
+ p_2^2 \left(p_1^2 + p_3^2 - p_2^2\right)
\ln\!\left(\frac{p_2^4}{p_1^2 p_3^2}\right)
\nonumber\\
&
+ p_3^2 \left(p_1^2 + p_2^2 - p_3^2\right)
\ln\!\left(\frac{p_3^4}{p_1^2 p_2^2}\right)
\Bigg] \, .
\end{align}

\begin{equation}
I_{1\{000\}}
=
\frac{1}{2\sqrt{-J^2}}
\left[
\frac{\pi^2}{6}
- 2 \ln\!\left(\frac{p_1}{p_3}\right)
  \ln\!\left(\frac{p_2}{p_3}\right)
+ \ln X \ln Y
- \operatorname{Li}_2 X
- \operatorname{Li}_2 Y
\right] .
\end{equation}

\begin{equation}
X =
\frac{-p_1^2 + p_2^2 + p_3^2 - \sqrt{-J^2}}
{2 p_3^2} \, ,
\qquad
Y =
\frac{-p_2^2 + p_1^2 + p_3^2 - \sqrt{-J^2}}
{2 p_3^2} \, .
\end{equation}

\begin{equation}
J^2 =
(p_1 + p_2 + p_3)
(-p_1 + p_2 + p_3)
(p_1 - p_2 + p_3)
(p_1 + p_2 - p_3)
\end{equation}

\begin{equation}
= -p_1^4 - p_2^4 - p_3^4
+ 2 p_1^2 p_2^2
+ 2 p_2^2 p_3^2
+ 2 p_3^2 p_1^2 \, .
\end{equation}

\section{Non covariant TT decomposition of the Fierz--Pauli action and the anomaly-induced scalar channel}
\label{app:TT_anomaly_reorganized}
The analysis presented in the Section \ref{closing}  can be recovered using a non-covariant (1+3) decomposition of the metric fluctuations. The derivation is lengthier compared to the covariant one, but provides a different perspective on the separation of the propagating from the non-propagating degrees of freedom in the general $h_{\mu \nu}$ as identified from the quadratic FP action coupled to an external conserved source. We first analyze the ordinary, non-anomalous case and show that only the TT sector propagates, as expected, while the vector and scalar sectors are constrained. We then include the anomaly pole and identify precisely which part of the decomposition is modified. The conclusion is that the pole present in the source does not affect the TT and vector sectors, but activates a specific nonlocal scalar channel, which can be described either in the \(3+1\) decomposition using two variables $\Theta$ and $\Xi$, defined below, or covariantly through the unique gauge-invariant scalar \(\Phi_{cov}\equiv h-\Box w\).\\
We begin from the quadratic action \eqref{FP} and decompose the metric fluctuation on the spatial slices as
\begin{equation}
h_{00}=2\Phi,
\label{eq:h00_reorg}
\end{equation}
\begin{equation}
h_{0i}=u_i+\partial_i B,
\qquad
\partial_i u_i=0,
\label{eq:h0i_reorg}
\end{equation}
and introduce the tensor decomposition 
\begin{equation}
h_{ij}
=
h_{ij}^{TT}
+\partial_i v_j+\partial_j v_i
+\left(\partial_i\partial_j-\frac13\delta_{ij}\nabla^2\right)E
+\frac13\delta_{ij}H,
\label{eq:hij_reorg}
\end{equation}
with
\begin{equation}
\partial_i v_i=0,
\qquad
\partial_i h_{ij}^{TT}=0,
\qquad
\delta^{ij}h_{ij}^{TT}=0.
\label{eq:constraints_reorg}
\end{equation}
Under the linearized diffeomorphisms \eqref{eq:lin_diff_ped}
with
\begin{equation}
\label{decs}
\xi_i=\xi_i^T+\partial_i\xi,
\qquad
\partial_i\xi_i^T=0,
\end{equation}
the variables transform as
\begin{equation}
\delta h_{ij}^{TT}=0,
\qquad
\delta v_i=\xi_i^T,
\qquad
\delta E=2\xi,
\qquad
\delta H=2\nabla^2\xi,
\end{equation}
\begin{equation}
\delta u_i=\partial_0\xi_i^T,
\qquad
\delta B=\xi_0+\partial_0\xi,
\qquad
\delta\Phi=\partial_0\xi_0.
\end{equation}
Notice that $\xi$ in \eqref{decs} is the scalar appearing in the longitudinal-transverse decomposition of the 
gauge variation $\xi_i$.
Natural gauge-invariant combinations are
\begin{equation}
V_i\equiv u_i-\partial_0v_i,
\qquad
\partial_iV_i=0,
\label{eq:Vi_reorg}
\end{equation}
and
\begin{equation}
\Theta\equiv \Phi-\partial_0B+\frac12\partial_0^2E,
\qquad
\Xi\equiv H-\nabla^2E.
\label{eq:ThetaXi_reorg}
\end{equation}

\subsection{TT, vector, and scalar sectors in the non-anomalous case}
Using
\begin{equation}
h=\eta^{\mu\nu}h_{\mu\nu}=2\Phi-H.
\label{eq:trace_reorg}
\end{equation}
and 
\begin{equation}
\partial_\mu h^{\mu0}=2\partial_0\Phi-\nabla^2B,
\label{eq:div0_reorg}
\end{equation}

\begin{equation}
\partial_\mu h^{\mu i}
=
-\partial_0u_i+\nabla^2v_i
+\partial_i\left(
-\partial_0B+\frac23\nabla^2E+\frac13H
\right),
\label{eq:divi_reorg}
\end{equation}

\begin{equation}
\partial_\mu h^{\mu i}
=
-\partial_0V_i
+\partial_i\left(
-\partial_0B+\frac23\nabla^2E+\frac13H
\right),
\end{equation}
after substitution into \eqref{FP}, one rewrites the Lagrangian in terms of the TT, vector, and scalar components as
\begin{equation}
\mathcal{L}_{\mathrm{FP}}
=
\mathcal{L}_{TT}
+
\mathcal{L}_{V}
+
\mathcal{L}_{S},
\label{eq:split_reorg}
\end{equation}
with
\begin{equation}
\mathcal{L}_{TT}
=
\frac14\,\partial_\lambda h_{ij}^{TT}\partial^\lambda h_{ij}^{TT},
\label{eq:LTT_reorg}
\end{equation}
\begin{equation}
\mathcal{L}_{V}
=
\frac12\,V_i\nabla^2V_i,
\label{eq:LV_reorg}
\end{equation}
and
\begin{equation}
\mathcal{L}_{S}
=
\frac43\,\Theta\,\nabla^2\Xi
+\frac13(\partial_0\Xi)^2
-\frac19(\partial_i\Xi)^2.
\label{eq:LS_reorg}
\end{equation}

Equivalently, up to total derivatives, the latter takes the form
\begin{equation}
\mathcal{L}_{S}
=
\frac43\,\Theta\,\nabla^2\Xi
-\frac13\,\Xi\,\partial_0^2\Xi
+\frac19\,\Xi\,\nabla^2\Xi.
\end{equation}

The full free quadratic action is therefore
\begin{equation}
S_{\mathrm{FP}}
=
\int d^4x\,
\left[
\frac14\,\partial_\lambda h_{ij}^{TT}\partial^\lambda h_{ij}^{TT}
+\frac12\,V_i\nabla^2V_i
+\frac43\,\Theta\,\nabla^2\Xi
+\frac13(\partial_0\Xi)^2
-\frac19(\partial_i\Xi)^2
\right],
\label{eq:free_action_reorg}
\end{equation}
giving the equations of motion
\begin{equation}
\Box h_{ij}^{TT}=0,
\label{eq:TT_free_reorg}
\end{equation}
\begin{equation}
\nabla^2V_i=0,
\label{eq:V_free_reorg}
\end{equation}
\begin{equation}
\nabla^2\Xi=0.
\label{eq:Xi_free_reorg}
\end{equation}
Hence, for fields vanishing suitably at spatial infinity,
\begin{equation}
\Xi=0,
\end{equation}
and the remaining scalar equation determines \(\Theta\) without introducing a propagating scalar mode. Therefore, in the non-anomalous case, the only radiative sector is the TT one, here retrieved in a different decomposition.

\subsection{Coupling to a conserved source}

We now include the coupling to a conserved source,
\begin{equation}
S_{\mathrm{int}}
=
\frac{\kappa}{2}\int d^4x\,h_{\mu\nu}J^{\mu\nu},
\qquad
\partial_\mu J^{\mu\nu}=0.
\label{eq:source_reorg}
\end{equation}

We decompose the source as
\begin{equation}
J^{00}=\rho,
\end{equation}
\begin{equation}
J^{0i}=J_T^i+\partial^iJ_L,
\qquad
\partial_iJ_T^i=0,
\end{equation}
\begin{equation}
J^{ij}
=
J_{TT}^{ij}
+\partial_iW_j+\partial_jW_i
+\left(\partial_i\partial_j-\frac13\delta_{ij}\nabla^2\right)\sigma
+\frac13\delta_{ij}\tau,
\end{equation}
with
\begin{equation}
\partial_iW_i=0,
\qquad
\partial_iJ_{TT}^{ij}=0,
\qquad
\delta_{ij}J_{TT}^{ij}=0.
\end{equation}

Current conservation implies
\begin{equation}
\partial_0\rho+\nabla^2J_L=0,
\label{eq:cons_rho_reorg}
\end{equation}
\begin{equation}
\partial_0J_T^i+\nabla^2W_i=0,
\label{eq:cons_vec_reorg}
\end{equation}
\begin{equation}
\partial_0J_L+\frac23\nabla^2\sigma+\frac13\tau=0.
\label{eq:cons_scal_reorg}
\end{equation}

Using these relations and integrating by parts, the interaction term becomes
\begin{equation}
\mathcal{L}_{\mathrm{int}}
=
\frac{\kappa}{2}\,h_{ij}^{TT}J_{TT}^{ij}
+
\kappa V_iJ_T^i
+
\kappa\Theta\,\rho
+
\frac{\kappa}{6}\,\Xi\,\tau.
\label{eq:Lint_reorg}
\end{equation}

Combining \eqref{eq:LTT_reorg}, \eqref{eq:LV_reorg}, \eqref{eq:LS_reorg} with \eqref{eq:Lint_reorg}, one gets
\begin{equation}
\mathcal{L}_{TT}^{(2)}
=
\frac14\,\partial_\lambda h_{ij}^{TT}\partial^\lambda h_{ij}^{TT}
+
\frac{\kappa}{2}\,h_{ij}^{TT}J_{TT}^{ij},
\end{equation}
\begin{equation}
\mathcal{L}_{V}^{(2)}
=
\frac12\,V_i\nabla^2V_i
+
\kappa V_iJ_T^i,
\end{equation}
and using \eqref{eq:ThetaXi_reorg}
\begin{equation}
\mathcal{L}_{S}^{(2)}
=
\frac43\,\Theta\,\nabla^2\Xi
+\frac13(\partial_0\Xi)^2
-\frac19(\partial_i\Xi)^2
+\kappa\Theta\,\rho
+\frac{\kappa}{6}\,\Xi\,\tau.
\label{eq:LSsourced_reorg}
\end{equation}
The corresponding equations are
\begin{equation}
\Box h_{ij}^{TT}=-\kappa J_{TT}^{ij},
\end{equation}
\begin{equation}
\nabla^2V_i=-\kappa J_T^i,
\end{equation}
\begin{equation}
\frac43\nabla^2\Xi+\kappa\rho=0,
\end{equation}
\begin{equation}
\frac43\nabla^2\Theta
-\frac23\partial_0^2\Xi
+\frac29\nabla^2\Xi
+\frac{\kappa}{6}\tau=0.
\end{equation}
Thus the TT sector propagates, whereas the vector and scalar sectors remain constrained even in the presence of an ordinary conserved source. The presence of a term of the form
\begin{equation}
(\partial_0\Xi)^2
\end{equation}
is necessary, but not sufficient, for \(\Xi\) to become dynamical. What matters is whether \(\Xi\) satisfies an independent second-order hyperbolic equation once all constraints are taken into account.

Indeed, in the ordinary scalar sector one has
\begin{equation}
\mathcal L_S
=
\frac43\,\Theta\,\nabla^2\Xi
+\frac13(\partial_0\Xi)^2
-\frac19(\partial_i\Xi)^2
+\cdots,
\label{eq:LS_Xi_dyn}
\end{equation}
where \(\Theta\) appears without time derivatives. Therefore \(\Theta\) is a Lagrange multiplier, not an independent propagating scalar. Varying with respect to \(\Theta\) gives
\begin{equation}
\nabla^2\Xi=\text{source},
\label{eq:Xi_constraint_dyn_full}
\end{equation}
which is a constraint equation rather than a wave equation. Thus, although the term \((\partial_0\Xi)^2\) is present, \(\Xi\) is not an independent propagating field: it is fixed by the constraint imposed by \(\Theta\).

For \(\Xi\) to become genuinely dynamical, the \(\Theta\)-equation must cease to be a pure constraint, so that \(\Xi\) obeys an independent hyperbolic equation. Equivalently, after integrating out the nondynamical fields, the reduced action for \(\Xi\) must contain a genuine inverse propagator of the form
\begin{equation}
\Xi\,\bigl(\alpha\,\partial_0^2-\beta\,\nabla^2+\cdots\bigr)\,\Xi,
\end{equation}
with no remaining relation of the type \eqref{eq:Xi_constraint_dyn_full} fixing \(\Xi\) instantaneously.

This can be expressed in terms of the scalar quadratic form. Writing the scalar sector as
\begin{equation}
S_S
=
\frac12
\begin{pmatrix}
\Theta & \Xi
\end{pmatrix}
K_S
\begin{pmatrix}
\Theta\\
\Xi
\end{pmatrix}
+
\begin{pmatrix}
\Theta & \Xi
\end{pmatrix}
J_S,
\label{eq:scalar_kernel_general_dyn}
\end{equation}
the ordinary Einstein--Hilbert theory has a constrained scalar kernel of reduced rank, precisely because \(\Theta\) has no independent kinetic term. In practical terms, \(\Xi\) becomes dynamical only if the anomaly modifies the scalar kernel in such a way that the \(\Theta\)-constraint is lifted. Equivalently, after inclusion of the anomaly, the reduced scalar kernel must develop a genuine propagating pole rather than remaining of purely constrained type.\\
This point is even clearer in the covariant formulation. \\
In the \(3+1\) decomposition, \(\Theta\) and \(\Xi\) are both gauge invariant, but they do not represent two independent covariant scalar modes.  
\subsection{Anomaly-induced pole term}

We now introduce the anomaly-induced contribution with a nonlocal current. We split the source as
\begin{equation}
J_{\mu\nu}(q)=J_{\mu\nu}^{\mathrm{pole}}(q)+J_{\mu\nu}^{\mathrm{reg}}(q),
\label{eq:Jsplit_anom_reorg}
\end{equation}
and define the pole part as
\begin{equation}
J_{\mu\nu}^{\mathrm{pole}}(q)
=
\frac{1}{q^2}\,\hat{\pi}_{\mu\nu}(q)\,u(q),
\qquad
\hat{\pi}_{\mu\nu}(q)=\eta_{\mu\nu}q^2-q_\mu q_\nu,
\label{eq:Jpole_reorg}
\end{equation}
where \(u(q)\) denotes the tensor structure \(u^{\alpha\beta}\) defined in \eqref{uab}, with the spectator Lorentz indices suppressed. We use
\begin{equation}
q^\mu=(\omega,\mathbf q),
\qquad
q^2=q_\mu q^\mu=\omega^2-\mathbf q^2.
\label{eq:q2_reorg}
\end{equation}
Equivalently,
\begin{equation}
\pi_{\mu\nu}(q)
=
\eta_{\mu\nu}-\frac{q_\mu q_\nu}{q^2}
=
\frac{\hat{\pi}_{\mu\nu}(q)}{q^2},
\qquad
J_{\mu\nu}^{\mathrm{pole}}(q)=\pi_{\mu\nu}(q)\,u(q).
\label{eq:pi_reorg}
\end{equation}

The components of the pole current are
\begin{equation}
J_{00}^{\mathrm{pole}}(q)
=
-\frac{\mathbf q^2}{q^2}\,u(q),
\qquad
J_{0i}^{\mathrm{pole}}(q)
=
-\frac{\omega q_i}{q^2}\,u(q),
\qquad
J_{ij}^{\mathrm{pole}}(q)
=
-\frac{\delta_{ij}q^2+q_iq_j}{q^2}\,u(q).
\label{eq:Jpole_components_reorg}
\end{equation}
Its trace is local,
\begin{equation}
J^{\mathrm{pole}\,\mu}{}_{\mu}(q)=3\,u(q).
\label{eq:Jpole_trace_reorg}
\end{equation}

To investigate the changes induced by TT decomposition, we introduce the spatial transverse and TT projectors 
\begin{equation}
P^T_{ij}=\delta_{ij}-\frac{q_iq_j}{\mathbf q^2},
\qquad
\Lambda_{ij,kl}
=
\frac12\left(
P^T_{ik}P^T_{jl}
+
P^T_{il}P^T_{jk}
-
P^T_{ij}P^T_{kl}
\right).
\label{eq:projectors_reorg}
\end{equation}
The projected vector and tensor sources are
\begin{equation}
J_T^i(q)=P^T_{ij}J_{0j}(q),
\qquad
J_{TT}^{ij}(q)=\Lambda_{ij,kl}J_{kl}(q).
\label{eq:proj_sources_reorg}
\end{equation}
Since
\begin{equation}
P^T_{ij}q_j=0,
\qquad
\Lambda_{ij,kl}\delta_{kl}=0,
\qquad
\Lambda_{ij,kl}q_kq_l=0,
\end{equation}
the pole contribution satisfies
\begin{equation}
J_T^{i,\mathrm{pole}}(q)=0,
\qquad
J_{TT}^{ij,\mathrm{pole}}(q)=0.
\label{eq:noTTpole_reorg}
\end{equation}
Therefore the anomaly does not modify the TT and vector sectors. Their equations remain those of the ordinary case, with the only difference that the regular part of the source replaces the full one. Thus the only sector modified by the anomaly is the scalar one.

\subsection{Scalar response in the \((\Phi,B,\Psi)\) basis}

To display the anomaly-induced scalar response explicitly, it is convenient to work in the basis
\begin{equation}
\chi_{\bar A}=(\Phi,B,\Psi),
\qquad
\bar A,\bar B\in\{\Phi,B,\Psi\},
\label{eq:PhiBpsi_basis_reorg}
\end{equation}
where we have redefined 
\beq
\Psi\equiv \frac{H}{6}.
\eeq
We define the scalar source components by
\begin{equation}
J_\Phi(q)\equiv J_{00}(q),
\qquad
J_B(q)\equiv iq_iJ_{0i}(q),
\qquad
J_\Psi(q)\equiv \delta_{ij}J_{ij}(q).
\label{eq:scalar_sources_reorg}
\end{equation}
For the pole part one finds
\begin{equation}
J_\Phi^{\mathrm{pole}}(q)
=
-\frac{\mathbf q^2}{q^2}\,u(q),
\qquad
J_B^{\mathrm{pole}}(q)
=
-\,i\omega\,\frac{\mathbf q^2}{q^2}\,u(q),
\qquad
J_\Psi^{\mathrm{pole}}(q)
=
\frac{2\mathbf q^2-3\omega^2}{q^2}\,u(q).
\label{eq:scalar_pole_sources_reorg}
\end{equation}

The scalar kernel in this basis is
\begin{equation}
K_{\bar A\bar B}
=
\begin{pmatrix}
0 & -2\nabla^2\partial_0 & -2\nabla^2\\
2\nabla^2\partial_0 & -2\nabla^4 & 2\nabla^2\partial_0\\
-2\nabla^2 & -2\nabla^2\partial_0 & -6\partial_0^2+2\nabla^2
\end{pmatrix}_{\bar A\bar B},
\label{eq:kernel_PhiBpsi_reorg}
\end{equation}
which in momentum space becomes
\begin{equation}
K_{\bar A\bar B}(\omega,\mathbf q)
=
\begin{pmatrix}
0 & -2i\omega\mathbf q^2 & 2\mathbf q^2\\
2i\omega\mathbf q^2 & -2\mathbf q^4 & 2i\omega\mathbf q^2\\
2\mathbf q^2 & -2i\omega\mathbf q^2 & 6\omega^2-2\mathbf q^2
\end{pmatrix}_{\bar A\bar B}.
\label{eq:kernel_PhiBpsi_mom_reorg}
\end{equation}
The scalar equations are
\begin{align}
-2i\omega\mathbf q^2 B+2\mathbf q^2\Psi+J_\Phi&=0,\notag\\
2i\omega\mathbf q^2\Phi-2\mathbf q^4B+2i\omega\mathbf q^2\Psi+J_B&=0,\notag\\
2\mathbf q^2\Phi-2i\omega\mathbf q^2B+(6\omega^2-2\mathbf q^2)\Psi+J_\Psi&=0.
\label{eq1}
\end{align}
Substituting \eqref{eq:scalar_pole_sources_reorg}, one finds
\begin{equation}
B(q)
=
-\frac{3 i\,u(q)\,\omega^3}
{4 q^2\left(\mathbf q^4+3\omega^2\mathbf q^2-3\omega^4\right)},
\label{eq2}
\end{equation}
\begin{equation}
\Psi(q)
=
\frac{u(q)\left(2\mathbf q^4+6\omega^2\mathbf q^2-3\omega^4\right)}
{4 q^2\left(\mathbf q^4+3\omega^2\mathbf q^2-3\omega^4\right)},
\label{eq3}
\end{equation}
\begin{equation}
\Phi(q)
=
-\frac{3 u(q)\,\omega^2\left(\mathbf q^2+\omega^2\right)}
{4 q^2\left(\mathbf q^4+3\omega^2\mathbf q^2-3\omega^4\right)},
\label{eq:Phisol_reorg}
\end{equation}

These expressions show separate dependence on \(\omega\) and \(\mathbf q^2\). This is expected: \(\Phi\), \(B\), and \(\Psi\) are variables adapted to the \(3+1\) foliation, not covariant scalars. The covariant meaning of the anomaly-induced excitation emerges only after combining them into the unique gauge-invariant scalar.\\
The gauge-invariant scalar combinations of the \(1+3\) decomposition are $\Theta$ and $\Xi$ as defined in \eqref{eq:ThetaXi_reorg}.
In momentum space,
\begin{equation}
\Theta(q)=\Phi(q)+i\omega B(q)-\frac12\omega^2E(q),
\qquad
\Xi(q)=H(q)+\mathbf q^2E(q).
\label{eq:ThetaXi_mom_reorg}
\end{equation}
Connecting these variables to the gauge-invariant scalar combination $\Phi_{cov}$ is also rather straightforward. We can use the four-dimensional transverse projector acting on $h_{\mu\nu}$
\begin{equation}
\Phi_{cov}(q)=\pi^{\mu\nu}(q)h_{\mu\nu}(q),
\label{eq:Phicov_proj_reorg}
\end{equation}
and apply the scalar part of the metric decomposition,
\begin{equation}
h_{00}=2\Phi,
\qquad
h_{0i}=iq_iB,
\qquad
h_{ij}
=
\left(-q_iq_j+\frac13\delta_{ij}\mathbf q^2\right)E
+\frac13\delta_{ij}H,
\label{eq:metric_scalar_mom_reorg}
\end{equation}
to obtain
\begin{equation}
\Phi_{cov}(q)
=
-\frac{2\mathbf q^2}{q^2}\Phi
-\frac{2i\omega\mathbf q^2}{q^2}B
+\frac{2}{3}\frac{\mathbf q^4}{q^2}E
-\left(1+\frac{\mathbf q^2}{3q^2}\right)H.
\label{eq:Phicov_raw_reorg}
\end{equation}
Rewriting this in terms of \(\Theta\) and \(\Xi\), the dependence on \(E\) cancels and one finds
\begin{equation}
\Phi_{cov}(q)
=
-\frac{2\mathbf q^2}{q^2}\,\Theta(q)
-\frac{3\omega^2-2\mathbf q^2}{3q^2}\,\Xi(q).
\label{eq:Phicov_ThetaXi_reorg1}
\end{equation}
This relation shows that \(\Theta\) and \(\Xi\) are gauge invariant within the \(1+3\) decomposition, while the anomaly pole selects only one specific covariant scalar combination, namely \(\Phi_{cov}\). The separate \(\omega\)-dependence of the noncovariant variables \(\Phi\), \(B\), and \(\Psi\) is therefore not in conflict with covariance; rather, it is an artifact of the \(3+1\) parametrization and not a property of the physical scalar channel itself. When the anomaly pole is included, its effect is entirely concentrated in the scalar sector.\\
In the \(3+1\) decomposition this effect is encoded in the nonlocal responses \eqref{eq1}--\eqref{eq3}, while in covariant form it is captured by the unique gauge-invariant scalar \(\Phi_{cov}\), related to the \(1+3\) variables by \eqref{eq:Phicov_ThetaXi_reorg1}. In this sense, the anomaly activates a single scalar channel without modifying the fact that the TT sector remains the only ordinary propagating sector directly governed by the tree-level action. The covariant and non-covariant decompositions are therefore fully consistent at the kinematical level, even though they organize the same content in different ways.\\
Moreover, from \eqref{eq1}, \eqref{eq2}, and \eqref{eq3} one can derive hyperbolic equations for the scalar components, but only in the anomaly-induced case, namely in the presence of the \(u(q)\) term. This is seen by passing to coordinate space and acting with the d'Alembertian operator \(\Box\) on both sides of the equations. Among the resulting relations, the only gauge-invariant one is the equation associated with \eqref{eq:Phicov_ThetaXi_reorg1}, or equivalently with \eqref{cov}. Accordingly, these equations should be interpreted as describing the effective dynamics of the gauge-invariant scalar mode \(\Phi_{cov}\) induced by the anomaly, activating a scalar channel in the interaction.

\section{Collection of derivatives of the Riemann curvatures and their contractions in momentum space} 
\begin{align}
\left[ R^{(1)}_{\mu\alpha\nu\beta} \right]^{\mu_1 \nu_1}(p)
&=
\frac{1}{2}
\left\{ \delta^{(\mu_1}_{\alpha}\,\delta^{\nu_1)}_{\beta}\,p_{\mu} p_{\nu} + \delta^{(\mu_1}_{\mu}\,\delta^{\nu_1)}_{\nu}\,p_{\alpha} p_{\beta} - \delta^{(\mu_1}_{\beta}\,\delta^{\nu_1)}_{\mu}\,p_{\alpha} p_{\nu} - \delta^{(\mu_1}_{\alpha}\,\delta^{\nu_1)}_{\nu}\,p_{\beta} p_{\mu} \right\}.
\\[6pt]
\left[ R^{(1)}_{\mu\nu} \right]^{\mu_1 \nu_1}(p)
&=
\eta^{\alpha\beta}
\left[ R^{(1)}_{\mu\alpha\nu\beta} \right]^{\mu_1 \nu_1}(p) = \frac{1}{2}
\left\{
\eta^{\mu_1 \nu_1}\,p_{\mu} p_{\nu} + \delta^{(\mu_1}_{\mu}\,\delta^{\nu_1)}_{\nu}\,p^2 - p^{(\mu_1}\,\delta^{\nu_1)}_{\mu}\,p_{\nu} - p^{(\mu_1}\,\delta^{\nu_1)}_{\nu}\,p_{\mu}
\right\}.
\\[6pt]
\left[ R^{(1)} \right]^{\mu_1 \nu_1}(p)
& = \eta^{\mu\nu}
\left[ R^{(1)}_{\mu\nu} \right]^{\mu_1 \nu_1}(p) =
p^2 \eta^{\mu_1 \nu_1} - p^{\mu_1} p^{\nu_1} = p^2 \,\pi^{\mu_1 \nu_1}(p).
\end{align}

\begin{align}
\left[
R^{(1)}_{\mu\alpha\nu\beta} R^{(1)\,\mu\alpha\nu\beta}
\right]^{\mu_1 \nu_1 \mu_2 \nu_2}(p_1,p_2)
&\equiv
\left[ R^{(1)}_{\mu\alpha\nu\beta} \right]^{\mu_1 \nu_1}(p_1)\,
\left[
R^{(1)\,\mu\alpha\nu\beta}
\right]^{\mu_2 \nu_2}(p_2)
\nonumber\\
&=
(p_1 \cdot p_2)^2\,
\eta^{\mu_1(\mu_2}\eta^{\nu_2)\nu_1} -
2 (p_1 \cdot p_2)\,
p_1^{(\mu_2}\eta^{\nu_2)(\nu_1} p_2^{\mu_1)} + p_1^{\mu_2} p_1^{\nu_2} p_2^{\mu_1} p_2^{\nu_1}.
\\[6pt]
\left[
R^{(1)}_{\mu\nu} R^{(1)\,\mu\nu}
\right]^{\mu_1 \nu_1 \mu_2 \nu_2}(p_1,p_2)
&\equiv
\left[ R^{(1)}_{\mu\nu} \right]^{\mu_1 \nu_1}(p_1)\,
\left[ R^{(1)\,\mu\nu} \right]^{\mu_2 \nu_2}(p_2)
\nonumber\\
&=
\frac{1}{4} p_1^2
\left( p_2^{\mu_1} p_2^{\nu_1} \eta^{\mu_2 \nu_2} -
2\, p_2^{(\mu_1}\eta^{\nu_1)(\nu_2} p_2^{\mu_2)} \right)
+ \frac{1}{4} p_2^2 \left( p_1^{\mu_2} p_1^{\nu_2} \eta^{\mu_1 \nu_1} - 2\, p_1^{(\mu_2}\eta^{\nu_2)(\nu_1} p_1^{\mu_1)} \right)
\nonumber \\
&\quad+
\frac{1}{4} p_1^2 p_2^2\,
\eta^{\mu_1(\mu_2}\eta^{\nu_2)\nu_1}
+
\frac{1}{4} (p_1 \cdot p_2)^2\,
\eta^{\mu_1 \nu_1}\eta^{\mu_2 \nu_2}
+
\frac{1}{2}
p_1^{(\mu_1} p_2^{\nu_1)}
p_1^{(\mu_2} p_2^{\nu_2)}
\nonumber \\
&\quad+
\frac{1}{2} (p_1 \cdot p_2)
\left(
p_1^{(\mu_1}\eta^{\nu_1)(\nu_2} p_2^{\mu_2)}
-
\eta^{\mu_1 \nu_1} p_1^{(\mu_2} p_2^{\nu_2)}
-
\eta^{\mu_2 \nu_2} p_1^{(\mu_1} p_2^{\nu_1)}
\right).
\end{align}

\begin{table}[]
\centering
\small
\renewcommand{\arraystretch}{1.12}
\setlength{\tabcolsep}{4.5pt}
\begin{tabular}{l l >{$}l<{$} @{\hspace{1.0em} \vrule width 0.75pt\hspace{1em} } l l >{$}l<{$}}
\toprule
Momentum $p_i$ & Polarization $\epsilon_j$ & p_i \cdot \epsilon_j &
Momentum $p_i$ & Polarization $\epsilon_j$ & p_i \cdot \epsilon_j \\
\midrule
$p_1$ & $\epsilon(p_1,\pm)$ & 0 &
$p_3$ & $\epsilon(p_1,\pm)$ & -E\sin\theta/\sqrt{2} \\

$p_1$ & $\epsilon(p_2,\pm)$ & 0 &
$p_3$ & $\epsilon(p_2,\pm)$ & -E\sin\theta/\sqrt{2} \\

$p_1$ & $\epsilon(p_3,\pm)$ & E\sin\theta/\sqrt{2} &
$p_3$ & $\epsilon(p_3,\pm)$ & 0 \\

$p_1$ & $\epsilon(p_4,\pm)$ & -E\sin\theta/\sqrt{2} &
$p_3$ & $\epsilon(p_4,\pm)$ & 0 \\

$p_2$ & $\epsilon(p_1,\pm)$ & 0 &
$p_4$ & $\epsilon(p_1,\pm)$ & E\sin\theta/\sqrt{2} \\

$p_2$ & $\epsilon(p_2,\pm)$ & 0 &
$p_4$ & $\epsilon(p_2,\pm)$ & E\sin\theta/\sqrt{2} \\

$p_2$ & $\epsilon(p_3,\pm)$ & -E\sin\theta/\sqrt{2} &
$p_4$ & $\epsilon(p_3,\pm)$ & 0 \\

$p_2$ & $\epsilon(p_4,\pm)$ & E\sin\theta/\sqrt{2} &
$p_4$ & $\epsilon(p_4,\pm)$ & 0 \\
\bottomrule
\end{tabular}
\caption{Dot products of momenta and polarization vectors $p_i \cdot \epsilon(p_j,\lambda_j)$}
\end{table}

\begin{table}[]
\centering
\small
\renewcommand{\arraystretch}{1.12}
\setlength{\tabcolsep}{4.5pt}
\begin{tabular}{cc c @{\hspace{1.0em} \vrule width 0.75pt\hspace{1em} } cc c}
\toprule
$\epsilon_i$ & $\epsilon_j$ & $\epsilon_i\!\cdot\!\epsilon_j$ &
$\epsilon_i$ & $\epsilon_j$ & $\epsilon_i\!\cdot\!\epsilon_j$ \\
\midrule
$\epsilon(p_1,+)$ & $\epsilon(p_1,+)$ & $0$ &
$\epsilon(p_2,-)$ & $\epsilon(p_2,-)$ & $0$ \\

$\epsilon(p_1,+)$ & $\epsilon(p_1,-)$ & $-1$ &
$\epsilon(p_2,-)$ & $\epsilon(p_3,+)$ & $-\frac{t}{s}$ \\

$\epsilon(p_1,+)$ & $\epsilon(p_2,+)$ & $-1$ &
$\epsilon(p_2,-)$ & $\epsilon(p_3,-)$ & $-\frac{s+t}{s}$ \\

$\epsilon(p_1,+)$ & $\epsilon(p_2,-)$ & $0$ &
$\epsilon(p_2,-)$ & $\epsilon(p_4,+)$ & $\frac{s+t}{s}$ \\

$\epsilon(p_1,+)$ & $\epsilon(p_3,+)$ & $-\frac{t}{s}$ &
$\epsilon(p_2,-)$ & $\epsilon(p_4,-)$ & $\frac{t}{s}$ \\

$\epsilon(p_1,+)$ & $\epsilon(p_3,-)$ & $-\frac{s+t}{s}$ &
$\epsilon(p_3,+)$ & $\epsilon(p_3,+)$ & $0$ \\

$\epsilon(p_1,+)$ & $\epsilon(p_4,+)$ & $\frac{s+t}{s}$ &
$\epsilon(p_3,+)$ & $\epsilon(p_3,-)$ & $-1$ \\

$\epsilon(p_1,+)$ & $\epsilon(p_4,-)$ & $\frac{t}{s}$ &
$\epsilon(p_3,+)$ & $\epsilon(p_4,+)$ & $1$ \\

$\epsilon(p_1,-)$ & $\epsilon(p_1,-)$ & $0$ &
$\epsilon(p_3,+)$ & $\epsilon(p_4,-)$ & $0$ \\

$\epsilon(p_1,-)$ & $\epsilon(p_2,+)$ & $0$ &
$\epsilon(p_3,-)$ & $\epsilon(p_3,-)$ & $0$ \\

$\epsilon(p_1,-)$ & $\epsilon(p_2,-)$ & $-1$ &
$\epsilon(p_3,-)$ & $\epsilon(p_4,+)$ & $0$ \\

$\epsilon(p_1,-)$ & $\epsilon(p_3,+)$ & $-\frac{s+t}{s}$ &
$\epsilon(p_3,-)$ & $\epsilon(p_4,-)$ & $1$ \\

$\epsilon(p_1,-)$ & $\epsilon(p_3,-)$ & $-\frac{t}{s}$ &
$\epsilon(p_4,+)$ & $\epsilon(p_4,+)$ & $0$ \\

$\epsilon(p_1,-)$ & $\epsilon(p_4,+)$ & $\frac{t}{s}$ &
$\epsilon(p_4,+)$ & $\epsilon(p_4,-)$ & $-1$ \\

$\epsilon(p_1,-)$ & $\epsilon(p_4,-)$ & $\frac{s+t}{s}$ &
$\epsilon(p_4,-)$ & $\epsilon(p_4,-)$ & $0$ \\

$\epsilon(p_2,+)$ & $\epsilon(p_2,+)$ & $0$ &
$\epsilon(p_2,+)$ & $\epsilon(p_2,-)$ & $-1$ \\

$\epsilon(p_2,+)$ & $\epsilon(p_3,+)$ & $-\frac{s+t}{s}$ &
$\epsilon(p_2,+)$ & $\epsilon(p_3,-)$ & $-\frac{t}{s}$ \\

$\epsilon(p_2,+)$ & $\epsilon(p_4,+)$ & $\frac{t}{s}$ &
$\epsilon(p_2,+)$ & $\epsilon(p_4,-)$ & $\frac{s+t}{s}$ \\
\bottomrule
\end{tabular}
\caption{Dot products of polarization vectors.}
\end{table}

\subsection{Regularizations}
The two-dimensional Fourier transform
\begin{equation}
I(b) \;=\; \int \frac{d^2q}{(2\pi)^2}\,\frac{e^{i \vec{q} \cdot \vec b}}{q^2},
\end{equation}
which enters the eikonal phase can be handled by a cutoff as well as in dimensional regularization (DR). The integral is logarithmically divergent in the infrared (small $q\equiv |\vec{q}|$). We regularize it in two ways.
We consider the integral
\begin{equation}
I(b;\mu,\Lambda) \;=\; \frac{1}{2\pi}\int_{\mu}^{\Lambda} \frac{dq}{q}\, J_0(qb),
\end{equation}
with $\mu$ an infrared cutoff (momentum) and $\Lambda$ an ultraviolet cutoff (momentum), that we rewrite as
\begin{equation}
I(b;\mu,\Lambda) \;=\; \frac{1}{2\pi}\int_{\mu b}^{\Lambda b} \frac{dx}{x}\, J_0(x).
\end{equation}
Introduce $x=1$ as an intermediate point to separate the small-$x$ (IR) and large-$x$ (UV) contributions:
\begin{equation}
I(b;\mu,\Lambda) \;=\; \frac{1}{2\pi}\left[\int_{\mu b}^{1} \frac{dx}{x}\, J_0(x)
+ \int_{1}^{\Lambda b} \frac{dx}{x}\, J_0(x)\right].
\end{equation}
For $x \ll 1$
\begin{equation}
\frac{J_0(x)}{x} = \frac{1}{x} - \frac{x}{4} + \mathcal{O}(x^3),
\end{equation}
therefore
\begin{equation}
\int_{\mu b}^{1} \frac{dx}{x} J_0(x)
= \ln\!\frac{1}{\mu b} - \frac{1}{8}\left(1 - (\mu b)^2\right). 
\end{equation}
The IR contribution contains the universal term $\ln\!\frac{1}{\mu b}$
which provides the $-\ln b$ dependence.\\
For $x \gg 1$,
\begin{equation}
J_0(x) \;\sim\; \sqrt{\frac{2}{\pi x}} \cos\!\Big(x - \tfrac{\pi}{4}\Big).
\end{equation}
Hence
\begin{equation}
\frac{J_0(x)}{x} \;\sim\; \sqrt{\frac{2}{\pi}} \frac{\cos(x-\pi/4)}{x^{3/2}}.
\end{equation}
The corresponding integral
\begin{equation}
\int_{1}^{\Lambda b} \frac{dx}{x} J_0(x)\to \int_1^\infty  \frac{dx}{x} J_0(x), \qquad  (\Lambda \to \infty) 
\end{equation}
is convergent as $\Lambda b \to \infty$ since the integrand decays like $1/x^{3/2}$. The oscillations ensure that the limit approaches a {finite constant}, independent of $b$.
Combining both regions we get
\begin{equation}
I(b;\mu) \;=\; \frac{1}{2\pi}\left[\ln\!\frac{1}{\mu b} \;+\; C \right],
\end{equation}
where $C$ is a finite constant depending on the UV details but {independent of $b$}. Equivalently, we can write
\begin{equation}
I(b) \;=\; \frac{1}{2\pi} \ln\!\left(\frac{L}{b}\right),
\end{equation}
where $L$ is an effective length scale absorbing the constants and the regulator dependence.\\
Alternatively, we can use dimensional regularization (DR).
Now compute in $d=2+\epsilon$ transverse dimensions
\begin{equation}
I(b) = \mu^\epsilon \int \frac{d^{2 +\epsilon}q}{(2\pi)^{2 +\epsilon}} \,\frac{e^{i \vec q \cdot \vec b}}{q^2},
\end{equation}
where $\mu$ is an arbitrary mass scale introduced to keep the integral dimensionless.

The general Fourier transform formula is
\begin{equation}
\int \frac{d^d q}{(2\pi)^d}\, \frac{e^{i \vec q \cdot \vec b}}{(q^2)^\alpha}
= \frac{1}{(4\pi)^{d/2}} \,\frac{\Gamma\!\left(\tfrac{d}{2}-\alpha\right)}{\Gamma(\alpha)}
\left(\frac{4}{b^2}\right)^{\tfrac{d}{2}-\alpha}.
\end{equation}
Here $d=2 + \epsilon$, with $\epsilon>0$ since we are regulating an IR divergence, and $\alpha=1$. Thus
\begin{equation}
I(b) = \frac{1}{(4\pi)^{1+\epsilon/2}} \,\Gamma\!\Big(\tfrac{\epsilon}{2}\Big)\,
\left(\frac{4}{b^2}\right)^{\epsilon/2} \,\mu^{-\epsilon}.
\end{equation}
Expand for small $\epsilon$ and collecting terms we obtain
\begin{equation}
I(b) = \frac{1}{4\pi}\left[\frac{2}{\epsilon} - \gamma_E + \ln 4\pi - 2\ln(\mu b) + \mathcal{O}(\epsilon)\right].
\end{equation}
We subtract the pole and scheme-dependent constants (in $\overline{\text{MS}}$) to obtain
\begin{equation}
I(b) = -\frac{1}{2\pi}\ln(\mu b)
\end{equation}
in agreement with the result obtained by the cutoff regularization.


\providecommand{\href}[2]{#2}\begingroup\raggedright\endgroup

\end{document}